\documentclass[11pt,a4paper]{article}
\usepackage[dvipsnames]{xcolor}
\usepackage{pgf}
\usepackage{jheppub}
\usepackage[T1]{fontenc}
\usepackage{amsmath}
\usepackage{amssymb}
\usepackage{tikz-cd}
\usepackage{tikz}
\usetikzlibrary{arrows.meta,graphs}
\usetikzlibrary{calc,math}
\usepackage{tikz-feynman}
\usepackage{pgfplots}
\tikzfeynmanset{compat=1.1.0}
\usetikzlibrary{arrows.meta, calc}
\usetikzlibrary{decorations.pathreplacing}
\usepackage{amsfonts}
\usepackage{bm}
\usepackage{bbold}
\usepackage{amsthm}
\usepackage{slashed}
\usepackage{cleveref}
\crefrangeformat{equation}{(#3#1#4)--(#5#2#6)}
\usepackage{hyperref}
\usepackage{mathtools}
\usepackage{mathrsfs}
\usepackage{array}
\newcolumntype{C}[1]{>{\centering\arraybackslash}m{#1}}

\newcommand{\la}{\langle}
\newcommand{\ra}{\rangle}

\newcommand{\lexp}{\langle\!\langle}
\newcommand{\rexp}{\rangle\!\rangle}

\newcommand{\lexpBig}{{\Big\langle}\!\!{\Big\langle}}
\newcommand{\rexpBig}{{\Big\rangle}\!\!{\Big\rangle}}

\newcommand{\mbf}{\mathbf}
\newcommand{\ang}{\langle \mbf{12}\rangle}
\newcommand{\sq}{[\mbf{12}]}
\newcommand{\angsq}{\langle \mbf{12}]}
\newcommand{\sqang}{[\mbf{12}\rangle}
\newcommand{\SD}{\langle \mbf{1}3\rangle\langle 3\mbf{2}]}
\newcommand{\ASD}{[\mbf{1}3\ra\la 3\mbf{2}\ra}

\newcommand{\al}{\alpha}
\newcommand{\dt}{\dot{\al}}

\makeatletter
\newcommand{\doublewidetilde}[1]{{%
  \mathpalette\double@widetilde{#1}%
}}
\newcommand{\double@widetilde}[2]{%
  \sbox\z@{$\m@th#1\widetilde{#2}$}%
  \ht\z@=.9\ht\z@
  \widetilde{\box\z@}%
}
\makeatother

\newcommand{\singlebox[1]}{%
\tikz[baseline={(current bounding box.center)}, scale = #1,transform shape]{
  \draw (0,0) rectangle (0.6,0.6);
}}

\newcommand{\doublebox[1]}{%
\tikz[baseline={(current bounding box.center)}, scale = #1,transform shape]{
  \draw (0,0) rectangle (0.6,0.6);
  \draw (0,0) rectangle (0.6,-0.6);
}}
\newcommand{\TSlabelnew[1]}{%
\tikz[baseline={(current bounding box.center)}, scale = #1,transform shape]{
  \node at (0,0) {$\singlebox[1]^{\,2s}$};
}}
\newcommand{\SDlabelnew}[1]{%
\tikz[baseline=-0.5ex, scale = #1,transform shape]{
  \node at (0,0) {$\doublebox[1]^{\,n}$};
  \node at (1.4,0.32) {$\singlebox[1]^{\,s-n}$};
}}

\author{
Lucile Cangemi}
\author{\& Iustin Surubaru}

\affiliation{
School of Mathematics and Maxwell Institute for Mathematical Sciences,\\ University of Edinburgh,
EH9 3FD, U.K. }

\emailAdd{lucile.cangemi@ed.ac.uk}
\emailAdd{iustin.surubaru@ed.ac.uk}

\title{Five-dimensional Geometry from Spinning Amplitudes}

\abstract{Massive spinor-helicity variables in four dimensions are a useful tool for studying amplitudes between higher-spin fields and gravitons. Among them a special, simple set of amplitudes reproduces the linearized stress-energy tensor of a Kerr black hole in the classical limit. In this work we initiate the study of the classical limit of three-point spinor-helicity amplitudes in five dimensions. We introduce the map between the spinor invariants and the expectation values of spin operators and match the amplitude building blocks with those of the multipole expansion. Interestingly, in order to take the classical limit of a general amplitude, we need to augment the multipole structures with the Hodge dual of the classical spin tensor. We study the classical limit of alternative spinning states not described by fully-symmetric products of polarisations and conclude that they can describe the same spacetimes. Finally, by relaxing the orthogonality condition of the spin tensor we are able to model spacetimes with a single rotational isometry and match these to the classical limit of amplitudes allowing for an internal spin shift. Along the way we also identify the class of amplitudes describing the Myers-Perry black hole and comment on its generalization to arbitrary dimensions.} 

\makeatletter
\gdef\@fpheader{}
\makeatother

\begin{document}

\maketitle
\section{Introduction}

In the context of scattering amplitudes, spinor-helicity variables appear as an efficient tool for deriving properties that would otherwise be obscured by the Feynman diagram computations \cite{Dixon:1996wi}. In four-dimensional massless scattering of gluons and gravitons they streamlined the derivation of remarkable formulae such as the Parke-Taylor \cite{Parke:1986gb} and Hodges \cite{Hodges:2012ym} expressions for tree level MHV scattering to arbitrary multiplicity. Amplitude factorization, soft and collinear limits become transparent in spinor-helicity. In a supersymmetric context, they simplify the SUSY Ward identities and connect to twistor geometry \cite{Witten:2003nn,Mason:2010yk,Boels:2006ir}, the positive Grassmannian \cite{Arkani-Hamed:2012zlh} and the gauge/gravity duality \cite{Bern:2008qj}. 
\par Massive spinor-helicity variables have played an increasingly important role in amplitude-based approaches to classical gravitational dynamics. While the focus of massless spinor-helicity variables has often been on describing spin-1 or spin-2 interactions, the massive spinor-helicity formalism provides a way of constructing consistent amplitudes of particles of arbitrary spin, bypassing the Lagrangian approach. The external spin-$s$ states can be constructed directly as little-group covariant tensors using massive spinors, which makes it possible to classify the low-point amplitudes using only Lorentz invariance, factorization, locality, and little-group covariance \cite{Arkani-Hamed:2017jhn, Conde:2016vxs, Conde:2016izb, Penrose:1972ia}. In this way, massive higher-spin amplitudes provide a natural on-shell parametrisation of the data entering effective field theories of compact gravitating objects.
\par This formalism is particularly useful for black hole physics as the spin dependence of a massive amplitude is directly tied to the multipole expansion of the corresponding classical source. For a spinning compact object, gravitational observables are organised in powers of the spin tensor, which is a covariant repackaging of the mass and current multipole moments. Amplitudes involving two massive spinning particles and one or more gravitons therefore encode the couplings of these multipoles to the gravitational field. Already at low orders in spin, massive higher-spin amplitudes were shown to reproduce the structures relevant for the spin-induced multipole moments of black holes \cite{Vaidya:2014kza}. Subsequent work has developed this idea into a systematic programme in which three-point amplitudes, Compton amplitudes, and higher-point amplitudes are used as on-shell input for extracting the classical potential or the classical observables directly.

\par In this context, the classical limit of a quantum amplitude means isolating the part of the amplitude that survives when the intrinsic spin and orbital angular momentum are large compared to $\hbar$, while the classical quantities such as the masses, impact parameter, angular momenta, and spin tensors are held fixed \cite{Vines:2017hyw,Guevara:2017csg,Chung:2018kqs,Guevara:2019fsj,Maybee:2019jus,Chung:2020rrz, Aoude_2020, Bern_2021, Aoude:2021oqj}. The resulting classical amplitude data can then be mapped to the effective Hamiltonian~\cite{
Chung:2019duq,Chung:2020rrz, Bern_2021, Chen:2021kxt,Bern_2023,Alaverdian:2024spu,Alaverdian:2025jtw} or directly to observables through an eikonal/radial-action calculation \cite{KoemansCollado:2019ggb, Bjerrum_Bohr_2020,Aoude:2020ygw, Bern:2020uwk,DiVecchia:2021bdo,Heissenberg:2021tzo,Haddad:2021znf,Adamo:2021rfq,Menezes:2022tcs,DiVecchia:2022piu,Bellazzini:2022wzv,Damgaard_2022,
Luna:2023uwd,Gatica:2023iws,Georgoudis:2023eke,DiVecchia:2023frv,Heissenberg_2023,Heissenberg:2023uvo,
Fernandes:2024xqr,Chen:2024mmm,Chen:2024bpf,Du:2024rkf,Akpinar:2024meg, Akpinar:2025bkt,Aoude:2025xxq,Kim:2025hpn}, a KMOC-type expectation-value formalism \cite{
Maybee:2019jus,
Cristofoli:2021vyo,Cristofoli:2021jas,
Cristofoli:2022phh,Adamo:2022rmp,Adamo:2022qci,
Bohnenblust:2024hkw,
Aoude:2023dui,De_Angelis_2024,Brandhuber_2024,
Alessio:2025flu,Doran:2026bng}, or by matching onto a worldline effective theory \cite{Goldberger:2004jt,Porto:2005ac,Porto:2006bt,Goldberger:2007hy,Kol:2007bc,Porto:2008tb,Porto:2008jj,Levi:2008nh,Goldberger:2009qd,Porto:2010tr,Porto:2010zg,Levi:2010zu,Levi:2011eq,Porto:2012as,Foffa:2013qca,Levi:2014gsa,Levi:2014sba,Levi:2015msa,Levi:2015uxa,Levi:2015ixa,Levi:2016ofk,Foffa:2016rgu,Maia:2017yok,Maia:2017gxn, 
Kalin:2020mvi,Levi:2020kvb,Levi:2020uwu,Mogull:2020sak,
Liu:2021zxr,Jakobsen:2021smu,Jakobsen:2021lvp,Jakobsen:2021zvh,
Jakobsen_2022,Comberiati:2022cpm,Ben-Shahar:2023djm,Damgaard:2023vnx,Jakobsen_2023,Bohnenblust:2023qmy,
Gonzo_2024,Haddad:2024ebn,Mogull:2025cfn,Bonocore:2025stf,Hoogeveen:2025tew,Bohnenblust:2026ujk}.


\par In the latter approach, the Wilson coefficients in the worldline action encode the spin-induced multipole moments of the compact object. Massive higher-spin amplitudes give an on-shell way to determine the same data: the spin dependence of the amplitude fixes the allowed multipole couplings, while matching to known solutions such as Kerr fixes the corresponding Wilson coefficients. From this perspective, the amplitude and worldline descriptions are complementary. The worldline action gives a spacetime EFT interpretation of the multipoles, whereas the massive spinor-helicity amplitude packages the same information into compact on-shell building blocks.

\par While a great deal of effort has been put into the analysis of four-dimensional massive higher-spin amplitudes and their classical limits, much less is known about their higher-dimensional analogues. In five dimensions, a classification of arbitrary spin three-point spinor-helicity amplitudes involving either massive or massless particles has been performed in refs.~\cite{Chiodaroli:2022ssi,Pokraka:2024fao}. For massless scattering, a similar analysis has been performed in six dimensions \cite{Cheung:2009dc, Monteiro:2018xev}. Although higher-dimensional gravitational theories are of less direct observational relevance, they play an important role for the theoretical understanding of gravitational dynamics. They exhibit black holes and black objects with no four-dimensional analogue, including Myers-Perry black holes \cite{Myers:1986un}, black rings \cite{Emparan:2001wn}, and more general multi-horizon configurations \cite{Elvang:2007rd}. They also arise naturally in string theory, supergravity, and compactifications, where higher-dimensional scattering data can descend to lower-dimensional EFT couplings. Understanding higher-dimensional massive spinor-helicity amplitudes is therefore useful both as a structural extension of the on-shell programme and as a way of probing the multipole data of higher-dimensional gravitational solutions. Work on understanding the multipole expansion of gravitational solutions in an arbitrary number of dimensions has been performed in \cite{Heynen:2023sin,Gambino:2024uge,Gambino:2025iyx,Bianchi:2024shc,Bianchi:2025xol, Monteiro:2026rby}. The authors of \cite{Gambino:2024uge,Bianchi:2025xol, Campanella:2026wqt} also took the first steps towards recovering the leading multipole data of higher-dimensional black-hole geometries from three-point amplitudes of massive $S\leq 1$ fields coupled to a graviton.
\par In this paper, we provide a systematic classical map of the five-dimensional three-point spinor-helicity amplitudes from \cite{Pokraka:2024fao}. On the amplitudes side, we outline the map from the 5D spinor-helicity basis of the amplitudes to the expectation value of the spin operator where the classical limit becomes clear. On the gravitational side, we construct a classical massive spinning worldline coupled linearly to the Riemann tensor and, by varying it with respect to the metric, match the stress-energy tensor of the massive worldline to the expressions derived in \cite{Bianchi:2024shc}. The explicit maps provided in this work are between the amplitude coefficients, often denoted $g_{v,i}$ and the worldline Wilson coefficients $C_{v,i}$.

\begin{table}[!t]
\centering
{
\renewcommand{\arraystretch}{1.8}
\setlength{\tabcolsep}{8pt}

\begin{tabular}{|C{5.4cm}|C{4cm}|C{4cm}|}
\hline
\textbf{Schematic} &
\textbf{little group labels $(n_{L,i},n_{R,i})$, $i \in \{1,2\}$} &
\textbf{necessary d.o.f in $S_{EFT}$}
\\
\hline\hline

\makebox[5.4cm][c]{%
\begin{tikzpicture}[scale=0.8, transform shape, baseline={(current bounding box.center)}]
\begin{feynman}
    \vertex (a) at (-2,1);
    \vertex (b) at (2,1);
    \vertex (c) at (0,-1.4);
    \vertex (v) at (0,0);

    \diagram* {
        (a) -- [fermion, thick, edge label'={{\TSlabelnew[0.6]}}] (v),
        (b) -- [fermion, thick, edge label={{\TSlabelnew[0.6]}}] (v),
        (v) -- [photon, thick] (c),
    };
\end{feynman}
\end{tikzpicture}
}
&
$\displaystyle n_{L,i} = n_{R,i} = 2s$
&
$\displaystyle S_{EFT}(S^{\mu\nu})$
\\
\hline

\makebox[5.4cm][c]{%
\begin{tikzpicture}[scale=0.8, transform shape, baseline={(current bounding box.center)}]
\begin{feynman}
    \vertex (a) at (-2,1);
    \vertex (b) at (2,1);
    \vertex (c) at (0,-1.4);
    \vertex (v) at (0,0);

    \diagram* {
        (a) -- [fermion, thick, edge label'={{\SDlabelnew{0.6}} {\tiny mSD}}] (v),
        (b) -- [fermion, thick, edge label={{\SDlabelnew{0.6}} {\tiny mASD}}] (v),
        (v) -- [photon, thick] (c),
    };
\end{feynman}
\end{tikzpicture}
}
&
$\displaystyle
\begin{aligned}
n_{L,1} &= n_{R,2} = 2s\\
n_{R,1} &= n_{L,2} = 2s-2n
\end{aligned}$
&
$\displaystyle S_{EFT}(S^{\mu\nu})$
\\
\hline

\makebox[5.4cm][c]{%
\begin{tikzpicture}[scale=0.8, transform shape, baseline={(current bounding box.center)}]
\begin{feynman}
    \vertex (a) at (-2,1);
    \vertex (b) at (2,1);
    \vertex (c) at (0,-1.4);
    \vertex (v) at (0,0);

    \diagram* {
        (a) -- [fermion, thick, edge label'={{\SDlabelnew{0.6}} {\tiny mSD}}] (v),
        (b) -- [fermion, thick, edge label={{\SDlabelnew{0.6}} {\tiny mSD}}] (v),
        (v) -- [photon, thick] (c),
    };
\end{feynman}
\end{tikzpicture}
}
&
$\displaystyle
\begin{aligned}
n_{L,i} &= 2s\\
n_{R,i} &= 2s-2n
\end{aligned}$
&
$\displaystyle S_{EFT}(S^{\mu\nu},K^{\mu})$
\\
\hline
\end{tabular}
}

\caption{In five dimensions the massive irreducible representations are labelled by $(p,n_{L},n_{R})$, with momentum $p$ and $n_{L,R}$ count spin in the left/right little group $SU(2)$ copies. We always consider mass-conserving scattering such that $p_1^2 = p_2^2 =m^2$ and $p_1 + p_2 + k =0$, where $k^2=0$ is the graviton momentum.}
\label{tab:Results}
\end{table}

\par When scattering massive fields in four dimensions, the little group $SU(2)$ allows a unique fully-symmetric on-shell state at spin $s$. The same massive little group classification differs in five dimensions, where the spin representation is characterised by two labels $(n_L, n_R)$, corresponding to factorisation of the little group into the left and right $SU(2)$ groups. The five-dimensional states include a generalization of the four-dimensional totally-symmetric states that have equal amounts of left and right little group spinors. Unequal little group contributions, $n_{L}\neq n_{R}$ lead to mixed symmetry tensors that split into self-dual (SD) and anti-self-dual (ASD) sectors. For this category of states, scattering can interpolate between a mixed-SD and a mixed-ASD state or two mixed-(A)SD ones\footnote{We work in an all incoming configuration where an incoming SD field is an outgoing ASD one and vice versa. Hence, what we call SD-ASD amplitudes preserve the little group spin of the massive field, while SD-SD swap the two components.}. Thus, we consider three main sectors as summarised in Table \ref{tab:Results}. While physically different, the obvious classical limit of the fully-symmetric amplitude involving $s\rightarrow\infty$ and one of the classical limits of the SD-ASD scenario, involving both the $n_L$ and $n_R$ diverging, lead to the same multipole expansion in five dimensions. We find, however, that for unequal spins, fixing one of the little group factors, while scaling the second one (for example $n_L \to \infty, n_R\sim \mathcal{O}(1)$) leads to a truncated multipole expansion in the infinite spin limit. 
\par In both cases the classical object is described by its spin tensor $S^{\mu\nu}$ and its momentum $p_\mu$. However, the scattering of two self-dual or anti-self-dual states leads, in the classical limit, to multipole moments involving the boost component of the classical angular momentum $K^\mu$. In four dimensions, this boost component has been identified when considering three-point functions involving scattering states in a superposition of a spin $s$ and a spin $s-1$ field. The overlap between the higher and lower spin states leads to the boost vector expectation value \cite{Bern:2023ity}. In five dimensions, the analogue of the overlap between different spin states is the overlap of states with the same total spin, but a reshuffling of the left and right little group components between the two fields, the SD-SD case being part of this category in our all-incoming conventions. This is not very surprising since a spin-1 state in five dimensions decomposes exactly into a superposition of a four-dimensional spin-1 and a scalar under dimensional reduction. 
\par One of the interesting features of the five-dimensional three-point amplitudes is their behaviour under parity. In four dimensions, parity is known to relate the amplitude coefficients between the scattering of a positive and a negative helicity graviton. For a fixed helicity, however, the classical limit leads to a parity-covariant classical stress-energy tensor, i.e. the Levi-Civita tensors always come in pairs. In five dimensions, the natural classical limit of the spinor-helicity amplitudes leads to a more general expansion than the one in \cite{Bianchi:2024shc} that involves not only the spin tensor $S^{\mu\nu}$, but also its Hodge dual $\tilde{S}^{\mu\nu}=\varepsilon^{\mu\nu\rho\tau\sigma}S_{\rho\tau}p_\sigma$. We are able to extend the worldline action to include couplings involving $\tilde{S}^{\mu\nu}$ and match these to the unconstrained amplitude coefficients. 
\par For the fully symmetric amplitudes, a simple constraint between its coefficients imposes parity invariance. For the unequal spin amplitudes, parity is more subtle as its action maps between different amplitudes instead of relating the coefficients in a single expansion, much like in four dimensions. We will see however that we can still lift the classical parity invariance to a relation between the leading order terms of the amplitude coefficients in their infinite spin expansion. 
\par For all of the above analysis, the classical limit has been taken for scattering of definite spin states. While a general worldline EFT involving $S^{\mu\nu},\ \tilde{S}^{\mu\nu}$ can be recovered from the infinite spin limit of such states, the classical data in the SD-SD scattering scenario only recovers a single $K^\mu$ term in the multipole expansion. It turns out that an infinite expansion in $K^\mu$ can be recovered from the classical limit of a coherent state, i.e. an infinite superposition of definite spin states. We make this precise and discuss the relevant maps in the last section of the paper.

We start, in section \ref{sec:littlegroupsec}, by reviewing the physical states, constructing explicit representations of the little group relevant for massive and massless states in four and five spacetime dimensions. In section \ref{sec:review4d}, we review the four-dimensional scattering story, introducing two choices of kinematic variables relevant to massive-massive-massless three-point scattering: $4$D spinor-helicity variables and spin-operator variables. The latter variables are motivated by the  classical variables of the worldline EFT reviewed in section \ref{sec:4dWL}. In section \ref{sec:5dStates}, we construct the spinor-helicity and spin-operator variables valid in five-dimensions and introduce the equivalent to the mass- and spin-conserving scattering studied in $4$d. Section \ref{sec:5dCWL} discusses the construction of the classical worldline action $S_{EFT}$ in general dimensions, how to extend it to cover a larger class of solution in $5$d. We also mention how the Wilson coefficients are matched to specific solutions, with results for Myers-Perry and a limit of the black ring. Sections \ref{sec:5dTotSymm} and \ref{sec:5dSDscatt} detail the map from the three-point amplitudes to the classical multipole expansion and what degrees of freedom are required in the corresponding $S_{EFT}$. Finally in section \ref{sec:KVecScat} we discuss a looser definition of spin-conserving scattering, allowing internal spin changes in the massive little group. In this case, we can construct coherent states of massive spinning particles with fixed total spin but distributed arbitrarily in little group.

\section{A review of little group representations in $D=4$ and $D=5$}
\label{sec:littlegroupsec}

Since we would like to study scattering amplitudes in $D=5$, in particular those between two massive particles and a massless graviton, we should first spend some time discussing which particles we can even scatter. This corresponds to studying the irreducible unitary representations of the Poincar\'e group, labelled by their momentum $p$ and their transformation under a representation of the little group \cite{Wigner:1939cj, Weinberg:1964cn, PhysRev.134.B882}. 

Since the little group is defined as the set of Lorentz transformations that leave the momentum invariant, it will differ based on whether the state is massless or not. The little group for massive states, $p^2=m^2$, is $SO(D-1)$, while for finite representations of massless states, $p^2 =0$, the little group is $SO(D-2)$. 

The irreducible representations of $SO(N)$ are characterised by the Young diagrams $\lambda$ which satisfy $h^{\lambda} \leq \lfloor N/2 \rfloor$ where $h^{\lambda}$ is the maximum height of the Young diagram. For example, valid Young diagrams for $SO(3)$ can have height at most $1$ and have the form
\begin{equation}
\label{youngsym}
\lambda = \underbrace{
\begin{tikzpicture}[baseline={(current bounding box.center)}]
  \draw (0,0) rectangle (0.6,0.6);
  \draw (0.6,0) rectangle (1.2,0.6);
  \draw (1.2,0) rectangle (1.8,0.6);
  \node at (2.2,0.3) {$\cdots$};
  \draw (2.8,0) rectangle (3.4,0.6);
\end{tikzpicture}
}_{s}\,, \qquad h^{\lambda} = 1\,.
\end{equation}

One can read off the symmetry properties of the states from the diagram. In the case of $SO(3)$, the valid states are totally symmetric and labelled by the number of symmetrised indices $s$, also called spin. The Lorentz covariant polarisations for the state labelled by momentum $p$ can then be constructed from transverse, null spin-$1$ polarisations $\varepsilon(p)$, 
\begin{equation}
    \varepsilon(p)^{\mu_1 \dots \mu_s}_{(i_1 \dots i_s)} = \varepsilon(p)^{\mu_1}_{i_1} \dots \varepsilon(p)^{\mu_s}_{i_s} \,, \quad p \cdot \varepsilon(p) = 0 \,.
\end{equation}
Since $SO(3)$ is the little group for the massive particles in $D=4$ and the massless particles in $D=5$, they are both totally symmetric states, labelled by a spin $s$, but with differing on-shell constraints.

There is a slight caveat for even group dimension, $N = 2n$, where the representations saturating the bound $h^{\lambda}_1 = n$ are no longer irreducible; they admit a self-dual and anti-self-dual decomposition using the $N$-dimensional antisymmetric tensor $\epsilon$. For example, consider the little group for massless $D=4$ states, $SO(2)$. The only non-trivial representation is the totally symmetric representation, however this can be further reduced by the action of the antisymmetric tensor $\epsilon_{ij}$, $\epsilon_{12} = -i= - \epsilon_{21}$. In the spin-$1$ case we are left with the two helicities of the photon/gluon 
\begin{equation}
    \varepsilon_{-}\sim \varepsilon^{\mu}_{1} + \epsilon_{12} \, \varepsilon^{\mu}_2\,, \quad \varepsilon_{+}\sim \varepsilon^{\mu}_{2} + \epsilon_{21} \, \varepsilon^{\mu}_1\,. 
\end{equation}
Note that in $D=4$, if we parametrise a photon momentum $k^{\mu} = (1,0,0,-1)$ then the equation above defines the two complex circular polarisations for a photon, in terms of the two real polarisations  $\varepsilon^{\mu}_{1} = (0,1,0,0)$ and $\varepsilon^{\mu}_{2} = (0,0,1,0)$. Higher-spin representations are still totally symmetric combinations of these two polarisations, so there are still two possible irreducible representations at a given spin $s$ labelled by the helicity $h = \pm s$. 

The massive states in $D=5$, which have the little group $SO(4)$ need similar treatment. In this case the non-trivial Young diagrams have either height $h^{\lambda}_1 = 1$ or  $h^{\lambda}_1 = 2$. The first case reproduces the totally symmetric states, familiar from $D=4$ and labelled by spin $s$.

The latter, representations of height two are labelled by the length of the top row, $s$ and the number of columns with height $2$, i.e. pairs of anti-symmetrised indices, which we label $n$. The corresponding Young diagram

\begin{equation}
\label{youngsplit}
\lambda =
\begin{tikzpicture}[baseline={(current bounding box.center)}]
  \draw (0,0) rectangle (0.6,0.6);
  \draw (0.6,0) rectangle (1.2,0.6);
  \node at (1.6,0.3) {$\cdots$};
  \draw (2.0,0) rectangle (2.6,0.6);
  \draw (2.6,0) rectangle (3.2,0.6);
  \node at (3.6,0.3) {$\cdots$};
  \draw (4.0,0) rectangle (4.6,0.6);
  \draw (0,0) rectangle (0.6,-0.6);
  \draw (0.6,0) rectangle (1.2,-0.6);
  \draw (2.0,0) rectangle (2.6,-0.6);
  \draw[decorate,decoration={brace,amplitude=5pt}]
  (0,0.75) -- (4.6,0.75)
  node[midway,above=6pt] {$s$};
  \draw[decorate,decoration={brace,amplitude=5pt,mirror}]
    (0,-0.82) -- (2.6,-0.82)
    node[midway,below=5pt] {$n$};
\end{tikzpicture}
\ , \qquad h^{\lambda} = 2\,, \sim \varepsilon^{[\mu_1 \nu_1]\dots [\mu_n, \nu_n]\mu_{n+1}\dots \mu_{s}}\,.
\end{equation}
However these states are once again the states that saturate the bound for an even $SO(2n)$ group, so are not irreducible and admit a further decomposition. The first case where we see the decomposition is for $s=n=1$. The states are described by tensors with antisymmetric Lorentz indices $\zeta^{[\mu\nu]}_{ij}$. As little group tensors with indices $(ij)$, they are antisymmetric $SO(4)$ tensors. The situation is identical to that of irreducible components of a two-form in four dimensions. We know then that such an antisymmetric tensor can be decomposed into its self-dual (SD) and anti-self-dual (ASD) components using the four-dimensional Levi-Civita tensor. We can then define the SD and ASD parts of our tensor polarisation
\begin{equation}\label{SDASDsplit}
    \zeta_{\pm}^{[\mu_1 \nu_1]} \sim \varepsilon_{ij}^{[\mu_1 \nu_1]} \pm \epsilon_{ijkl} \varepsilon^{kl\, [\mu_1 \nu_1]}\,.
\end{equation}
These $\zeta_{\pm}$ correspond to the polarisation tensors for the massive antisymmetric tensor fields that appear from $D>4$ \cite{Chiodaroli:2022ssi, Townsend:1983xs} \footnote{The self-duality terminology used in our work refers to the little group $SO(4)$, while in five dimensions the authors in refs.~\cite{Chiodaroli:2022ssi,Townsend:1983xs} impose self-duality as a constraint on the Lorentz indices $B^{\mu\nu}=\frac{i}{3!m}\epsilon^{\mu\nu\rho\tau\sigma}\partial_\rho B_{\tau\sigma}$. For on-shell states the two notions are equivalent.}. Using these polarisations, we can extend the decomposition to arbitrary $s=n$ states, which we will refer to as SD with polarisation $(\zeta_+)^n$ or ASD with $(\zeta_-)^n$ as they only contain symmetric copies of the $s=n=1$ tensors.
\par The most general states are those having $s\neq n$ with $n>0$. Their polarisation tensors will be a mixture of the $\zeta^{\mu\nu}_\pm$ and $\epsilon^\mu$ introduced earlier. We will refer to these as mixed-SD or mixed-ASD to differentiate them from the $s=n$ case of purely $\zeta_\pm$ tensors:
\begin{equation}\label{mixedSD}
    \begin{cases}
        \zeta_{+}^{[\mu_1 \nu_1]}\dots \zeta_{+}^{[\mu_n \nu_n]} \varepsilon^{\mu_{n+1}}\dots \varepsilon^{\mu_s}\,,\text{ mixed-SD}\\
        \zeta_{-}^{[\mu_{1} \nu_{1}]} \dots \zeta_{-}^{[\mu_{n} \nu_{n}]} \varepsilon^{\mu_{n+1}}\dots \varepsilon^{\mu_s}\,, \text{ mixed-ASD}
    \end{cases}
\end{equation} 
where $n$ counts the (anti-) self-dual tensors $\zeta_{\pm}$ and $s-n$ counts the remaining symmetric polarisations $\varepsilon$.

Note that the analysis we did so far only generates integer spin states, but it can be extended to include spinor states, such that $s$ can take half-integer values. In such scenarios one should include a spinor $u(p)$ in the tensor polarisation.

To summarise, in $D=4$ the scattered fields are totally symmetric states labelled by the spin-$s$, where the massless particles are further labelled by helicity. However in $D=5$, while the massless states remain the totally symmetric states labelled by spin-$s$, which are now irreducible, we have various choices to scatter massive spinning states. We can still scatter the totally symmetric states as in $D=4$ but we also have access to mixed tensor states, which admit a self-dual and an anti-self-dual decomposition.

Although we used polarisation vectors and tensors throughout this discussion, the majority of the paper relies on a different set of variables, spinor-helicity variables. The scattering amplitudes often have more compact forms using these variables. The spinor-helicity approach also provides small building blocks which form a basis and can construct general scattering amplitudes. However, the form of the spinor-helicity variables is sensitive to the spacetime dimension. The building blocks are constructed from the matrix
\begin{equation}
    \slashed{p} = p_{\mu} \Gamma^{\mu}\,,
\end{equation}
which depends on a representation of Dirac algebra where $\Gamma$ satisfies $\{\Gamma^{\mu},\Gamma^{\nu}\} =2 \eta^{\mu\nu}$. We will start by first revisiting the $D=4$ setup.

\section{A review of scattering in $D=4$}\label{sec:review4d}

\subsection{Review of $D=4$ spinor variables}\label{sec:4dSH}

While spinor variables in $D=4$ have been extensively studied \cite{Dixon:1996wi,Parke:1986gb,Hodges:2012ym,Conde:2016izb,Arkani-Hamed:2017jhn},  we will take this opportunity to set up our conventions. The starting point for constructing spinor-helicity variables are the spin-$1/2$ solutions to the Dirac equation. In momentum space, the Dirac equation corresponds to a constraint 
\begin{equation}
    (p_\mu \gamma^\mu - m) u(p) = 0\,.
\end{equation}
where the on-shell degrees of freedom for the particle with momentum $p^2 = m^2$ are captured by the Dirac spinor $u(p)$. Since the spacetime dimension is even, we work with the Weyl representation Gamma matrices,
\begin{equation}
    \gamma^{0} = (\sigma^1 \otimes  \sigma^{0})\,, \, \gamma^{i=1,2,3} =  (i \sigma^2 \otimes  \sigma^{i})\,,
\end{equation}
where  $\sigma^0$ is the two-dimensional identity matrix and  $\sigma^i$ are the three Pauli matrices. Usually we define the two matrices into $(\sigma^{\mu})^{\alpha \dot{\beta}} = (\mathbf{1}, \sigma^i)= (\bar{\sigma}_{\mu})_{\dot{\alpha}\beta}$ such that the Gamma matrices have the form
\begin{equation}
    \gamma^{\mu} = \begin{pmatrix}
        0 & \sigma^{\mu} \\ \bar{\sigma}^{\mu} & 0 
    \end{pmatrix}\,.
\end{equation}

If we first consider a massless spinor $u(k)$, $k^2 =0$, we see that in this representation the Dirac equation reduces to two decoupled equations solved by Weyl spinors
\begin{alignat}{3} \label{eq:4dMasslessDiraceq}
   (k \cdot \bar{\sigma}) |k\rangle =0\,, \quad (k \cdot \sigma)  |k] = 0\,, \quad u(k) = \begin{pmatrix}|k\rangle \\ |k]\end{pmatrix}\,.
\end{alignat}
For generic momentum these spinors are independent and live in the different Weyl representations.\footnote{If we require $k$ to be a real momentum the two spinors will be related since the Weyl representations are related by complex conjugation. However since we are interested in three-point amplitudes, in this paper we will work with complex kinematics so that the amplitudes do not trivially vanish.} They transform differently under the little group $SO(2)\simeq U(1)$
\begin{equation}
    |k\rangle \to \lambda |k \rangle \,, |k] \to \lambda^{-1}|k]\,,
\end{equation}
where $\lambda$ is a phase since the momentum $k^{\mu} = \langle k | \sigma^{\mu} | k]/2$ must remain invariant. We can identify the $|k\rangle$ spinor with the positive helicity solution and define a massless polarisation 
\begin{equation}
    {\varepsilon_{3}^{+}}^{\mu} = \frac{\langle k|\sigma^{\mu}|r]}{\sqrt{2}[kr]}\,,
\end{equation}
where $r^{\mu}$ is a null reference vector whose dependence drops out in gauge invariant quantities. Note that if we consider the momentum space self-dual field-strength tensor $f^{+ \mu \nu}_{3} = f_{3}^{\mu\nu} + 
\frac{i}{2}\epsilon^{\mu\nu\rho\sigma} f_{3\,\mu\nu}$ constructed from $f_{3}^{\mu\nu} = k^{[\mu}\varepsilon^{\nu]}$, it has a spinor form
\begin{equation}
   f^{+}_{3\,\mu\nu} (\sigma^{[\mu}\bar{\sigma}^{\nu]})^{\alpha}{}_{\beta} = {}^{\alpha}|k\rangle \langle k|_{\beta} \,.
\end{equation}
Thus, our Weyl spinors make explicit the decomposition of the massless little group into a self-dual and anti-self-dual representation, as seen in the previous section, and we identify  $|k\rangle$ with the self-dual (positive helicity) representation and $|k]$ with the anti-self-dual (negative helicity) representation.

From the previous section we do not expect the same decomposition for the massive representations. Indeed the massive spin-$1/2$ Dirac equation decomposed into two coupled equations
\begin{alignat}{3} \label{eq:4dDiraceq}
    (p \cdot \bar{\sigma}) |p^{a}\rangle  &= m |p^{a}]\,, \quad (p \cdot \sigma) |p^{a}]  &= m |p^{a}\rangle\,, \quad  u^a(p) = \begin{pmatrix}|p^a\rangle \\ |p^a ]\end{pmatrix}\,.
\end{alignat}
Note that the massive spinors have an additional $SU(2)$ index, $a,b\in \{1,2\}$, reflecting that the massive momentum is parameterised by the sum of two outer products of spinor pairs
\begin{align}
    (p \cdot \sigma) = |p^{1}\rangle [ p_{1}| +|p^{2}\rangle [ p_{2}|=|p^{a}\rangle [ p^b| \epsilon_{a b}\,.
\end{align}
A massless momentum requires only one pair as the matrix $(k \cdot \sigma)$ is not full rank. The $SU(2)$ index is a feature of the construction, it is the index corresponding to the massive little group $SO(3) \simeq SU(2)$ and the spinors transform as
\begin{equation}
    |i^a\rangle = U^{a}{}_{b}|i^b\rangle\,, \qquad |i^a] = U^{a}{}_{b}|i^b] \,,
\end{equation}
where $U^a{}_{b}$ is a unitary matrix, such that the momentum $p$ is left invariant. In general we will absorb the little group index with an little-group spinor $z_{a}$ and make use of bold spinors 
\begin{equation}
    |{\bm p}\rangle = | p^a \rangle z_a.
\end{equation}

Note that while the massless spinors $|k\rangle, |k]$ are independent, the massive spinors $|p^a\rangle, |p^a]$ are not, they are related by the mass term in the Dirac equation. So we can choose to parametrise everything with the spinor $|{\bm p}\rangle$, for a fixed little group spinor $z_a$.

We can further use the spinor to parametrise a massive spin-$1$ state
\begin{equation}
    {\bm\varepsilon}^{\mu}(p) = \frac{\langle {\bm p}| \sigma^\mu p|{\bm p}\rangle}{\sqrt{2} m^2} = \varepsilon^{\mu}_{+} (z_1)^2+ \sqrt{2} \varepsilon^{\mu}_{L} (z_1 z_2) + \varepsilon^{\mu}_{-} (z_2)^2\,.
\end{equation}
Note that the polarisation satisfies ${\bm\varepsilon}^2 =0 = p \cdot {\bm\varepsilon}$ and we see the three physical polarisations if we expand it in a polynomial of $z$'s. Note that since we use the bold spinors, the little group indices are symmetrised. Antisymmetrisation over the little group indices just generates a trivial $SU(2)$ matrix,
\begin{equation}
    {}_{\alpha}|i^{[a}\rangle {}_{\beta}|i^{b]}\rangle = m \epsilon^{ab} \epsilon_{\alpha \beta}\,.
\end{equation}
Thus, the only way to build non-trivial states is by taking strings of totally symmetrised spinors as they fully span the space of totally symmetric states available in $D=4$
\begin{equation}
    |\mathbf{i}\rangle^n  \sim \begin{cases}
        {\bm\varepsilon}^{\mu_1}\dots {\bm\varepsilon}^{\mu_{s}}\,, \text{ for } n=2s\,,\\
        {\bm\varepsilon}^{\mu_1}\dots {\bm\varepsilon}^{\mu_{s}}u(p)\,, \text{ for } n=2s+1\,.\\
    \end{cases}
\end{equation}

Thus in $D=4$, if we want to study three-point amplitudes between two massive spin-$s$ states, $p_{i= 1,2}^2 = m^2$ and a massless graviton, $k^2=0$, the amplitude will be a function of the spinors
\begin{equation}
    \mathcal{A}(1^{s},2^{s},3^{h} )\sim\begin{cases}
|{\bm 1}\rangle^{2s}|{\bm 2}\rangle^{2s} |3\rangle^2 \, &\text{ for } h = 2 \,,\\
|{ \bm 1}\rangle^{2s}|{\bm 2}\rangle^{2s} |3]^2 \, &\text{ for } h = - 2\,, \\ 
    \end{cases}
\end{equation}
where we use the shorthand notation $|\mathbf{i}\rangle = |{\bm p}_i\rangle$ and  $|3\rangle,|3] = |k\rangle,|k]$.

Given spinor traces inevitably collapse due to the Dirac algebra, the spinor variables allow for a construction of a simple basis. For positive helicity, $h=2$, the three independent spinors structures are
\begin{equation}
    x := \sqrt{2}  \frac{p_1 \cdot \varepsilon_{3}^{+}}{m}\,, \quad \langle \mathbf{1} \mathbf{2} \rangle\,, \quad x^{-1} \langle \mathbf{1} 3 \rangle\langle3\mathbf{2} \rangle\,.
\end{equation}
Since the amplitude must transform with an overall $\lambda^{h=2}$ and $x$ is the only variable that has a non-trivial transformation, the general amplitude has the form \cite{Arkani-Hamed:2017jhn}
\begin{equation}\label{eq:4dGenAmp}
\mathcal{A}(1^s , 2^s, 3^{+}) = (p_1 \cdot \varepsilon_{3}^{+})^2\sum_{n=0}^{2s} g_{n} \langle \mathbf{12}\rangle^{2s-n} \left( x^{-1} \langle \mathbf{1} 3 \rangle \langle 3 \mathbf{2} \rangle\right)^{n}\,.
\end{equation}
The couplings $g_n$ satisfy two constraints, the normalisation of the mass monopole and the universality of the dipole coupling in gravity \cite{Cangemi:2024apk}, such that, we have $2s-1$ free couplings for this scattering scenario. There is an analogous basis for $h=-2$, with a different set of $2s-1$ couplings. However, if we impose the constraint that the underlying theory is parity invariant then the $h=-2$ scattering is fixed by the $h=2$ amplitude.

\subsection{An aside on classical GR solutions in $D=4$} \label{sec:4dWL}

A notable feature of the three-point amplitudes in eq.~\eqref{eq:4dGenAmp} is that they encode the same degrees of freedom, and ultimately, kinematic information as the linear couplings of a generic compact spinning body in $D=4$. This feature will turn out to be true in higher dimensions as well, so we will spend some time outlining the correspondence in four dimensions. We first introduce the classical worldline set up by Levi-Steinhoff \cite{Levi_2015}. The minimal action for the compact body is  built from the point-particle action and the universal dipole coupling of its spin degrees of freedom, encoded in the spin tensor $S^{\mu\nu}$, to gravity,
\begin{equation}\label{Smin}
    S_{\text{min}}=\int \mathrm{d}\sigma\left[-m\sqrt{-u^2}-\frac{1}{2}S_{\mu\nu}\Omega^{\mu\nu}\right],\ u=\dot{x},\ \Omega^{\mu\nu}\equiv \Lambda_A^\mu\frac{D \Lambda^{A\nu}}{D\sigma}.
\end{equation}
Here $\Lambda_A^\mu$ is a local tetrad, but its exact definition will not be necessary for our results, for more details see refs.~\cite{Levi_2015,Ben-Shahar:2023djm}.

A generic spinning compact body can have higher order couplings to the background, captured by non-minimal operators,
\begin{equation}\label{SEFT}
    S_{\text{EFT}}=S_{\text{min}}+S_{\text{non-min}}.
\end{equation}
In the next sections we will be interested in fixed solutions to GR. From a worldline perspective, these are fixed by specifying all the worldline couplings that are linear in the Riemann tensor.

The linear non-minimal couplings will involve Lorentz invariant contractions between the Riemann tensor, its derivatives and the spin tensor. In $D=4$, we can decompose the Riemann tensor into the electric $E_{\mu\nu} = R_{\mu \rho\nu\sigma}u^{\rho}u^{\sigma}$ and magnetic $B_{\mu \nu} = \epsilon_{\mu \alpha \beta \gamma} R_{\nu \rho}{}^{\alpha \beta} u^{\gamma} u^{\rho}$ components.\footnote{Strictly speaking these are usually defined in terms of the Weyl tensor but since we consider vacuum solutions the Ricci scalar vanishes and we can remove any dependence on the Ricci tensor via a suitable field transformation.} The resulting independent set of non-minimal interactions is 
\begin{align}\label{4dworldline}
    S_{\text{non-min.}} = &\sum_{n=1}^{\infty} \frac{(-1)^n}{(2n)!}\frac{C_{\mathrm{ES}^{2n}}}{m^{2n-1}} D_{\mu_{2n}}\dots D_{\mu_{3}} E_{\mu_2 \mu_1}S^{\mu_1}S^{\mu_{2}}\dots S^{\mu_{2n-1}}S^{\mu_{2n}}\nonumber\\
    +&\sum_{n=1}^{\infty} \frac{(-1)^n}{(2n+1)!}\frac{C_{\mathrm{BS}^{2n+1}}}{m^{2n}} D_{\mu_{2n+1}}\dots D_{\mu_{3}} B_{\mu_2 \mu_1}S^{\mu_1}S^{\mu_{2}} \dots S^{\mu_{2n}}S^{\mu_{2n+1}}\,,
\end{align}
written in terms of the spin vector $S^{\mu} = \epsilon^{\mu \nu \rho\sigma}u_{\nu}S_{\rho \sigma}$ and where $D_\mu$ is the spacetime covariant derivative.

From the action we can define an effective linearised stress-energy tensor for the solution, 
\begin{equation}\label{TEFT}
    T^{\mu\nu} = 2\frac{\delta S_{EFT}}{\delta g_{\mu\nu}}\Bigg|_{g_{\mu\nu}=\eta_{\mu\nu}}\,,
\end{equation}
that can be matched to an explicit linearised solution $g_{\mu \nu} = \eta_{\mu\nu} + \kappa h_{\mu \nu}$ via a Fourier transform
\begin{equation}
    h_{\mu\nu}(\Vec{x})=\frac{\kappa}{2}\int\frac{\mathrm{d}^3\vec{k}}{(2\pi)^3}\frac{e^{-i\Vec{k}\cdot \Vec{x}}}{\vec{k}^2}P_{\mu\nu,\rho\tau}T^{\rho\tau}(\Vec{k})\,,
\end{equation}
We describe this matching procedure in more detail in Appendix \ref{sec:5dMultipoles}. For example, the Kerr solution corresponds to $C_{ES^{2n}} =1 = C_{BS^{2n+1}}$. The on-shell degrees of freedom of $T_{\mu\nu}$ are encoded by contracting it with a positive helicity graviton, 
\begin{multline}
    T_{\mu\nu}\varepsilon^{+ \mu\nu} = (u \cdot \varepsilon^{+})^2 \left(1 +  (k \cdot S) +\sum_{n=1}^{\infty} \frac{1}{(2n)!}C_{ES^{2n}} (k \cdot S)^{2n}\right. \\
    + \left.\sum_{n=1}^{\infty}\frac{1}{(2n+1)!} C_{BS^{2n+1}} (k \cdot S)^{2n+1}\right)\,.
\end{multline}
Up to spin multipole $S^{n}$ there are $n-1$ free parameters, just as in eq.~\eqref{eq:4dGenAmp} the scattering of spin-$s$ particles generates $2s-1$ free couplings. This motivates the claim in \cite{Vaidya:2014kza} that scattering of spin-$s$ particles can generate up to $2s$ spin-multipoles of a classical solution. For example, the linearised Kerr solution can be reproduced by setting $g_n = \delta_{n}^0$. This choice corresponds to the family of AHH amplitudes, named after the authors of ref.\cite{Arkani-Hamed:2017jhn} where they were originally highlighted to their high-energy behaviour. However since the map between the couplings  $g_n$ and the classical multipoles only depends on the leading behaviour of $g_n$ in the spinor-helicity amplitude, the set of amplitudes reproducing a classical solution is not unique. In order to define a unique map, one need to impose additional constraints like spin-universality.

To make the map more explicit one needs to introduce a new set of variables for the QFT amplitudes, the spin operator basis. Expanding the general amplitude in this basis,
\begin{equation}\label{3pt4dspinhel}
    \mathcal{A}(1^s,2^s,3^{+}) = (p \cdot \varepsilon^+)^2 \sum_{n}^{2s} \tilde{g}_n \lexp (k \cdot \hat{S})^n \rexp\,,
\end{equation}
one can then identify the operator products $\lexp \hat{S}^{(\mu_1}\dots\hat{S}^{\mu_n)}\rexp$ with the classical variables $S^{\mu_1}\dots S^{\mu_n}$ and the couplings $\tilde{g}_n$ with the Wilson coefficients $C_{ES^{2n}}$, $C_{BS^{2n+1}}$.

\subsection{Spin Operator basis in $D=4$}\label{sec:4DSOp}

Above we saw how the classical worldline encodes the spin dependence of an effective source in the couplings of the spin tensor $S^{\mu\nu}$ to the Riemann tensor. While the spin dependence in the QFT amplitudes depends on the number of massive spinors and which spinor blocks appear, we saw that up to a fixed spin multipole the two effective approaches have the same $2s-1$ degrees of freedom. The map can be made explicit by introducing the spin operator $\mathbb{S}^{\mu\nu}$ which we will be able to map to the classical spin tensor $S^{\mu\nu}$. 

The spin operator is defined with respect to the Lorentz generator $\mathbb{M}^{\rho \sigma}$,
\begin{equation} \label{eq:spinOpLorentz}
    \mathbb{S}^{\mu\nu} = P^{\mu}_{\rho}P^{\nu}_{\sigma} \mathbb{M}^{\rho \sigma}\,,
\end{equation}
where the projector $P^{\mu}{}_{\nu} = \eta^{\mu}{}_{\nu} - \frac{p^{\mu}p_{\nu}}{m^2}$ projects out the components of the Lorentz generator in the direction of momentum $p^{\mu}$. If $p^\mu$ corresponds to the momentum of a massive spin-$s$ state, we can define the spin operator acting on its little group via contraction with its massive spinors
\begin{equation}
    (S^{\mu\nu})^{\Vec{a}}{}_{\Vec{b}} = (\langle p^a|)^{\odot 2s} \mathbb{S}^{\mu\nu}  (| p_b\rangle)^{\odot 2s} \,,
\end{equation}
where the spin-$s$ Lorentz generator is given by $\mathbb{M}^{(s)\mu \nu} = \frac{1}{2}[\sigma^{\mu},\bar{\sigma}^{\nu}] \otimes\mathbb{1}^{\otimes (2s-1)}$. We use the notation $\lexp \ldots \rexp$ to imply that we absorb free little group indices with the spinors $z_a$ and $\bar{z}^b$, 
\begin{equation}
    \lexp S^{\mu\nu} \rexp =(S^{\mu\nu})^{\Vec{a}}{}_{\Vec{b}} \,\bar{z}_{a_1}\dots \bar{z}_{a_{2s}}\, z^{b_1}\dots z^{b_{2s}} \,.
\end{equation}
We can interpret $\lexp S^{\mu\nu} \rexp$ as the expectation value of the spin operator with respect to an initial state $|{\bm p}\rangle^{2s}$ and a final state $|\bar{{\bm p}}\rangle^{2s}$, which share a little group. The $z_a$, $\bar{z}^b$ spinors encode the general direction of the spin-axis in the classical space of $SO(3)$ rotations. In the rest-frame the spin-tensor is 
\begin{align}
    \lexp S^{\mu\nu} \rexp = \frac{1}{2}\begin{pmatrix}
        0 & 0 & 0 &0\\
        0 & 0 & s_3 & s_2\\
        0 & -s_3 & 0 &s_1\\
        0 & -s_2 &-s_1 &0
    \end{pmatrix}(\bar{z}^a z_a)^{2s-1},\nonumber\\
    s_1 = \left(\bar z^1 z_2 + \bar z^2 z_1\right),\ s_2 = i\left(\bar z^1 z_2 - \bar z^2 z_1\right),\ s_3 = (\bar z^1 z_1 - &\bar z^2 z_2)\,.
\end{align}
Aligning the spin-axis with axis $i = 1,2,3$ corresponds to choosing $z$, $\bar{z}$ such that $s_{j}=\delta_{ij}$. This is clearest if we introduce the Hodge-dual of the spin-operator,\footnote{Note this operator acts on the little group and can be defined from the Pauli-Lubanski operator that acts on the Lorentz representations, in the same way that $(S^{\mu\nu})^{\Vec{a}}{}_{\Vec{b}}$ was defined with respect to $\mathbb{S}$.} which is a vector in $D=4$
\begin{equation}\label{eq:PauliLubanski}
    (S^{\mu})^{\Vec{a}}{}_{\Vec{b}} = \epsilon^{\mu\nu\rho\sigma} p_{\nu}(S_{\rho\sigma})^{\Vec{a}}{}_{\Vec{b}}\,, \qquad \lexp S^{\mu} \rexp =(S^{\mu})^{\Vec{a}}{}_{\Vec{b}} \,  \bar{z}_{a_1}\dots \bar{z}_{a_{2s}} \, z^{b_1}\dots z^{b_{2s}}.
\end{equation}
We can align the spin with the $x^3$-axis by choosing $z_{1} = \bar{z}^1 =1$, $z_{2} = \bar{z}^2 =0$ such that $\lexp S^{\mu} \rexp = \frac{s}{2} \delta^{\mu}_{3}$.

However our scattering amplitudes $\mathcal{A}(1^s , 2^s, 3)$ are not functions of $|{\bm p}\rangle^{2s}$ and $|\bar{\bm p}\rangle^{2s}$ but depend on states $|\mathbf{1}\rangle^{2s}$ and $|\mathbf{2}\rangle^{2s}$. In order to express the amplitude in terms of these spin expectation values we need to: $(i)$ parametrise the kinematics of particle $2$ in terms of particle $1$ and $3$, $(ii)$ repackage the spinors $|1^a\rangle \langle 1_b|$ into covariant variables $p^{\mu}\delta^a_b$ and $(s^{\mu \nu})^a_b$ and finally $(iii)$ express the amplitude in terms of the spin-$s$ spin tensor.

Step $(i)$ is implemented by a composition of a reflection (since $p_2$ is incoming) and a boost, $\Lambda$ such that $p_2 = \Lambda p_1 = - p_1 - k$. The boost acts on the spinors 
\begin{equation}
|2^a\rangle = \Lambda |1^a \rangle  = -|1^a \rangle - \frac{1}{2m} \slashed{k} |1^a]\,.
\end{equation}
In $D=4$ the reflection just corresponds to a sign $| (-1)^a\rangle = -|1^a\rangle$. Note that we also identify the little groups of particle $1$ and particle $2$.

Step $(ii)$ is implemented by a simple map between the symmetric and antisymmetric combinations of the little group and the spin-$1/2$ spin tensor $\hat{s}^{\mu\nu}:= \langle 1^{(a} |\mathbb{S}^{\mu\nu}|1^{b)}\rangle\big|_{s=1/2}$
\begin{equation}
    |1^{[a}\rangle\langle1^{b]}|= m \epsilon^{ab}\,, \qquad |1^{(a}\rangle\langle1^{b)}|= - \frac{1}{4}(\hat{s}^{\mu\nu} \sigma_{\mu\nu})^{ab}\,.
\end{equation}

The first two steps allow us to map the spinor building blocks to functions of $\hat{s}^{\mu\nu}$, for example
\begin{align}
    \langle \mathbf{1} \mathbf{2} \rangle &= - m \bar{z}\cdot z -\frac{1}{4m^2} i \epsilon[p k \hat{s}] \,, \quad x^{-1}\langle \mathbf{1} 3 \rangle\langle3\mathbf{2} \rangle =  -\sqrt{2}x^{-1} k \cdot \hat{s} \cdot \varepsilon^{+}_{3}\,,
\end{align}
where $\epsilon[ab\hat{s}]=\epsilon^{\mu\nu\rho\sigma}a_\mu b_\nu \hat{s}_{\rho\sigma}$. However, these spin-tensor structures are not independent in four-dimensions. The relations are apparent when using the spin-$1/2$ pseudo-vector operator, $\hat{s}^{\mu}$, defined in eq.~\eqref{eq:PauliLubanski},
\begin{align}
    k \cdot \hat{s} =- \frac{1}{m} \epsilon[p k \hat{s}] = -2^{3/2} (x^{-1}\,k \cdot \hat{s} \cdot \varepsilon^{+}_{3})\,.
\end{align}
Note that this monomial is related to the spin-tensor by the simple quadratic formula, $k\cdot \hat{s}\cdot \hat{s}\cdot k = \frac{1}{4}(k \cdot \hat{s})^2$.

The monomials in the general amplitude \eqref{eq:4dGenAmp} can then be written in terms of the spin structure $(k \cdot \hat{s})$,
\begin{align}
     \ang^{2s-n}(x^{-1} \la {\bm 1} 3\ra \la 3 \bm {2} \ra)^{n} = (-1)^{2s+n}m^{2s}\sum_{r=n}^{2s}\binom{2s-n}{r-n} (-1)^{n} 2^{n-2r}\left(\frac{k \cdot \hat{s}}{m} \right)^{r}\,,
\end{align}

The final step $(iii)$, expressing the spin-$s$ amplitude in terms of the spin-$s$ operators, requires us to change from the spin-$1/2$ representation to general spin-$s$. This could be done at the spin-tensor level, for example, using
\begin{align}
  (k \cdot \hat{s}\cdot \hat{s}\cdot k)^{n} &=  \frac{(2s-n)!}{(2s)!} \left \langle (k \cdot S \cdot S \cdot k)^{n}  \right \rangle\,.
\end{align}
However, in this case we have several spin-tensor structures to work out. Therefore it's easiest to change representation at the spin-vector level, where there is a single monomial,
\begin{align}
  (k \cdot \hat{s})^{n} &=  \frac{(2s-n)!}{(2s)!} \left \langle (k \cdot S)^{n}  \right \rangle\,.
\end{align}
Implementing these three steps on the general three-point amplitudes we can write it in a spin-multipole expansion
\begin{align}
    \mathcal{A}(1^s , 2^s, 3^{\pm}) &= m^2 \sum_{r=0}^{2s} G_r\left \langle \Big(\frac{k \cdot S}{m}\Big)^{r}  \right \rangle \,, \quad G_{r}  = \sum_{n=0}^{r} g_{n}  \frac{(-1)^{r}2^{n}(2s-n)!}{(r-n)!(2s)!}\,.
\end{align}
We see then that starting from a three-point amplitude expressed in spinor-helicity, we recover the general spin operator basis expansion with $\tilde{g}_n=G_n$. Note that if $g_{n}=\delta_{n}^{0}$ in the spinor basis we recover the ``minimal-coupling'' amplitudes in ref.~\cite{Arkani-Hamed:2017jhn}. In the spin-operator basis we recover the classical multipoles of the Kerr black hole discussed in section~\ref{sec:4dWL}.  
\begin{equation}
    G_r = \begin{cases}
        \frac{1}{(2n)!}\, \text{ for } r = 2n\,,\\
        \frac{1}{(2n+1)!}\, \text{ for } r = 2n+1\,,
    \end{cases}
\end{equation}
where we can read off the Kerr Wilson coefficients $C_{ES^{2n}}=1=C_{BS^{2n+1}}$ from the numerators.

\section{5D Massive spinning states}
\label{sec:5dStates}

\subsection{$D=5$ spinor variables}\label{sec:5dSH}

Spinor variables in five dimensions can be constructed in a similar fashion to in four-dimensions and indeed have been studied in detail in recent years, see refs.~\cite{Chiodaroli:2022ssi, Pokraka:2024fao}. The construction and notation in this paper is most similar to Pokraka et al. \cite{Pokraka:2024fao}, however we differ notably on metric conventions, choice of Gamma representations and explicit spinor parametrisations. A detailed list of our conventions on the Dirac representations and algebra in five dimensions can be found in Appendix \ref{appconv}. 

In $D=5$, we use the following representation for the Clifford algebra, 
\begin{equation}
    \Gamma^{\mu} = \begin{cases}
        (\sigma^1 \otimes \sigma^{0})\,, &\text{ if } \mu= 0\\
        (i\sigma^2 \otimes \sigma^{i})\,, &\text{ if } \mu=i= 1,2,3\\
        (-i\sigma^3 \otimes \sigma^{0})\,, &\text{ if } \mu= 4
    \end{cases}\,,
\end{equation}
which once again can be constructed from the Pauli matrices.  Since the Lorentz group in five dimensions is $SO(1,4) \simeq USp(2,2)$ the Dirac indices of $(\Gamma^{\mu})_{A}{}^{B}$ can be raised and lowered with the symplectic metric $C_{AB}= - C^{AB} $\cite{Chiodaroli_2022}
\begin{equation}
    C_{AB} = \begin{pmatrix}
        0 &1 &0&0 \\ -1 &0&0&0 \\ 0&0&0&1\\ 0&0&-1&0
    \end{pmatrix}\,. 
\end{equation}
Note that the Gamma matrices are symmetric if both the Dirac indices are lowered $(\Gamma^{\mu})_{AB}= (\Gamma^{\mu})_{A}{}^{D} C_{DA}  = (\Gamma^{\mu})_{BA}$ and equivalently if both raised.

Since we are in an odd dimension, the representation is irreducible, so our spinor variables will be Dirac spinors. Consider the massless spinor variables first, since $k$ is null, the matrix $\slashed{k}$ is rank-$2$ and can be parametrised by two ket-spinors $| k^{I=1,2}\rangle$,
\begin{equation}
(k\cdot \Gamma) = {}_{A}|k^{I}\rangle \langle k_{I}|^{B}\,, 
\end{equation}
where we raise and lower the Dirac index by $C_{AB}$ and the $SU(2)$ index with $\epsilon_{IJ}$ such that $\epsilon_{12}= -1$,
\begin{equation}
    \langle k^{I}|^{A} =  {}_{B} |k^{I}\rangle \, C^{BA}\,, \qquad {}_{A}|k_{I}\rangle =  {}_{A}|k^{J}\rangle \, \epsilon_{JI}\,.
\end{equation}
The  $SU(2)$ index $I=1,2$ corresponds to the little group $SO(3) \simeq SU(2)$ under which the spinors transform in the spin-$1/2$ representations via the unitary matrix $U_{J}{}^{I}$\, $ | k^{I} \rangle =  | k^J \rangle U_{J}{}^I$. The massless Dirac equation and on-shell constraint translate to the following spinor identities
\begin{equation}
     \slashed{k}| k^{I} \rangle = 0\,, \quad \langle k^{I} k^{J} \rangle = 0\,.
\end{equation}
These massless five-dimensional spinors behave similarly to the massive $D=4$ spinors as they are related by a trivial compactification. Using bold spinors, $|{\bm k}\rangle = |k^{I}\rangle v_I$, we can parametrise the massless spin-$1$ polarisation 
\begin{equation}
    \varepsilon^{\mu}_{k} = \frac{\langle {\bm k} | \slashed{r} \Gamma^{\mu} | {\bm k} \rangle}{ 2 k \cdot r}\,,
\end{equation}
which contains the three polarisation states for a massless $5D$ particle. The higher spin massless polarisation tensors just correspond to the tensor products, e.g. the spin-$2$ polarisation is $\varepsilon^{\mu\nu}_{k} = \varepsilon^{\mu}_{k}\varepsilon^{\nu}_{k}$.

The treatment of the massive states is similar, we can decompose the Dirac operator $\slashed{p}$ into outer-products of four spinors $\{|i^{1}\rangle, |i^{2}\rangle, |i^{1}], |i^{2}]\}$
\begin{equation}
    (p_i \cdot \Gamma)_A{}^{B} =  {}_{A}|i^{a}\rangle \langle i_{a}|^{B}+ {}_{A}|i^{\dot{a}}][i_{\dot{a}}|^{B}\,.
\end{equation}
The operator factors over the massive little group $SO(4)\simeq SU(2) \times SU(2)$. The angle spinors, carrying the undotted index $a=1,2$ , transform under the left $SU(2)$, while the square spinors, carrying the dotted index $\dot{a}=1,2$, transform under the right $SU(2)$. The bra and ket spinors are related via the conjugation matrix $C$ as was the case for the massless spinors. However the angle and square spinors are linearly independent and the Dirac equation does not introduce a mixing,
\begin{equation}
    \slashed{p}|i^{a}\rangle  = m |i^{a}\rangle\,, \qquad \slashed{p}|i^{\dot{a}}]  = - m |i^{\dot{a}}]\,.
\end{equation}

As with the massless states, we will parametrize the little group by spinors $z_a,\ w_{\dot{a}}$ and define the bolded Dirac spinors as $|{\bm p}\rangle=|{\bm p}^a\rangle z_a,\ |{\bm p}]=|{\bm p}^{\dot{a}}]w_{\dot{a}}$. We can still construct higher spin states by taking tensor products of the spinors. However, since the angle and square spinors are linearly independent, we have two parameters to adjust, $n_{L,R}$, which count the number of spinors from each copy of the $SU(2)$,
\begin{equation}
    | {\bm p}\rangle^{n_{L}}|{\bm p}]^{n_{R}}\,.
\end{equation}
Note, if we set $n_{L}= n_{R}=1$ we can construct the vector
\begin{equation}
    \varepsilon^{\mu}(p) = \frac{\langle{\bm p} | \Gamma^{\mu} |{\bm p}]}{m\sqrt{2}}\,,
\end{equation}
which satisfies all the conditions for a spin-$1$ polarisation state $p \cdot \varepsilon =0 = \varepsilon^2$. In fact, the states satisfying $n_{L}= n_{R}=S$ correspond to the totally symmetric spin-$S$ states with polarisation tensors $\varepsilon^{\mu_1 \dots \mu_{S}} = \varepsilon^{\mu_1}\dots \varepsilon^{\mu_S}$. In the discussion of our previous section these states correspond to the Young diagram in eq.~\eqref{youngsym}.

Now consider the states satisfying $n_L \neq n_R$. The simplest such states are those with $n_L=2$ and $n_R=0$ and vice versa. We can construct the polarisation tensors for these two states 
\begin{equation}
    \zeta_+^{\mu \nu}(p) = \frac{\langle {\bm p}| \Gamma^{\mu\nu}|{\bm p}\rangle}{4 m \sqrt{2}}\, \text{ or }   \zeta_-^{\mu \nu}(p) = \frac{[ {\bm p}| \Gamma^{\mu\nu}|{\bm p}]}{4 m \sqrt{2}},\,\mathrm{ where }\,\, \Gamma^{\mu\nu} = \frac{1}{2}[\Gamma^{\mu},\Gamma^{\nu}]\,,
\end{equation}
which satisfy $(p \cdot \zeta_+)^{\mu} = 0 = (\zeta_+ \cdot \zeta_+)^{\mu\nu}$ and similarly for $\zeta_-^{\mu\nu}$. These states correspond to the Young diagrams in eq.~\eqref{youngsplit} with $s=n=1$. As in eq.~\eqref{SDASDsplit}, we refer to $\zeta_+$ and $\zeta_-$ as the SD and ASD polarisations, respectively: as little-group representations, they transform in the symmetric representations of the left and right $SU(2)$ factors. This terminology mirrors the four-dimensional decomposition of Lorentz two-forms into self-dual and anti-self-dual components. In five dimensions, they correspond to the physical polarisations for the massive SD/ASD tensor states $B^{\mu\nu}$ and $\tilde{B}^{\mu\nu}$ \cite{Chiodaroli:2022ssi,Townsend:1983xs}.

In section \ref{sec:littlegroupsec} we discussed that generic states labelled by the Young diagrams lengths $(s,n)$ further decompose into two inequivalent pieces characterised by SD or ASD tensors $\zeta^{\mu\nu}$. The distinction can be made precise at the level of the $(n_L,n_R)$ labels such that we have the map
\begin{equation}\label{eq:labeldecomp}
\begin{cases}
        n_L = 2s\,, \, n_R = 2s -2n\,, &\text{mSD states}\,,\\
        n_L = 2s-2n\,,\  n_R = 2s\,, &\text{mASD states}\,,
    \end{cases}
\end{equation}
where we refer to eq.~\eqref{mixedSD} for the equivalent presentation in terms of polarisation tensors. We use the shorthand notation $\text{mSD}\equiv\text{mixed-SD}$ and $\text{mASD}\equiv\text{mixed-ASD}$. Throughout the subsequent sections, we will refer to the scattering states either through their $(s,n)$ or $(n_L,n_R)$ labels, where the map should be understood as above.

\subsection{Mass and spin-conserving scattering}\label{sec:5dGenAmp}

One of the interesting aspects of scattering in five dimensions is that one can scatter various different combinations of massive spinning states. Depending on the chosen spins, we can make properties such as parity more or less manifest. In the classical limit, some of the structural differences between the combinations will be washed out, but some combinations require new building blocks as tabulated in Table~\ref{tab:Results}. The details will be discussed in further sections; here we will simply organise the possible scattering by their spin-conserving properties.

The systematic construction of general five-dimensional three-point amplitudes between generic mass and spin states was completed in ref.~\cite{Pokraka:2024fao}. In this paper, we will focus on amplitudes that satisfy some notion of spin conservation. In the $D=4$ analysis, conservation of spin is unambiguous given that the massive states are labelled by a single quantum number $s$, so the spin-conserving amplitudes correspond to $\mathcal{A}(1^s,2^s,3^{\pm})$. However, as we saw in section~\ref{sec:littlegroupsec}, five-dimensional massive states are labelled by $(s,n)$.\footnote{Strictly speaking the quantum numbers that uniquely label our state are $(n_L, n_R)$, since for $n \neq 0$, the $(s,n)$-representation decomposes into a self-dual and an anti-self-dual state, see eq.~\eqref{eq:labeldecomp}. However, we find that we can define a suitable notion of spin-conservation at the level of the $(s,n)$ labels.} A suitable notion of spin conservation could then require that our initial and final states have the same spin quantum numbers $(s,n)$.

First we can consider the scattering of totally symmetric massive states in $5D$. Spin-conservation requires the two massive states are both labelled by $(s,n=0)$, in this case the quantum numbers are unambiguous $n_{L,i} = n_{R,i} = s/2$ for both $i= 1,2$. These amplitudes are studied in detail in section \ref{sec:5dTotSymmSH}; here we will introduce their spinor-helicity building blocks. The amplitude is a polynomial in the following monomials,
\begin{equation}\label{eq:SHBuildingBlocks}
    \{p_1\cdot \varepsilon_{3}, \sqang \SD,\angsq \ASD, \ang\sq\}\,.
\end{equation}
The first three variables carry the spin dependence of the graviton. In general, the coefficients of the monomials $\sqang \SD$ and $\angsq \ASD$ could be unrelated, introducing parity-violating terms in the amplitude. This is easiest to see if we rewrite the monomials in terms of the on-shell massive polarisations ${\bm \varepsilon}_{i=1,2}$ and the linearised field strength $f^{\mu\nu}_3 = k^{[\mu}\varepsilon_{3}{}^{\nu]}$,
\begin{align}\label{eq:paritybreakingterms}
   \sqang \SD& = m^2 {\bm\varepsilon}_1\!\cdot\! f_3\!\cdot\!{\bm\varepsilon}_2 + \frac{p_1{\cdot}\varepsilon_3}{2} {\bm\varepsilon}_1\!\cdot\! k{\bm\varepsilon}_2\!\cdot\!k+ \frac{i m}{2}\epsilon[p_{1}k\varepsilon_{3}{\bm\varepsilon}_{1}{\bm\varepsilon}_{2}]\,,\nonumber\\
   \angsq \ASD &= m^2 {\bm\varepsilon}_1\!\cdot\! f_3\!\cdot\!{\bm\varepsilon}_2 + \frac{p_1{\cdot}\varepsilon_3}{2} \,{\bm\varepsilon}_1\!\cdot\! k{\bm\varepsilon}_2\!\cdot\!k - \frac{i m}{2}\epsilon[p_{1}k\varepsilon_{3}{\bm\varepsilon}_{1}{\bm\varepsilon}_{2}]\,,
\end{align}
where we use the shorthand $\epsilon[abcde]=\epsilon^{\mu\nu\rho\sigma\delta} a_{\mu}b_{\nu}c_{\rho}d_{\sigma}e_{\delta}$. For scattering of totally symmetric states, these two monomials are the source of any five-dimensional Levi-Civita dependence. So the amplitudes will generically not be parity covariant unless we impose a cancellation of the Levi-Civita contributions. This is discussed further in sections~\ref{sec:5dTotSymm} and ~\ref{sec:5dSDscatt}.

Besides the difference in how parity acts in five-dimensions, this scattering is the closest analogue to the $D=4$ scattering. Indeed, we will find that the space of linearised gravitational solutions spanned by this class of amplitudes is similar, in that they are characterised by a massive worldline with spin degrees of freedom.

\par We can also scatter tensor states with $(s,n)$, such that $n\neq 0$. While the full details are discussed in section~\ref{sec:5dSDscatt}, we will introduce their general features here. Firstly, for $n \neq 0$, the label $(s,n)$ does not specify the state uniquely but the states decompose into the mSD and mASD sectors \eqref{eq:labeldecomp}. The scattering amplitudes can then be classified into two distinct cases.
\begin{itemize}
    \item The first category consists of sector-preserving amplitudes. In this case an incoming mSD state remains mSD after scattering. However we use all-incoming conventions and reflection acts non-trivially on the little group such that the outgoing mSD state is mapped to an incoming mASD state. Therefore this scattering scenario corresponds to states with quantum numbers $n_{L,1} = 2s = n_{R,2}$, $n_{R,1} = 2s-2n= n_{L,2}$.
    
    These amplitudes are constructed out of the same spinor monomials listed in eq.~\eqref{eq:SHBuildingBlocks} with the addition of the lone spinor variable $\angsq$. In particular, we will see that the $n\neq 0$ dependence is captured by an overall factor of $\angsq^{2n}$ in the amplitude. While this means the spinor variable form of the amplitude is markedly different in structure to the totally symmetric scattering, we will find that it washes out in the classical limit and will not introduce any new classical degrees of freedom! These amplitudes still capture the same degrees of freedom as a massive spinning worldline.
        
    \item Our definition of spin-conservation also allows for sector-flipping scattering. In contrast to the previous case an incoming mSD particle can become mASD after scattering. In 
    our all-incoming conventions this corresponds to states with quantum numbers satisfying $n^{L}_{i=1,2} = 2s$, $n^{R}_{i=1,2} = 2s-2n$.

    These amplitudes are constructed from the monomials in eq.~\eqref{eq:SHBuildingBlocks} and, this time, the lone spinor variable $\ang$. Similarly to the sector-preserving scattering, the amplitudes will exhibit an overall $\ang^{2n}$ factor. However, this spinor structure leads to non-trivial classical longitudinal degrees of freedom captured by the boost vector $K^{\mu}$. Therefore the space of classical solutions is broader than the previous two, and can include non-biaxially symmetric solutions too, see section~\ref{sec:nonbiaxal} for details.
\end{itemize} 

Since both sector-preserving and sector-flipping amplitudes are constructed from the mSD and mASD monomials, they are not automatically parity conserving. The parity breaking goes beyond the $f_3$ terms listed in eq.~\eqref{eq:paritybreakingterms}, since the polarisations of the mixed states have inherent Levi-Civita dependence. Indeed, we find that imposing parity conservation on this type of scattering introduces non-trivial constraints linking different amplitudes and is discussed in detail in section \ref{sec:5dSDscatt}. 

\par So far our definition of spin conservation depends on preserving the $SO(4)$ representations, hence why we allow a change in sector. However we can choose to loosen our definition of spin conservation further and only impose that the sum of the quantum numbers is conserved
\begin{equation}\label{eq:totspincons}
    n_{L,1}+n_{R,1} = n_{L,2}+n_{R,2}\,.
\end{equation}
We call this ``total-spin'' conservation. Clearly the previous amplitudes are ``total-spin'' conserving. However now we can allow shifts between the left and right little group factors,
\begin{align}
    n^{L}_2 = n^{R}_1 + \Delta S\,, \quad n^{R}_2 = n^{L}_1 - \Delta S\,,
\end{align}
where the spin-shift $\Delta S$ is constrained by our ``total-spin'' constraint. We will study such amplitudes in section~\ref{sec:CohAmp} in the context of coherent spin states. Remarkably, even though these amplitudes involve some notion of spin change, they do not need any more degrees of freedom than those captured by $S^{\mu\nu}$ and $K^{\mu}$, where the latter corresponds to the extra longitudinal degrees of freedom.

\par While we will introduce and study the general amplitudes separately in their respective sections, the spin-dependence of the classical amplitudes can be very simply traced to their spinor building blocks: the powers of the building block $\ang\sq$ map to the mass multipole tower in five dimensions, while the blocks involving the graviton spinor map to the current or stress multipoles. Note, if we do not restrict ourselves to parity-conserving scattering, then we will need to include a new degree of freedom on the classical worldline, $\tilde{S}^{\mu\nu} = \epsilon^{\mu\nu}{}_{\rho\sigma\tau} u^{\rho}S^{\sigma\tau}$.

\subsection{$D=5$ spin operator variables}\label{sec:5dSOp}

As in $D=4$, we can express these scattering amplitudes in terms of expectations of the spin operator. The definition of $\mathbb{S}^{\mu\nu}$ in terms of the Lorentz generator given in eq.~\eqref{eq:spinOpLorentz} is still valid. The spin-$1/2$ representation is given by 
\begin{equation}
    \mathbb{S}^{\mu\nu}\big|_{s=1/2}  = \mathbb{P}^{\mu}{}_{\rho}\mathbb{P}^{\nu}{}_{\sigma}\Gamma^{\rho\sigma} = \frac{1}{2}[\Gamma^{\mu},\Gamma^{\nu}] - \frac{1}{m^2}p^{[\mu}[\slashed{p},\Gamma^{\nu]}]\,.
\end{equation}
We can once again use the massive spinors to project onto the little group, but for a spin-$1/2$ particle the initial state is either $|p^a \rangle$ or $|p^{\dot{a}}]$ depending on whether the spin is in the left or right copy of $SU(2)$, since the little group is  $SU(2)\times SU(2)$. So we have two spin operators which act either on the left or the right $SU(2)$,
\begin{equation}
    {s^{\mu\nu}_{L}}^{a}{}_{b}=\frac{1}{m}\la p^{a}|\Gamma^{\mu\nu}|p_b\ra\,, \qquad \  {s^{\mu\nu}_{R}}^{\dot{a}}{}_{\dot{b}}=-\frac{1}{m}[p^{\dot{a}}|\Gamma^{\mu\nu}|p_{\dot{b}}].
\end{equation}
Note that due to the symmetry of the lower Gamma matrices, $(\Gamma^{\mu\nu})= (\Gamma^{\nu\mu})^{T}$, where the transpose is defined on the Dirac indices, the spinor trace satisfies the symmetry $\la p^{a}|\Gamma^{\mu\nu}|p_b\ra = \la p_{b}|\Gamma^{\mu\nu}|p^{a}\ra$ and likewise for the square spinors. The extension of the spin-operators to the purely chiral $(n_L,0)$ and anti-chiral $(0,n_R)$ representations is simply
\begin{equation}\label{eq:LRspinops}
    {S^{\mu\nu,}_{L}}^{\Vec{a}}{}_{\Vec{b}}=n_L s_{L}^{\mu\nu,a}{}_{b}(\delta^{a}{}_{b})^{n_L-1}\,, \qquad \  {S^{\mu\nu,}_{R}}^{\Vec{\dot{a}}}{}_{\Vec{\dot{b}}}=n_R s_{R}^{\mu\nu,\dot{a}}{}_{\dot{b}}(\delta^{\dot{a}}{}_{\dot{b}})^{n_R-1}\,,
\end{equation}
where the symmetrisation over $\Vec{a}$ ($\Vec{b}$) is implied by not indexing the $n_L$ ($n_R$) indices. For such representations the initial and final states are either all angles or square brackets and the contributions of the $p^{\mu}$ terms in the projectors vanish since
\begin{equation}
   \la p^{a}|[\slashed{p},\Gamma^{\mu}]|p^b\ra = 0 =  [ p^{\dot{a}}|[\slashed{p},\Gamma^{\mu}]|p^{\dot{b}}]\,.
\end{equation}
However the projectors contribute non-trivially for mixed states and allow us to define a spin operator that acts diagonally on the little group representations,
\begin{equation}\label{eq:SO4spinop}
    S^{\mu\nu}=\begin{pmatrix}
        S^{\mu\nu}_{L} & 0\\
        0 & S^{\mu\nu}_{R}
    \end{pmatrix}\,.
\end{equation}

We once again absorb the little group indices with the little-group (LG) spinors, but in this case we will distinguish between $S_{L/R}$ 
\begin{equation}
    \langle S_{L}^{\mu\nu}\rangle = (S_L)^{\Vec{a}}{}_{\Vec{b}} (\bar{z}_{a}z^{b})^{n_L}\,, \qquad [ S_{R}^{\mu\nu}] = (S_R)^{\Vec{\dot{a}}}{}_{\Vec{\dot{b}}} (\bar{w}_{\dot{a}}w^{\dot{b}})^{n_R}\,.
\end{equation}

In five dimensions, the massive little group matches the group of spatial rotations transverse to the momentum in its rest frame. We can make this identification explicit by choosing the two Cartans of the LG $SU(2)$ factors to generate rotations in the  $(x_1,x_2)$ and $(x_3,x_4)$ planes, with classical spin parameters $(a_1,a_2)$. The choice of directions in the $SU(2)$ factors is made by a choice of LG spinors $(z,\bar{z})$ and $(w,\bar{w})$. When taking the expectation values of the spin operator in states specified by these choices of LG spinors we obtain
\begin{equation}
    \langle S_{L}^{ij}\rangle= S^{ij} - \frac{1}{2}\epsilon^{ij}{}_{kl}S^{kl} = (a_1 - a_2) S_-^{ij}\,, \qquad [S_{R}^{ij}]= S^{ij} +\frac{1}{2} \epsilon^{ij}{}_{kl}S^{kl} = (a_1+a_2) S_+^{ij}\,,
\end{equation}
where $S_-^{ij}$ and $S_+^{ij}$ span a basis for the ASD/SD two forms. Thus, with our orientation convention, the $SU(2)_L$ spin operator corresponds to the
anti-self-dual combination of the classical spin tensor, while the $SU(2)_R$ spin operator selects the self-dual combination. These combinations correspond to simultaneous rotations of the two rotation planes with either the same or opposite orientations.

We will find that the spin-tensor fully captures the little-group degrees of freedom for totally symmetric state scattering and the sector-preserving scattering. In these cases, the amplitudes are constructed from building blocks that reduce to the spin-$1/2$ blocks
    \begin{equation}\label{eq:Sopbuildingblocks}
        \{p_1\cdot \varepsilon_{3},\ \varepsilon_{3} \cdot \langle s_{L}\rangle \cdot k ,\ \varepsilon_{3} \cdot [s_{R}] \cdot k ,\ k \cdot \langle s_{L}\rangle \cdot [s_{R}] \cdot k\}\,.
    \end{equation}

However the spin-operator is not enough to fully encode the sector-flipping scattering, since the spinor building block $\sqang$ alone does not admit an expansion in either $S_{L,R}$. An equivalent, covariant, statement is that the contraction of two spin-$1$ SD polarisations $\zeta^{+}_2\cdot\zeta^{+}_1$ does not admit an expansion in the operators $S^{\mu\nu}_{L,R}$. However the solution is to consider the degrees of freedom of the full Lorentz generator.

Using our definition of the spin-tensor, the full Lorentz generator decomposes into
\begin{equation}
    \mathbb{M}^{\mu\nu} = \mathbb{S}^{\mu\nu} + 2\frac{p^{[\mu} \mathbb{K}^{\nu]}}{m}\,,
\end{equation}
where $\mathbb{K}^{\mu}$ encodes the degrees of freedom along $p^{\mu}$. This operator is referred to as the boost operator in the literature \cite{Alaverdian:2024spu,Alaverdian:2025jtw} and we take the decomposition above, along with the definition of the spin-tensor in terms of the projectors, to define $\mathbb{K}$. If we consider the spin-1/2 Dirac representation of $\mathbb{M}^{\mu\nu} = \Gamma^{\mu\nu}$, it follows that $\mathbb{K}^{\mu} = \frac{\slashed{p}\Gamma^{\mu}}{m}$.

\par We define the operators acting on the little group in analogy to $S_{L,R}$
\begin{align}\label{eq:kappaops}
    (\kappa^\mu)^{a}{}_{\dot{b}}&=\frac{1}{m} \langle p^{a}| \Gamma^{\mu}|p_{\dot{b}}]\,, \quad (\bar{\kappa}^\mu)^{\dot{a}}{}_{b}=\frac{1}{m}  [ p^{\dot{a}}| \Gamma^{\mu}|p_{b}\rangle\,.
\end{align}
We can relate these to the expectation values of $\mathbb{K}$ by contracting in by the relevant little-group spinors, 
\begin{equation}
    \langle \kappa^{\mu}] = (\kappa^\mu)^{a}{}_{\dot{b}} \bar{z}_a w^{\dot{b}} = \la \bar{\bm 1} |\mathbb{K}^{\mu}| {\bm 1}]\,, \quad  [ \kappa^{\mu}\rangle = (\bar{\kappa}^\mu)^{\dot{a}}{}_{b} \bar{w}_{\dot{a}} z^{b} =- [ \bar{\bm 1} |\mathbb{K}^{\mu}| {\bm 1}\ra\,.
\end{equation} 
Since these operators effectively convert between left and right $SU(2)$ indices we can consider $(\kappa^\mu)$ as the operator that maps an $n_{L}=0, n_{R}=1$ state to $n_{L}=1, n_{R}=0$, while $(\bar{\kappa}^\mu)$ provides the inverse map. The only non-trivial vector operators for higher representations are
\begin{align}
    (\kappa^\mu)^{\Vec{a}}{}_{\Vec{b}\,\dot{b}} = (\kappa^\mu)^{a}{}_{\dot{b}} (\delta^{a}_{b})^{n_{L}-1}\,, \quad \text{and} \quad    (\bar{\kappa}^\mu)^{\Vec{a}\,\dot{a}}{}_{\Vec{b}} = (\kappa^\mu)^{\dot{a}}{}_{b} (\delta^{a}_{b})^{n_{R}-1}\,,
\end{align}
which map $(n_L-1,1) \mapsto (n_L,0)$ and $(1,n_R-1) \mapsto (0,n_R)$. Note that while this operator shifts exactly one quantum of spin between the left and right $SU(2)$ copies of the initial state. The index structure is constraining enough that if we consider an initial $(n_{L}, n_{R})$-state and a final $(\bar{n}_{L}, \bar{n}_{R})$-state only one string of boost operators is possible
\begin{equation}
    \begin{cases}
      (\langle \kappa^{\mu}])^{\odot(n_L - \bar{n}_L)}  &\text{ for } n_L - \bar{n}_L = \bar{n}_R - n_{R} > 0 \,,\\
      ([\bar{\kappa}^{\mu}\rangle)^{\odot(\bar{n}_L- n_L)}  &\text{ for } \bar{n}_L- n_L  = n_{R} - \bar{n}_R > 0 \,.
    \end{cases}
\end{equation}
Therefore for scattering where $|n_L - \bar{n}_L|>0$ there is a non-trivial contribution from the boost generators, which we see is encoded, at least in the sector-flipping scattering, in the $\ang^{n_L - \bar{n}_L}$ factor in the general amplitude.

We can now set up the explicit maps, following the same strategy as in $D=4$. Namely we need to $(i)$ relate the spinors of particles $1$ and $2$ with the initial states characterised by $\{ |{\bm p}\ra^{n_L}|{\bm p}]^{n_R}\}$ and final states characterised by $\{ |\bar{\bm p}\ra^{\bar{n}_L}|\bar{\bm p}]^{\bar{n}_R}\}$, $(ii)$ resolve the spin-$1/2$ operator expectation values and finally $(iii)$ express the final amplitude in terms of the $SO(4)$ representation of the spin operator in eq.~\eqref{eq:SO4spinop}. 

For step $(i)$, we choose to identify $|\mathbf{1}\rangle,|\mathbf{1}]$ with the initial state spinors $|{\bm p}\rangle,|{\bm p}]$, and with little-group spinors $z_{a}, w_{\dot{a}}$. While the final state spinors $|\bar{{\bm p}}\rangle, |\bar{{\bm p}}]$ are identified by boosting the spinors of $p_2$. We can parameterise $p_2$ as a Lorentz transformation of $p_1$,
 \begin{equation}\label{eq:p2Boost}
    p_{2\mu} = \Lambda_{\mu}{}^{\nu} p_{1 \nu} =-p_{1 \mu} - k_{\mu}\,.
\end{equation}
However, due to our choice of all-incoming kinematics, $p_1+p_2+k =0$, $\Lambda$ is an improper transformation and we take it to be a composition of a reflection $\Lambda_{\mathrm{refl.}}$ and a boost $\Lambda_{\mathrm{bst.}}$ such that
\begin{align}
    p_{2} &= \Lambda_{\mathrm{refl.}}{}_{\mu}{}^{\nu} (-p_{2})_{\nu}\,, \qquad (-p_2)_{\nu} = \Lambda_{\mathrm{bst.}}{}_{\mu}{}^{\nu} p_{1 \nu} =p_{1 \mu} + k_{\mu}\,.
\end{align}

We can solve for the boost exactly in the vector representation and find the boost for a general representation is 
\begin{equation}\label{eq:boost}
    \Lambda_{\mathrm{bst.}}{}_{\tilde{\mu}}{}^{\tilde{\nu}} = {\exp\left[\frac{ k_{\rho}p_{\sigma}}{m^2}\mathbb{M}^{\rho \sigma}\right]_{\tilde{\mu}}}^{\tilde{\nu}}\,,
\end{equation}
where we use the appropriate representation of the five-dimensional Lorentz generator $(\mathbb{M}^{\rho\sigma})_{\tilde{\mu}}{}^{\tilde{\nu}}$ and $\tilde{\mu}, \tilde{\nu}$ are the appropriate indices. For example, if we want to act on the massive spinors, we use the appropriate spin-$\frac{1}{2}$ representation of the Lorentz generators. The exponential in the boost truncates such that the massive spinors for $(-p_2)$ are
\begin{equation} \label{eq:spinorboost_part1}
    |\mathbf{-2}\rangle = \Lambda_{\mathrm{bst.}} |\mathbf{\bm p}\rangle = \left(|\bar{\bm p}\rangle+\frac{1}{2m} \slashed{k}|\bar{\bm p}\rangle\right) \,, \qquad
    |\mathbf{-2}] = \Lambda_{\mathrm{bst.}} |\bar{\bm p}] =\left(|\bar{\bm p}]-\frac{1}{2m}\slashed{k}|\bar{\bm p}]\right)\,. 
\end{equation}
While we have absorbed the little group indices in the equation above, the Lorentz transformations could include a rotation of these indices. However we will make the identification $z_{b,2} = \bar{z}_b$, $w_{\dot{b},2} = \bar{w}_{\dot{b}}$. The bar notation indicates that we contract the spinors of leg $1$ with a new set of barred spinors $\bar{z}^a$, $\bar{w}^{\dot{a}}$. 
\begin{equation}
    \langle\bar{\bm p}{\bm p}\rangle =   m |z|^2\,, \qquad [\bar{\bm p}{\bm p}]= - m |w|^2,
\end{equation}
where $|z|^2 = -\bar{z}\cdot z = \epsilon^{ab} \bar{z}_a z_b  = \bar{z}_1 z_2 - \bar{z}_2 z_1$ and similarly for the dotted spinors.
\par The reflection is simply $(\Lambda_{\mathrm{refl.}})_{\mu}{}^{\nu} = - \delta_{\mu}{}^{\nu}$ in the vector representation. Reflections, however, act non-trivially on the little group of the massive spinors, switching the two copies of $SU(2)$. This manifests as a swap of the angle and square spinors\footnote{The reflection is necessary when using all incoming kinematics regardless of dimension, however in four dimensions it just corresponds to a sign $|\mathbf{2}\rangle = - |\mathbf{-2}\rangle$, $|\mathbf{2}] = |\mathbf{-2}]$.} 
\begin{equation}
    {}_A|\mathbf{2}\rangle = (\Lambda_{\mathrm{refl.}}){}_{A}{}^{B} {}_{B}|\mathbf{-2}] = - i\, {}_{A}|\mathbf{-2}]\,, \qquad
    {}_{A}|\mathbf{2}] = (\Lambda_{\mathrm{refl.}}){}_{A}{}^{B}   {}_{B}|\mathbf{-2}\rangle= i \, {}_{A}|\mathbf{-2}\rangle \,.
\end{equation}
Composing the two transformations $\Lambda = \Lambda_{\mathrm{refl.}}\Lambda_{\mathrm{bst.}}$, the full transformation of the spinors is then,
\begin{equation} \label{eq:spinorboost}
    |\mathbf{2}\rangle =  i \left(|\bar{\bm p}]-\frac{1}{2m}\slashed{k}|\bar{\bm p}]\right)\,, \qquad
    |\mathbf{2}] = i \left(  |\bar{\bm p}\rangle+\frac{1}{2m} \slashed{k}|\bar{\bm p}\rangle\right)\,. 
\end{equation}

There are four independent outer-products of massive spinors $|  {\bm p}\rangle \langle\bar{{\bm p}}|, | {\bm p}][ \bar{{\bm p}}|,| {\bm p}\rangle [ \bar{{\bm p}}|,| {\bm p}]\langle\bar{{\bm p}}|$, which can be expanded in terms of $\la s_L \ra, [s_{R}], \langle \kappa ], [ \bar{\kappa} \rangle$. We list two of them below:
\begin{align}
    {}_{A}|  {\bm p}\rangle \langle\bar{{\bm p}}|^{B} &= \frac{1}{4}\left[ -(\slashed{p}_{A}{}^{B} +m \delta_{A}{}^{B}) \,\bar{z}\cdot z  - m/2 \langle S_L^{\mu\nu}  \rangle (\Gamma_{\mu\nu})_{A}{}^{B} \right]\, \nonumber \\
    {}_{A}| {\bm p}\rangle [ \bar{{\bm p}}|^{B} &= \frac{1}{4} \left[ m [\bar{\kappa}^{\mu} \rangle (\Gamma_{\mu})_{A}{}^{B}  + p^{\mu} [ \bar{\kappa}^{\nu} \rangle (\Gamma_{\mu\nu})_{A}{}^{B}\right]\,,
\end{align}
the rest can be found in eq.~\eqref{eq:MSdecomp}. Using these relations we can map the building blocks into expressions in terms of $\la s_L \ra$ and $[s_{R}]$, 
\begin{alignat}{2}\label{Spin1/2Map}
    \langle \mathbf{1} \mathbf{2}][\mathbf{1} \mathbf{2}\rangle & = m^2 |z|^2 |w|^2\,, \qquad \qquad\qquad
    \langle \mathbf{1} \mathbf{2}\rangle[\mathbf{1} \mathbf{2}] &&= \frac{1}{8}k \cdot \la s_L \ra \cdot [s_{R}] \cdot k\,, \nonumber\\ 
    [\mathbf{1} \mathbf{2}\rangle\langle \mathbf{1} 3 \rangle \langle 3 \mathbf{2}] &= -\frac{m^2 |w|^2}{2}\langle S_{\varepsilon k} \rangle\,,\qquad \qquad \,\langle \mathbf{1} \mathbf{2}][\mathbf{1} 3 \rangle \langle 3 \mathbf{2}\rangle &&= \frac{m^2 |z|^2}{2} [S_{\varepsilon k}]\,, \\
    \angsq &= im|z|^2  \qquad \qquad \qquad \qquad\qquad \,\, \ang &&= \frac{i}{2} k \cdot [\bar{\kappa} \rangle 
\end{alignat}
where $\langle S_{\varepsilon k} \rangle =\varepsilon_{3} \cdot \la s_L \ra \cdot k$ and $[S_{\varepsilon k}]=\varepsilon_{3} \cdot [s_{R}] \cdot k $. The spin operators appearing in the above identities are in the $(\frac{1}{2},0)$ or $(0,\frac{1}{2})$ representations and in the amplitude they will appear as arbitrary powers of their spin-$\frac{1}{2}$ representations. However, in order to have a well-defined classical limit, we need to map powers of expectation values to expectation values of the spin operator. At three-points the necessary set of identities is
\begin{align}\label{expvalreln}
    \lexp (k \cdot S \cdot S \cdot k)^{v} \rexp = &\phantom{\big[}c_{0,0,v} (k \cdot \la s_L \ra \cdot [s_{R}] \cdot k)^{v}\,,\nonumber\\
    \lexp S_{\varepsilon k} (k \cdot S \cdot S \cdot k)^{v} \rexp = &\big[ c_{1,0,v} (\varepsilon \cdot \langle s_{L}\rangle \cdot k) + c_{0,1,v}  (\varepsilon \cdot [s_{R}]\cdot k)\big] (k \cdot \la s_L \ra \cdot [s_{R}] \cdot k)^{v}\,, \nonumber \\
    \lexp S_{\varepsilon k}S_{\varepsilon k}(k \cdot S \cdot S \cdot k)^{v} \rexp = &\big[ c_{2,0,v} (\varepsilon \cdot \langle s_{L}\rangle \cdot k)^2+ 2 c_{1,1,v}  (\varepsilon \cdot \langle s_{L}\rangle \cdot k)(\varepsilon \cdot [s_{R}]\cdot k) \nonumber\\
    + &\phantom{\big[}c_{0,2,v}  (\varepsilon \cdot [s_{R}]\cdot k)^2\big] (k \cdot \la s_L \ra \cdot [s_{R}] \cdot k)^{v}\,,
\end{align}
where we introduced the $\langle \!\langle \ldots\rangle \!\rangle$ notation for expectation values in an arbitrary spin representation and the combinatorics are captured by 
\begin{equation}\label{eq:combinatorialrepC}
    c_{i,j,v} = (-1)^j\, 2^{v}(v+i)!(v+j)!\binom{n_L}{v+i}\binom{n_R}{v+j} \,.
\end{equation}
These identities require a fixed relative factor between the expectation values $\la s_L \ra$ and $[s_{R}]$ in the left and right little group. A generic amplitude, where the relative factors are not fine-tuned, cannot be independently captured at the classical level by only spin tensor. The missing classical structure is the Hodge dual $\tilde{S}^{\mu\nu}=\epsilon^{\mu\nu\sigma\rho\tau}S_{\rho\tau}u_{\sigma}$. In practice, the decomposition of $\tilde{S}^{\mu\nu}$ into $S_{L,R}$ is the same and the only independent operator structures are $\lexp \tilde{S}_{\varepsilon k} (k \cdot S \cdot S \cdot k)^{v} \rexp$ and $\lexp \tilde{S}_{\varepsilon k}S_{\varepsilon k}(k \cdot S \cdot S \cdot k)^{v} \rexp$, which are given by applying the rule $c_{i,j\neq0,v}\to - c_{i,j,v}$ to the identities in eq.~\eqref{expvalreln}. As we will see in the next section, the $\tilde{S}$ multipoles are not parity invariant. However, the notions of parity at the level of the amplitude and the level of the classical solution are related such that parity-conserving amplitudes will generate only spin-tensor multipoles.

\section{Classical EFT actions and stress-energy tensors in general $D$}\label{sec:5dCWL}
In the previous section we introduced the spinning states and the general expressions for three-point amplitudes involving two massive higher-spin fields and a graviton. The amplitudes were written as an expansion in a basis of spinor-helicity blocks that we identified. From a classical perspective, the dynamical data encoded in the amplitude maps to the Wilson coefficients of the worldline action of a spinning object coupled linearly to gravity. In four dimensions, a general worldline action has been written in \cite{Levi_2015} and we discussed its relation to the amplitudes side in a previous section. The purpose of this section is to extend the ansatz from \cite{Levi_2015} to higher dimensions. We will further use the action to derive the stress-energy tensor of the massive source. The stress-energy tensor of a spinning field in higher dimensions has been derived from a set of assumptions in \cite{Bianchi:2024shc}. The general ansatz in \cite{Bianchi:2024shc} contains redundancies that do not encode physical multipole data, but can be eliminated. In contrast, the Wilson coefficients in the worldline expansion have a clear physical meaning modelling finite-size structures and can be mapped to microscopic physics. We refer the reader to \cite{Levi_2015} for further details and motivation.
\par The action will describe the multipole structure of a spinning object coupled to gravity. The gravitational part involves the metric $g_{\mu\nu}$ and the dynamics of gravity will be described by the Einstein-Hilbert action.\footnote{In order to couple the point-particle worldline to gravity, one needs the degrees of freedom of a tetrad too. We will treat these as implicit and more details can be found in \cite{Levi_2015}.} The worldline degrees of freedom involve its trajectory $x(\sigma)$, which is a function of an arbitrary affine parameter and the spin $S^{\mu\nu}(\sigma)$. We will also impose the SSC condition $S^{\mu\nu}p_\mu=0$, where $p_\mu$ is conjugate to the $x^\mu(\sigma)$ in the Lagrangian. The action we will introduce works in any number of dimensions. We saw, however, that in five dimensions the Hodge dual of the spin operator can appear in the classical expansion too. We expect that if we want to match the amplitude coefficients to the Wilson coefficients, we need to include $\tilde{S}^{\mu\nu}$ too and we will do so in the second part of the section. 
\par With the degrees of freedom listed as above, we decompose the action as in eq.~\eqref{SEFT}, where the minimal action is the same as the four-dimensional one (\ref{Smin}). The non-minimal couplings will involve Lorentz invariant contractions between the Riemann tensor, its derivatives and the spin tensor. In any number of dimensions, we can decompose the Riemann tensor into pieces that have a certain number of timelike indices and project the rest on the spatial indices:
\begin{equation}
    E_{\mu\nu}=P_{\mu}^{\rho}P_{\nu}^\tau R_{\rho\sigma\tau\chi}u^\sigma u^\chi,\ B_{\mu\nu\rho}=P_{\mu}^{\beta}P_{\nu}^\tau P_\rho^\gamma R_{\beta\tau\gamma\sigma}u^{\sigma},\ Q_{\mu\nu\rho\sigma}=P_{\mu}^{\beta}P_{\nu}^\tau P_\rho^\gamma P_\sigma^{\delta} R_{\beta\tau\gamma\delta},
\end{equation}
where the projector $P_{\mu}^\nu$ was defined below equation~\eqref{eq:spinOpLorentz}. From the spin tensor, we can write the operators $S^{\mu\nu},\ (S\cdot S)^{\mu\nu}$. The EFT expansion includes all possible contractions between the curvature components and powers of the spin tensor.
\begin{itemize}
    \item Higher powers of $S$ factorise into these building blocks.
    \item The symmetries of the tensors above forbid contractions such as $E_{\mu\nu}S^{\mu\nu}=B_{\mu\nu\rho}(S\cdot S)^{\mu\nu}=Q_{\mu\nu\rho\tau}(S\cdot S)^{\mu\nu}=0$.
    \item We also have that any free $u^\mu$ contracted into these tensors vanishes either by the SSC condition or the inherent projection used in defining the curvature tensors. 
    \item Traces of the curvature tensors also vanish as they lead to Ricci tensors appearing in the Lagrangian and we do not allow them for vacuum solutions.
    \item When adding derivatives, we do not allow contractions between derivatives $\nabla_\mu\nabla^\mu R_{\rho\tau\sigma\chi}$ as these lead to local terms which are not relevant to the long range sector that the EFT describes.
    \item Contracting derivatives with the Riemann curvature invariants is also not allowed as it can be shown that it reduces to Ricci tensors using the Bianchi identities.
    \item The constraints outlined above allow for more terms, but they can be reduced to the terms in eq.~\eqref{nonMinEFT} up to Ricci tensor or scalar contributions. An example would be
\begin{equation}
    \nabla_\beta B_{\nu\tau\mu}(S\cdot S)^{\mu\nu}(S\cdot S)^{\tau\beta},
\end{equation}
which, using the Bianchi identity, reduces to a term in the first line of eq.~\eqref{nonMinEFT}. 
\item Symmetries also allow terms such as $Q_{\mu\nu\rho\tau}(S\cdot S)^{\mu\rho}(S\cdot S)^{\nu\tau}$, but would require introducing a new dimensionful scale beyond $\frac{S}{m}$.
\end{itemize}

These constraints restrict the number of terms that can appear. The non-minimal expansion of the Lagrangian can be written as:
\begin{align}\label{nonMinEFT}
    L_{\text{non-min}}=&\sum_{n=1}^\infty \frac{C_{n,1}}{(2n)!}\nabla_{\rho_3}\ldots\nabla_{\rho_{2n}}E_{\rho_1\rho_2}\left(S\cdot S\right)^{\rho_1\rho_2}\ldots (S\cdot S)^{\rho_{2n-1}\rho_{2n}}\nonumber\\
    +&\sum_{n=1}^\infty \frac{C_{n,2}}{(2n+1)!}\nabla_{\rho_3}\ldots\nabla_{\rho_{2n}}\nabla_{\rho_2} B_{\mu\nu\rho_1}S^{\mu\nu}(S\cdot S)^{\rho_1\rho_2}\left(S\cdot S\right)^{\rho_3\rho_4}\ldots(S\cdot S)^{\rho_{2n-1}\rho_{2n}}\nonumber\\
    +&\sum_{n=1}^{\infty}\frac{C_{n,3}}{(2n)!}\nabla_{\rho_3}\ldots\nabla_{\rho_{2n}}Q_{\mu\rho_1\nu\rho_2}S^{\mu\rho_1}S^{\nu \rho_2}\left(S\cdot S\right)^{\rho_3\rho_4}\ldots(S\cdot S)^{\rho_{2n-1}\rho_{2n}}.
\end{align}
Comparing to the four-dimensional expansion in eq.~\eqref{4dworldline}, a third tower of spin operators is present in higher dimensions. If the first two towers lead to the mass and current multipoles of a spacetime, the last tower described by $Q_{\mu\nu\rho\tau}$ has been identified as the stress multipole tower in refs.~\cite{Gambino:2024uge,Bianchi:2024shc}. We will make this statement precise by deriving the stress-energy tensor from eq.~\eqref{nonMinEFT}) and comparing it to the expressions from the literature. 
\par From our action, we can derive the stress-energy tensor by expanding the metric $g_{\mu\nu}=\eta_{\mu\nu}+\kappa h_{\mu\nu}$, with $\kappa^2=32\pi G$, and picking the linear in $h_{\mu\nu}$ contribution:
\begin{equation}
    T^{\mu\nu}(x)=2\frac{\delta S_{\text{EFT}}}{\delta h_{\mu\nu}(x)}\bigg|_{h_{\mu\nu}=0}.
\end{equation}
We will also Fourier transform to momentum space with variable $k^\mu$. When expanding around the flat background, the covariant derivatives in eq.~\eqref{nonMinEFT} become flat derivatives and their effect on the Fourier transform is $\nabla_{\mu}\rightarrow -i k_{\mu}$. We can go to the frame where $u^{\mu}=(1,\mathbf{0})$, so all the covariant indices become spatial ones. We also consider the static limit with no dependence on time. The variation of the different Riemann tensor components with respect to the metric is then:
\begin{equation}
    \delta E_{ij}=\frac{1}{2}\partial_i\partial_j \delta g_{00},\ \delta B_{ijk}=\partial_k \partial_{[i} \delta g_{j]0},\ \delta Q_{ijkl}=\partial_k\partial_j\delta g_{il}|_{\text{RS}},
\end{equation}
where $\text{RS}$ means that the tensor variation has the Riemann tensor symmetries. It is not hard to see then that the stress-energy tensor generated by our worldline action is:
\begin{align}\label{TAxiAns}
   T_{\mu\nu}(k) = &m u_\mu u_\nu \Bigg( 1 + \sum_{n=1}^{\infty} \frac{C_{n,1}}{(2n)!}\left(\frac{k \cdot S \cdot S \cdot k}{-m^2}\right)^n \Bigg) \nonumber\\
   -i&u_{(\mu}(S \cdot k)_{\nu)}\Bigg( 1 + \sum_{n=1}^{\infty}\frac{2 C_{n,2}}{(2n+1)!}\left(\frac{k \cdot S \cdot S \cdot k}{-m^2}\right)^n \Bigg) \nonumber\\
   - &\frac{( S\cdot k)_\mu( S\cdot k)_\nu}{m} \sum_{n=0}^{\infty}\frac{4 C_{n+1,3}}{(2n+2)!}\left(\frac{k \cdot S \cdot S \cdot k}{-m^2}\right)^n\,.
\end{align}
The end result is exactly the invariant part of the stress-energy expansion from refs.~\cite{Heynen:2023sin,Gambino:2024uge,Bianchi:2024shc}, the map from our Wilson coefficients to their ansatz coefficients $F_{n,i}$ is  $C_{n,1} = (2n)! F_{2n,1}, C_{n,2} = \frac{(2n+1)!}{2} F_{2n+1,3}, C_{n+1,3} = - \frac{(2n+2)!}{4} F_{2n+2,2}$. 
\par We note that the momentum $k$ used above is off-shell. Our treatment differs from the one in ref.~\cite{Bianchi:2024shc} where a larger ansatz for $T^{\mu\nu}(k)$ was obtained, including degrees of freedom that can be eliminated by linearised residual diffeomorphisms. Our result is a specific representative of this equivalence class and is consistent when working at the linearised order in $G$. Higher order corrections to $T^{\mu\nu}(k)$, for example those needed for full metric reconstruction, can be derived from the Einstein-Hilbert action and the worldline action with only linear in $R_{\mu\nu\rho\tau}$ couplings \eqref{nonMinEFT} by computing higher-order graviton loop amplitudes.

\subsection{Restricting to $4$-dimensions}

\par It is instructive to see what happens when we restrict our ansatz \eqref{nonMinEFT} to four dimensions. Then we know that we can dualize the spin tensor to its spin vector:
\begin{equation}\label{spinDual4d}
    S^{\mu\nu}=\epsilon^{\mu\nu\rho\tau}u_\rho S_{\tau}.
\end{equation}
The tensor $Q_{ijkl}$ points only in the spatial directions, which will be three dimensional. These turn out to be not independent from the electric tensor $E_{ij}$ and we have the identity:
\begin{equation}
    Q_{ijkl}=\delta_{ik}E_{jl}+\delta_{jl}E_{ik}-\delta_{il}E_{jk}-\delta_{jk}E_{il}.
\end{equation}
This implies that in the action, the stress multipole tower collapses to the electric tower as:
\begin{equation}
    Q_{\mu\nu\rho\tau}S^{\mu\nu}S^{\rho\tau}=-4E_{\mu\nu}(S\cdot S)^{\mu\nu}.
\end{equation}
This means that the electric tower becomes:
\begin{equation}
    \sum_{n=1}^\infty \frac{C_{n,1}-4C_{n,3}}{(2n)!}\nabla_{\rho_3}\ldots\nabla_{\rho_{2n}}E_{\rho_1\rho_2}\left(S\cdot S\right)^{\rho_1\rho_2}\ldots (S\cdot S)^{\rho_{2n-1}\rho_{2n}}
\end{equation}
Moreover, from eq.~\eqref{spinDual4d}, we can derive:
\begin{equation}
    S^{\mu\nu}S_\nu{}^\rho=S^\mu S^\rho+S^2 P^{\mu\rho},
\end{equation}
so the quadratic in spin tensor invariants can be replaced by symmetrized copies of the spin vector. The tensor $E_{\mu\nu}$ reduces to the usual electric part of the Riemann tensor \cite{Levi_2015}. Lastly, we have the following identification:
\begin{equation}
    B_{\mu\nu\tau} S^{\mu\nu}\propto \tilde{B}_{\mu\tau}S^\mu\,,
\end{equation}
where $\tilde{B}_{\mu\nu}$ is the magnetic part of the Riemann tensor $\tilde{B}_{\mu\nu}=\epsilon_{\mu\beta\rho\tau}R^{\rho\tau\sigma\nu}u^\beta u^\sigma$. These identities ensure that our EFT expansion collapses to the usual one from ref.~\cite{Levi_2015}, so we have consistency with the four-dimensional results. 

\subsection{Extending the class of EFTs in $5$-dimensions}

\par As mentioned earlier, in order to match the classical expansion in five dimensions with the one coming from the amplitudes, we need to modify our initial Lagrangian ansatz by terms containing the Hodge dual of the spin tensor $\tilde{S}^{\mu\nu}=\epsilon^{\mu\nu\rho\tau\sigma}u_{\rho}S_{\tau\sigma}$. 

The building blocks containing $\tilde{S}^{\mu\nu}$ are limited as $(S\cdot \tilde{S})^{\mu\nu}\propto P^{\mu\nu}$ and $(\tilde{S} \cdot \tilde{S})\propto (S\cdot S)$ so including it in the towers $(S\cdot S)^n(S\cdot \tilde{S})^m$, reduces them either to terms that we set to 0 or the terms already present and dependent on $(S\cdot S)$. However, replacing the single $S^{\mu\nu}$ contractions into the Riemann tensor by its Hodge dual leads to new terms in the expansion. 

We need to include dipole operator $- \tilde{C}_{0,2}\tilde{S}^{\mu\nu} \Omega_{\mu\nu}$, which is zeroth order in the Riemann tensor, and further extend the linear-in-Riemann operators, given in eq.~\eqref{nonMinEFT}, to $ {L}_{\text{non-min}}+ \tilde{L}_{\text{non-min}}$ where
\begin{align}\label{nonMinEFTExt}
    \tilde{L}_{\text{non-min}}=&\sum_{n=1}^\infty \frac{\tilde{C}_{n,2}}{(2n+1)!}\nabla_{\rho_3}\ldots\nabla_{\rho_{2n}}\nabla_{\rho_2} B_{\mu\nu\rho_1}\tilde{S}^{\mu\nu}(S\cdot S)^{\rho_1\rho_2}\left(S\cdot S\right)^{\rho_3\rho_4}\ldots(S\cdot S)^{\rho_{2n-1}\rho_{2n}}\nonumber\\
    +&\sum_{n=1}^{\infty}\frac{\tilde{C}_{n,3}}{(2n)!}\nabla_{\rho_3}\ldots\nabla_{\rho_{2n}}Q_{\mu\rho_1\tau\rho_2}S^{\mu\rho_1}\tilde{S}^{\tau\rho_2}\left(S\cdot S\right)^{\rho_3\rho_4}\ldots(S\cdot S)^{\rho_{2n-1}\rho_{2n}}.
\end{align}
This subsequently leads to new terms in the stress-energy tensor
\begin{align}\label{THodgeAns}
    T_{\mu\nu}(k)\to T_{\mu\nu}(k) &-im \,  \frac{u_{(\mu}(\tilde{S} \cdot k)_{\nu)}}{m}\left(\sum_{n=0}^{\infty} 2\frac{\tilde{C}_{n,2}}{(2n+1)!}\left(\frac{k \cdot S \cdot S \cdot k}{-m^2}\right)^n \right)\nonumber \\
   &-4m  \sum_{n=0}^{\infty} \frac{\tilde{C}_{n+1,3}}{(2n+2)!}\frac{( \tilde{S}\cdot k)_{(\mu}( S\cdot k)_{\nu)}}{m^2}\left(\frac{k \cdot S \cdot S \cdot k}{-m^2}\right)^n\,.
\end{align}
The terms containing $\tilde{S}^{\mu\nu}$ in eq.~\eqref{nonMinEFTExt} are not parity covariant as they contain one Levi-Civita tensor, so clearly this effective stress-energy tensor does not describe a solution to pure GR. However, these parity breaking terms could arise in Chern-Simons type gravity, for instance by coupling gravity to Yang-Mills and including a topological term for the latter or choosing a background that explicitly breaks parity. We leave the investigation of these effects for future work.
\par A comment is in place for our definition of parity breaking in five (and generally odd number of) dimensions that will be used in the subsequent sections. In four dimensions, one can consider parity to change the sign of either one of the spatial axis or all three of them as both operators have a negative determinant. In five dimensions, we do not have the option of parity changing the sign of all the spatial coordinates as the action has positive determinant. Then we are limited to the first option. In both four and five dimensions, this notion of parity does not leave known gravitational solutions such as the Kerr and Myers-Perry black holes invariant, changing the sign of some of the angular momenta. Hence, the solutions break this notion of parity. However, as tensors, the metrics are parity covariant, reflected in the fact that their multipole expansions contain an even number of Levi-Civita tensors. Adding a $\tilde{S}$ term breaks parity covariance. It is this notion of \emph{parity breaking} and \emph{parity invariance} that will be used throughout the text. 

\subsection{Non-biaxially symmetric sources}\label{sec:nonbiaxal}
The linearized metrics described by the stress-energy tensors introduced in the previous section have a biaxial symmetry given by the two rotations acting on the asymptotic azimuthal angles $\phi_{1,2}$ as can be seen from the general expansions in eq.~\eqref{acmcmetric}. Important examples of black holes such as Myers-Perry, the black ring and supersymmetric examples \cite{Breckenridge:1996is} fall into this category. While most known solutions fall into this biaxially symmetric class, much less is known about geometries that break these rotational symmetries. At a linearized level, the symmetry is reflected in the vanishing of the mass multipoles for odd powers of spin and vice-versa for the current multipoles. Introducing explicit $\phi_{1,2}$ dependence in the metric requires some an additional piece of data parametrised by a vector. One possible natural origin for this spin data is the longitudinal part of the spin tensor before imposing the SSC condition. In this section we therefore introduce the $K^\mu$ vector and extend the worldline action \eqref{nonMinEFTExt} to include its contributions. We also derive the stress-energy tensor sourcing the linearized metric. 
We will see in \cref{sec:KVecScat,sec:SDSD}, that there are classes of spinor-helicity amplitudes whose classical limits require introducing the $K^\mu$ vector. Relaxing the SSC condition has been studied in four dimensions where the extra degrees of freedom have been linked to boost degrees of freedom \cite{Alaverdian:2025jtw, Ben-Shahar:2023djm, Alaverdian:2024spu, Bern:2023ity}.
\par  Let us start from the unconstrained spin tensor $M^{\mu\nu}$ and its decomposition
\begin{equation}\label{SpinDecomp}
    M^{\mu\nu}=S^{\mu\nu}+\frac{2}{m}p^{[\mu} K^{\nu]}.
\end{equation}
The vector $K^\mu$ satisfies $K\cdot p=0$. Just like the case of the spin tensor, the couplings of $K^\mu$ to the Riemann tensor and its derivatives in the worldline action are constrained. In order to have a smooth limit as $m\rightarrow0$, we require that every insertion of $K^\mu$ is accompanied by a derivative. This condition and the use of Bianchi identities allow for a string of $K^\mu$ vectors where the Lorentz index is contracted into a covariant derivative $\nabla_\mu$ acting on some piece of the Riemann tensor. Overall, the extension of the non-minimal Lagrangian to contributions of $K^\mu$ is:

\begin{align}\label{NaxisymAction}
    \tilde{L}_{\text{non-min}}^{(K)}=&\sum_{n=1,n'=0}^\infty \frac{C_{n,n',1}}{(2n)!(n')!}\nabla_{\rho_{3:2n}}\nabla_{\tau_{1:n'}}E_{S\cdot S}\left(S\cdot S\right)^{\rho_1\rho_2}\ldots (S\cdot S)^{\rho_{2n-1}\rho_{2n}}  K^{\tau_{1:n'}}\nonumber\\
    +&\sum_{n=1,n'=0}^\infty \frac{C_{n,n',2}}{(2n+1)!(n')!}\nabla_{\rho_{3:2n}} \nabla_{\tau_{1:n'}}(\nabla B_{SSS})\left(S\cdot S\right)^{\rho_3\rho_4}\ldots(S\cdot S)^{\rho_{2n-1}\rho_{2n}} K^{\tau_{1:n'}}\nonumber\\  
    +&\sum_{n=1,n'=0}^{\infty}\frac{C_{n,n',3}}{(2n)!(n')!}\nabla_{\rho_{1:2n}}\nabla_{\tau_{1:n'}} Q_{SS}\left(S\cdot S\right)^{\rho_3\rho_4}\ldots(S\cdot S)^{\rho_{2n-1}\rho_{2n}} K^{\tau_{1:n'}}\,,
\end{align}
where we introduce the following shorthand notation $\nabla_{\rho_{1:n}}=\nabla_{\rho_1}\ldots\nabla_{\rho_n}$, $K^{\tau_{1:n'}} = K^{\tau_1}\ldots K^{\tau_{n'}}$, $E_{S\cdot S}= E_{\rho_1 \rho_2} (S\cdot S)^{\rho_1 \rho_2} $, $\nabla B_{SSS} = \nabla_{\rho_2}B_{\mu \nu\rho_1}S^{\mu\nu}(S\cdot S)^{\rho_1 \rho_2}$ and $ Q_{SS}= Q_{\mu\rho_1\tau\rho_2}S^{\mu\rho_1}S^{\tau\rho_2}$ for compactness. When $n'=0$, we return to the usual non-minimal action in eq.~\eqref{nonMinEFT}. Note in principle we also need to introduce the parity breaking terms if our scattering amplitudes are not parity invariant, this would introduce Wilson coefficients $\tilde{C}_{n,n',2}$ and $\tilde{C}_{n,n',3}$ similar to in eq.~\eqref{nonMinEFTExt} but coupled to the full tower of $K$ multipoles.

The stress-energy tensor derived from eq.~\eqref{NaxisymAction} is
{\normalsize
\begin{align}\label{TNonAxi}
    T_{\mu\nu}&(k) = m u_\mu u_\nu \left( 1 + \sum_{n=1,n'=0}^{\infty} \frac{C_{n,n',1}}{(2n)!(n')!}\left(\frac{k \cdot S \cdot S \cdot k}{-m^2}\right)^n\left(\frac{k\cdot K}{im}\right)^{n'} \right)\nonumber\\
    &- im \,  \frac{u_{(\mu}(S \cdot k)_{\nu)}}{m}\left( 1 + \sum_{n=1,n'=0}^{\infty} 2\frac{C_{n,n',2}}{(2n+1)!(n')!}\left(\frac{k \cdot S \cdot S \cdot k}{-m^2}\right)^n\left(\frac{k\cdot K}{im}\right)^{n'} \right)\nonumber\\
   &-im \,  \frac{u_{(\mu}(\tilde{S} \cdot k)_{\nu)}}{m}\left(  \sum_{n=0,n'=0}^{\infty} 2\frac{\tilde{C}_{n,n',2}}{(2n+1)!(n')!}\left(\frac{k \cdot S \cdot S \cdot k}{-m^2}\right)^n \left(\frac{k\cdot K}{im}\right)^{n'}\right)\nonumber\\
   &-4m  \sum_{n, n'=0}^{\infty}\frac{C_{n+1,n',3}}{(2n+2)!(n')!}\frac{( S\cdot k)_\mu( S\cdot k)_\nu}{m^2}\left(\frac{k \cdot S \cdot S \cdot k}{-m^2}\right)^n\left(\frac{k\cdot K}{im}\right)^{n'}\nonumber\\
   &-4m  \sum_{n,n'=0}^{\infty} \frac{\tilde{C}_{n+1,n',3}}{(2n+2)!(n')!}\frac{( \tilde{S}\cdot k)_\mu( S\cdot k)_\nu}{m^2}\left(\frac{k \cdot S \cdot S \cdot k}{-m^2}\right)^n\left(\frac{k\cdot K}{im}\right)^{n'}\,.
\end{align}}
Fourier transforming this expansion introduces odd contributions in spin for the mass and stress multipoles and even ones for the current multipole tower, all involving the $K^\mu$ vector. More concretely, in appendix \ref{sec:5dMultipoles} we argue that the leading term at each order $n$ in the $r^{-1}$ expansion of the linearized metric has a $P_{n-2}(\cos 2\theta)$ dependence in the asymptotic spherical coordinates $(\theta,\phi_1,\phi_2)$. The presence of the $K^\mu$ vector then modifies the Legendre polynomial to the spherical harmonics $Y_{n-2,n'}(\theta,\phi_i)$ \cite{Bena:2020uup}.

\subsection{Explicit solutions}
Having discussed the features of the multipole expansion in five and higher dimensions earlier in the section, we now look at specific examples. In four dimensions, one of the most studied objects is the Kerr black hole. The natural generalization to five dimensions is the Myers-Perry black hole \cite{Myers:1986un}. Its metric is described by three conserved charges, its mass $m$ and two spin parameters $a_1,\ a_2$ and has an $S^3$ horizon topology. We then turn to a second example, the black ring, still described by two angular momenta, but with a horizon topology $S^1\times S^2$. While for Myers-Perry, a compact formula for the multipole coefficients is available \cite{Bianchi:2024shc}, for the black ring, we only compute the multipole terms up to $S^{12}$ and discuss the patterns observed.

\subsubsection{Myers-Perry solution}
The Myers-Perry metric is given by
\begin{equation}
    \begin{aligned}
d s^{2}= & -d t^{2}+\frac{\rho^{2}}{\Delta} d r^{2}+\left(a_1^{2}-a_2^{2}\right) \sin ^{4} \theta\ d \phi_{1}^{2}+\left(a_2^{2}-a_1^{2}\right) \cos ^{4} \theta\ d \phi_{2}^{2} \\
& +\frac{2 m}{\rho^{2}}\left(d t-a_1 \sin ^{2} \theta\ d \phi_{1}-a_2 \cos ^{2} \theta\ d \phi_{2}\right)^{2}+\rho^{2}\left(d \theta^{2}+\sin ^{2} \theta\ d \phi_{1}^{2}+\cos ^{2} \theta\ d \phi_{2}^{2}\right),
\end{aligned}
\end{equation}
where 
\begin{equation}
\rho^2=r^2+a_1^2\cos^2\theta+a_2^2\sin^2\theta,\quad \Delta=\frac{(r^2+a_1^2)(r^2+a_2^2)}{r^2}-2m.
\end{equation}
The mass and rotation parameters in the metric are related to the physical mass $M$ and the angular momenta in the $\phi_1$ and $\phi_2$ planes $\mathfrak{a}_{1,2}$ as
\begin{equation}
    M=\frac{6\pi}{8}m,\ \mathfrak{a}_1=\frac{\pi}{2}ma_1,\ \mathfrak{a}_2=\frac{\pi}{2}ma_2.
\end{equation}
Restricting to the spatial slice, the spin tensor $S^{\mu\nu}$ becomes
\begin{equation}
    S^{ij}=\begin{pmatrix}
        0 & \mathfrak{a}_1 & 0 & 0\\
        -\mathfrak{a}_1 & 0 & 0 & 0\\
        0 & 0 & 0 & \mathfrak{a}_2\\
        0 & 0 & -\mathfrak{a}_2 & 0
    \end{pmatrix}.
\end{equation}
Extracting the multipole moments from the expansion can be done using standard methods. Using these methods, the multipole moments have been computed in refs.~\cite{Bianchi:2024shc,Heynen:2023sin}. We will write the Wilson coefficients for the $D$-dimensional Myers-Perry black hole using $D = 4 + 2 \epsilon$,
\begin{equation}\label{MPmult}
    C^{MP}_{n,1}=\frac{n+1+\epsilon}{2n+1}C_{n,\epsilon} ,\qquad C^{MP}_{n,2}=\frac{1+\epsilon}{2}C_{n,\epsilon}\,, \qquad C^{MP}_{n,3}=-\frac{n}{4(2n+1)}C_{n,\epsilon},
\end{equation}
and the universal factor $C_{n,\epsilon} = 2 (-1)^{n}(1+\epsilon)^{2n} \binom{n+1+\epsilon}{1+\epsilon} \binom{2n+2+2\epsilon}{1+2\epsilon}^{-1}$. This rewriting makes deriving to the $D=4$ Kerr multipoles almost trivial since $\lim_{\epsilon\to 0}C_{n,\epsilon} =1$. However to recover the expected Kerr multipole moments, we have to remember that, in four-dimensions, the stress multipole tower is not an independent and combines with the mass multipole such that
\begin{equation}
    C^{MP}_{n,1}-4C^{MP}_{n,3}\underset{D=4}{\rightarrow}1,\,\quad 2C^{MP}_{n,2}\underset{D=4}{\rightarrow}1.
\end{equation}
While the multipole moments resum nicely to Bessel functions in Fourier space, their geometrical understanding is obscured. To observe the singular structure of the source one can analyse the stress-energy tensor in position space or the bending angle of a scalar probe \cite{Bjerrum-Bohr:2025lpw}. The four-dimensional Kerr source has the topology of a disk with the matter concentrated around the circumference \cite{Israel:1970kp}. A similar analysis has been performed for the Myers-Perry black hole in ref.~\cite{Bianchi:2024shc,Bianchi:2025xol} where it was shown that the source of the five-dimensional Myers-Perry is a 3-ellipsoid embedded in $\mathbb{R}^4$, with the two semi-axis depending on the angular momenta.

\subsubsection{A first look at the black ring solution}
It is known that in five dimensions, there is no black hole uniqueness theorem \cite{Emparan:2001wn}, meaning that for certain ranges of the asymptotic masses and angular momenta, one can construct black holes with different horizon topologies. The phenomenon can be observed by looking at a different class of black holes called \emph{black rings}. While they can also be described by their mass and two angular momenta, their horizon topology is $S^2\times S^1$. However, one does not need the global properties of the metric to observe the non-uniqueness. Indeed, as observed in ref.~\cite{Heynen:2023sin}, even at the linearized level, the multipole moments of the Myers-Perry black hole and a black ring can differ, despite being described by the same charges.
\par One can perform a similar analysis of the multipole moments of the black ring as has been done for Myers-Perry. The expressions for the metric are left in the Appendix \ref{BRapp}. Focusing on the black ring with one angular momentum, the main difference compared to Myers-Perry is that the Wilson coefficients are not simply numbers, but depend on a dimensionless parameter $\nu$ which appears in the metric. A few low order examples are
\begin{equation}
    C_{1,2}=-\frac{3!}{2}\frac{9(1+3\nu^2)}{16(1+\nu)^3},\, C_{2,2}=\frac{5!}{2}\frac{27(5 - \nu + 33 \nu^2 - 3 \nu^3 + 46 \nu^4)}{1280(1+\nu)^6},\, C_{1,3}=\frac{2!}{4}\frac{3\nu(3+\nu^2)}{8(1+\nu)^3},
\end{equation}
while a full expansion up to $S^{12}$ is found in \ref{BRapp}. From the expressions at hand, it is not clear what the general form of the $C_{n,i}$ coefficients is. One can get some intuition for the behaviour of the polynomials in the numerator by performing an expansion around the Myers-Perry regime ($\nu=1$). Choosing $\xi=1-\nu$, the zeroth order corrections resum to the usual Myers-Perry generating functions, while the first order corrections resum to a higher order Bessel function
\begin{align}
    &C^{(\mathrm{BR})}_{2}(\zeta)= \left(1 {-} \frac{3 \xi}{2} \right)C^{(\mathrm{MP})}_{2}(\zeta_{v}){-} \frac{\xi }{2} \zeta_{v} J_{3}(\zeta_{v})  {+}\mathcal{O}(\xi^2),\  C^{(\mathrm{BR})}_{3}(\zeta)= C^{(\mathrm{MP})}_{3}(\zeta_{v}){+} \frac{3\zeta }{2} J_{2}(\zeta_{v}){+} \mathcal{O}(\xi^2)\,, \nonumber\\
    &C^{(\mathrm{BR})}_{1}(\zeta) = C^{(\mathrm{BR})}_{2}(\zeta_v) {+}C^{(\mathrm{BR})}_{3}(\zeta_v){+} \mathcal{O}(\xi^2)\,, \quad\mathrm{where } \quad \zeta_{v}= \frac{3}{2}(2- \xi)^{-3/2} \zeta  \nonumber
\end{align}
Another approach would be to isolate the $\nu$ dependence from the numerical prefactors. Then, for the $C_{n,2}$ factors, it turns out that there is a recurrence relation of the form
\begin{equation}
    N_{l+1}(\nu)=\lambda_l(1+3\nu^2)N_l(\nu)+\nu(1-\nu)^3 R_l(\nu).
\end{equation}
However, more form factors would be needed to determine what the coefficient $\lambda_l$ and the residual polynomial $R_l(\nu)$ are. We leave this for further investigation.

\section{Scattering totally symmetric states}\label{sec:5dTotSymm}

In section \ref{sec:5dStates} we introduced the three types of amplitudes whose classical limit we will study. The space of classical solutions was described in the last two sections by parameterizing the most general worldline action and deriving the source stress-energy tensor. In the current section we will perform the required manipulations to match the two expansions. We start from the amplitudes describing fully symmetric states and take their classical limit. It turns out that the general multipole expansion \eqref{THodgeAns} can be reached by amplitudes in this particular class, without introducing antisymmetric states. As a concrete example, we obtain the amplitude coefficients whose classical limit describes the Myers-Perry black hole in five dimensions. As an aside, at the end of the section we explore a different expansion making use of the polarisation vectors and on-shell momenta themselves, instead of using their spinor-helicity counterparts. This allows us to extend the classical limit of fully symmetric states to arbitrary dimensions and we are able to extend our Myers-Perry amplitude beyond five dimensions.

\subsection{Totally symmetric states}\label{sec:5dTotSymmSH}
Our starting point are the three-point amplitudes given in spinor-helicity form in ref~\cite{Pokraka:2024fao}
\begin{align}\label{eq:gen3ptAmpSpinor}
    \mathcal{A}(1^{S}, \,2^{S}, 3^2) &= \sum^{S}_{v=0}  \Bigg\{  g_{v, 0} (p_1 \cdot \varepsilon_{3})^{2} (\langle \mathbf{1} \mathbf{2} ][ \mathbf{1} \mathbf{2} \rangle)^{S-v}  (\langle \mathbf{1} \mathbf{2}\rangle [ \mathbf{1} \mathbf{2}])^{v} \nonumber\\
    &+ (p_1 \cdot \varepsilon_{3})  \Big(g_{v, 1} [ \mathbf{1} \mathbf{2} \rangle \langle \mathbf{1} 3\rangle [\mathbf{2} 3\rangle +  \tilde{g}_{v, 1} \langle \mathbf{1} \mathbf{2}] [\mathbf{1} 3\rangle \langle \mathbf{2} 3\rangle \Big) (\langle \mathbf{1} \mathbf{2} ][ \mathbf{1} \mathbf{2} \rangle)^{S-v-1}  (\langle \mathbf{1} \mathbf{2}\rangle [ \mathbf{1} \mathbf{2}])^{v} \nonumber \\
    &+ \Big( g_{v, 2} ([ \mathbf{1} \mathbf{2} \rangle \langle \mathbf{1} 3\rangle [\mathbf{2} 3\rangle)^{2} +  \tilde{g}_{v, 2} (\langle \mathbf{1} \mathbf{2}] [\mathbf{1} 3\rangle \langle \mathbf{2} 3\rangle)^{2}  \Big)(\langle \mathbf{1} \mathbf{2} ][ \mathbf{1} \mathbf{2} \rangle)^{S-v-2}  (\langle \mathbf{1} \mathbf{2}\rangle [ \mathbf{1} \mathbf{2}])^{v}  \Bigg\} .
\end{align}
We note that the second and third rows are only present for $S\geq1$ and $S\geq2$, respectively and $g_{v,1},\,\tilde{g}_{v,1}=0$ for $v=S$ while $g_{v,2},\tilde{g}_{v,2}=0$ for $v=S-1,\,S$. The formalism and maps set up in the previous section relate the spinor-helicity invariants to expectation values of the spin-operator which will be mapped to the general multipole expansion in five dimensions. The amplitude is described in terms of two series of coefficients, $g$ and $\tilde{g}$. Without any constraints imposed, the classical limit does not land on the expansion written just in terms of $S^{\mu\nu}$ \eqref{TAxiAns}, but needs terms depending on $\tilde{S}^{\mu\nu}$ like in eq.~\eqref{THodgeAns}. 
\par Using eq.~\eqref{Spin1/2Map}, we can map each spinor structure to expectation values
\begin{align} \label{eq:gen3ptAmp}
    \mathcal{A}(1^{S}, \,2^{S}, 3^2) = &(-m^2)^{S}\Bigg\{(p_1 {\cdot} \varepsilon_{3})^2 \sum^{S}_{v=0}   g_{v,0}\left(\frac{k {\cdot} \la s_L \ra {\cdot} [s_{R}] {\cdot} k}{8m^2}\right)^{v}  \nonumber\\
    -&\sum^{S}_{v=0}p_1 {\cdot} \varepsilon_{3}\, \frac{g_{v,1}\,\varepsilon_{3} {\cdot} \la s_L \ra {\cdot} k -\tilde{g}_{v,1}\,\varepsilon_{3} {\cdot}  [s_{R}] {\cdot} k 
    }{2}\left(\frac{k {\cdot} \la s_L \ra {\cdot} [s_{R}] {\cdot} k}{8m^2}\right)^{v} \nonumber \\
    +&\sum^{S}_{v=0}\frac{g_{v,2}\, (\varepsilon_{3} {\cdot} \la s_L \ra {\cdot} k)^2 -  \tilde{g}_{v,2}\,(\varepsilon_{3} {\cdot}  [s_{R}] {\cdot} k)^2 }{4} \left(\frac{k {\cdot} \la s_L \ra {\cdot} [s_{R}] {\cdot} k}{8m^2}\right)^{v} \Bigg\} \,.
\end{align}
The individual powers of the expectation values can be converted into expectation values of products of operators using eq.~\eqref{expvalreln}. The final amplitude is then
\begin{multline} \label{eq:gen3ptAmpSpinTensor}
    \mathcal{A}(1^{S}, \,2^{S}, 3^2) = \sum^{S}_{v=0} (p_1 \cdot \varepsilon_{3})^{2}g^{cl}_{v,0} \lexpBig \left(\frac{k \cdot S \cdot S \cdot k}{m^2}\right)^{v} \rexpBig \\
    +(p_1 \cdot \varepsilon_{3})\left(g^{cl}_{v,1}\lexpBig \frac{S_{\varepsilon k}}{m}\left(\frac{k \cdot S \cdot S \cdot k}{m^2}\right)^v\rexpBig+\tilde{g}^{cl}_{v,1}\lexpBig \frac{\tilde{S}_{\varepsilon k}}{m}\left(\frac{k \cdot S \cdot S \cdot k}{m^2}\right)^v\rexpBig\right)\\
    +\left(g_{v,2}^{cl} \lexpBig\left(\frac{S_{\varepsilon k}}{m}\right)^2\left(\frac{k \cdot S \cdot S \cdot k}{m^2}\right)^v\rexpBig+\tilde{g}_{v,2}^{cl} \lexpBig\left(\frac{S_{\varepsilon k}}{m}\right)\left(\frac{\tilde{S}_{\varepsilon k}}{m}\right)\left(\frac{k \cdot S \cdot S \cdot k}{m^2}\right)^v\rexpBig\right) \,.
\end{multline}
where $g^{cl}_{v,i}$ are related to the quantum coupling coefficients via
\begin{align}
    g^{cl}_{v,0}&= \frac{1}{(-8)^v c_{0,0,v}}\left(g_{v,0} - \tilde{g}_{v-1,2}\frac{c_{1,1,v-1}}{4c_{0,0,v-1}c_{0,2,v-1}}\right)\nonumber,\\
    g^{cl}_{v,1}&= -\frac{1}{4\times (-8)^v}\left(\frac{g_{v,1}}{c_{1,0,v}}-\frac{\tilde{g}_{v,1}}{c_{0,1,v}}\right),\ \tilde{g}^{cl}_{v,1}=-\frac{1}{4\times (-8)^v}\left(\frac{g_{v,1}}{c_{1,0,v}}+\frac{\tilde{g}_{v,1}}{c_{0,1,v}}\right)\nonumber,\\
    g^{cl}_{v,2}&=\frac{1}{(-8)^{v+1}}\left(\frac{g_{v,2}}{c_{0,2,v}}-\frac{\tilde{g}_{v,2}}{c_{2,0,v}}\right),\ \tilde{g}^{cl}_{v,2}=\frac{1}{(-8)^{v+1}}\left(\frac{g_{v,2}}{c_{0,2,v}}+\frac{\tilde{g}_{v,2}}{c_{2,0,v}}\right).
\end{align}
We used the conversion conventions from Appendix \ref{appconv}. We can see that the expansion of the amplitude matches that in eq.~\eqref{THodgeAns}. In order to reach the classical multipole expansion, we need to take the $\hbar\rightarrow 0$ limit with $k=\hbar \bar{k}$ and $S\rightarrow \infty$ such that the pairing $(kS)$ is finite. As a consequence we can replace all the expectation values by their classical values in the above expression. This is the same procedure as in four dimensions when obtaining the Kerr expansion \cite{Arkani-Hamed:2019ymq,Cangemi:2022abk}. The amplitude \eqref{eq:gen3ptAmpSpinTensor} has a different behaviour in the classical limit in five dimensions compared to the four-dimensional case. In four dimensions, the analogue of the coefficients $g_{v,i}$ do not get multiplied by spin dependent prefactors such as $1/c_{0,0,v}$. For a non-zero limit then the amplitude coefficients need to asymptote to a constant in the $S\rightarrow\infty$ limit. Specifically for Kerr, we can choose them to be equal to one. However, we observe that this characteristic is different in eq.~\eqref{eq:gen3ptAmpSpinTensor} where the spin dependent prefactors have an overall spin dependence in the $S\rightarrow \infty$ limit. Thus, without any spin dependence in $g,\ \tilde{g}$, the amplitude will vanish. This requires the following scalings
\begin{equation}\label{eq:couplingscalings}
    g_{v,0}\overset{S\rightarrow \infty}{\sim} S^{2v};\quad\quad  g_{v,1},\ \tilde{g}_{v,1}\overset{S\rightarrow \infty}{\sim} S^{2v+1};\quad\quad 
    g_{v,2},\ \tilde{g}_{v,2}\overset{S\rightarrow \infty}{\sim} S^{2v+2}.
\end{equation}
\par Specializing this to Myers-Perry, we saw that the multipole expansion does not need any $\tilde{S}$ terms, so we can set $\tilde{g}_{v,1}=g_{v,1}$ and $\tilde{g}_{v,2}=-g_{v,2}$ in the large $S$ limit inside the amplitude. Then we get the following dictionary between amplitude coefficients and spacetime form factors
\begin{equation}
    g_{v,0} = (-8)^v c_{0,0,v} \frac{C_{v,1}}{(2v)!}, \,\, g_{v,1} = (-8)^{v+1} c_{1,0,v} \frac{2 C_{v,2}}{(2v+1)!},\,\, g_{v,2} = -(-8)^{v+2} \frac{C_{v+1,3}}{(2v+2)!},
\end{equation}
where $C_{v,i}$ are the Myers-Perry multipole coefficients \eqref{MPmult}.

\subsection{Aside on general dimension Lorentz covariant blocks}\label{sec:TotSymmCov}
The expansion of the amplitude into spinor-helicity blocks is not the only available choice. In four and five dimensions this serves as a convenient basis highlighting important properties of the black hole amplitude, the map to the spin operator expectation values also being simpler to implement. However, spinor-helicity variables are known to be less constraining in higher dimensions and it is the isomorphisms between the orthogonal groups and unitary or symplectic ones in lower dimensions that makes spinor-helicity variables powerful there.
\par An alternative, dimension-agnostic choice for parametrizing the spinning states is using Lorentz covariant polarisation tensors. For generic states, as discussed in section \ref{sec:littlegroupsec}, the choice of a covariant polarisation is not unique and is subject to various constraints that reduce the degrees of freedom to the on-shell ones. One can however, restrict to fully symmetric, traceless states, whose main building block is the spin-1 vector $\bm{\varepsilon}^\mu(p)$, without further constraints.
\par In this section we depart from spinor-helicity variables and consider three-point amplitudes of massive, fully symmetric states and a graviton in arbitrary number of dimensions described by covariant blocks. We then consider their classical limit and show that they can recover the expansion eq.~\eqref{TAxiAns}, available in any number of dimensions.
\par In our amplitude, the massive states are described by ${\bm \varepsilon}_{1,2}$. Their interaction with the graviton is expressed through contractions with its field strength. We introduce the (linearized) field strength of a gauge field $f_{3,\mu\nu}=\varepsilon_{3,[\mu}k_{\nu]}$ and take its double copy to the linearized curvature of the graviton. We repackage the covariant building blocks as 
\begin{equation}
    X := {\bm\varepsilon}_1\!\cdot\! k\, {\bm\varepsilon}_2\!\cdot\! k,
\qquad
Y := {\bm\varepsilon}_1\!\cdot\!{\bm\varepsilon}_2,
\qquad
Z := {\bm\varepsilon}_1\!\cdot\! f_3\!\cdot\!{\bm\varepsilon}_2,
\end{equation}
Then the covariant expansion for the three-point amplitude between fully symmetric states of spin $S$ and a graviton is

\begin{equation}
A_{\mathrm{cov}}(1^S,2^S,3^2)
= \sum_{v=0}^{S} X^vY^{S-v}\,
\mathcal{G}_v ,
\end{equation}
with 
\begin{equation}
\mathcal{G}_v
=
\hat g_{0,v}(p_1\!\cdot\!\epsilon_3)^2
+\hat g_{1,v}(p_1\!\cdot\!\epsilon_3)ZY^{-1}
+\hat g_{2,v}Z^2Y^{-2}.
\end{equation}
We choose our expansion to include only parity-invariant terms. However, in five dimensions, we can add terms involving the Levi-Civita tensor, such as $\epsilon^{\mu\nu\rho\tau\sigma}{\bm \varepsilon}_{1,\mu}{\bm \varepsilon}_{2,\nu} f_{3,\rho\tau}p_{1,\sigma}$. They will then map to the extended multipole expansion. We collect the maps between the covariant structures in five dimensions and the spinor-helicity blocks in Appendix \ref{Appcov}.

In order to take the classical limit, we perform the same series of steps as in the previous section. We can parameterise the momentum and polarisations of the second massive field in terms of the first via the Lorentz boosts (composed with a reflection) defined in eq.~\eqref{eq:boost}. We can then change to the spin operator basis, but in this case the expectation values are taken with initial and final states described by ${\bm\varepsilon}_1^{\mu, S}$ and ${\bm\bar{\varepsilon}}_1^{\mu, S}$. To bring arbitrary powers of the expectation values inside a single expectation, we need to extend the identities \eqref{expvalreln} to higher dimensions. One such identity is
\begin{equation*}
    \lexp(k {\cdot} S {\cdot} S {\cdot} k)^n \rexp = (-1)^{n} \frac{S!}{(S-n)!}\frac{(2S+ D -5)!!}{(2S+ D - 5 - 2n)!!} (\bm{\varepsilon}_1 {\cdot} k \bar{\bm{\varepsilon}}_1 {\cdot} k)^{n} (\bar{\bm{\varepsilon}}_1 {\cdot}\bm{\varepsilon}_1)^{S-n},
\end{equation*}
while the remaining formulae are in Appendix \ref{Appcov}. The amplitude expanded in spin-expectation values is
\begin{align}\label{covexpgenDpre}
    \mathcal{A}_{\text{cov}}(1^{S},2^{S},3^2)=\sum_{v=0}^S &\Big[c_{v,1}(S)(p_1\cdot \varepsilon_{3})^2  \lexp (k {\cdot} S {\cdot} S {\cdot} k)^v \rexp +c_{v,2}(S)\lexp (\epsilon {\cdot} S {\cdot} k) \, (k {\cdot} S {\cdot} S {\cdot} k)^v \rexp\nonumber \\
    +&c_{v,3}(S)\lexp (\epsilon {\cdot} S {\cdot}k)^2 \, (k {\cdot} S {\cdot} S {\cdot} k)^v \rexp\Big],
\end{align}
with the coefficients
\begin{align}
    c_{v,1}(S) &= (-1)^v\frac{(S-v)!(2S+D-5-2v)!!}{S!(2S+D-5)!!}\left(\hat{g}_{0,v}+\frac{1}{S-v-1}\hat{g}_{2,v}\right)\nonumber\\
    c_{v,2}(S) &= (-1)^{v+1}\frac{(S-v-1)!(2S+D-5-2v)!!}{S!(2S+D-5)!!}\hat{g}_{1,v} \\
    c_{v,3}(S) &= (-1)^v\frac{(S-v-2)!(2S+D-5-2v)!!}{S!(2S+D-5)!!}\hat{g}_{2,v}\nonumber\,.
\end{align}

One interesting application of the covariant expansion is in extending our results for the amplitude describing the Myers-Perry black hole to arbitrary number of dimensions. The multipole structure has been written in terms of the ansatz \eqref{TAxiAns} in ref.~\cite{Bianchi:2024shc} such that one can read off the Wilson coefficients $C^{MP}_{n,i}$ which we give in eq.~\eqref{MPmult}. Constraining the amplitude in eq~\eqref{covexpgenDpre} to reproduce these multipoles would require 
\begin{equation}
    \lim_{S\to \infty}c_{n,i}(S)=\frac{C_{n,i}^{MP}}{(2n)!}\,, \, \text{ for }i \in \{1,3\} \quad \text{ and } \quad \lim_{S\to \infty}c_{n,2}(S)=\frac{C_{n,2}^{MP}}{(2n+1)!}
\end{equation}
and thus fix the leading in $S$ dependence of the amplitude coefficients.

\par QFT amplitude with coefficients $c_{n,i}(S)$ that land exactly on the classical spin multipole coefficient for a given $S$ are called spin-universal. For $D=4$, this is a property of the three-point AHH amplitudes \cite{Arkani-Hamed:2017jhn} which reproduce the Kerr multipoles to all orders in spin, and whose higher-point and higher-loop analogues reproduce consistent results up to $S^4$. In the same vein, one could define a spin-universal amplitude for $D$-dimensional Myers-Perry, by constraining the coefficients 
\begin{equation}
    c_{n,i}(S)\overset{!}{=}\frac{C_{n,i}^{MP}}{(2n)!}\,, \, \text{ for }i \in \{1,3\} \quad \text{ and } \quad c_{n,2}(S)\overset{!}{=}\frac{C_{n,2}^{MP}}{(2n+1)!}\,,
\end{equation}
which in turn would constrain the couplings $g$ in the amplitude. Such a spin-universal amplitude, in the spin-operator basis, can then resum such that
\begin{multline}\label{covexpgenD}
    \mathcal{A}_{\text{cov}}(1^{S},2^{S},3^2)= \frac{2}{3} (p_1\cdot \varepsilon_{3})^2\lexp 3 \tilde{\zeta}^{-1} J_{1}(\tilde{\zeta})- J_{2}(\tilde{\zeta})\rexp \\+2 (p_1 \cdot \varepsilon_3)\lexp (\epsilon {\cdot} S {\cdot} k) \, \tilde{\zeta}^{-1} J_{1}(\tilde{\zeta}) \rexp
    -\frac{2}{3}\lexp (\epsilon {\cdot} S {\cdot}k)^2 \, J_{2}(\tilde{\zeta}) \rexp,
\end{multline}
where $\tilde{\zeta} = \frac{3}{2} (k \cdot S \cdot S \cdot k)$ is operator-valued. The Bessel-type generating functions can be found in refs.~\cite{Bianchi:2024shc, Akpinar:2025huz} but we write them out explicitly
\begin{align}
    F_{1}(\zeta) &
    = F_{2}(\zeta)+ F_{3}(\zeta) = \sum_{n=0}^{\infty} \frac{C_{n,1}}{(2n)!}\zeta^n\,, \nonumber\\
    F_{2}(\zeta) &= \frac{4}{3}  \zeta^{-1} J_{1}\left(\frac{3}{2} \zeta\right) = \sum_{n=0}^{\infty} 2\frac{C_{n,2}}{(2n+1)!} \zeta^n,\\
    F_{3}(\zeta) &= - \frac{2}{3}J_{2}\left(\frac{3}{2} \zeta\right) = \sum_{n=0}^{\infty} -4\frac{C_{n,3}}{(2n)!} \zeta^n\,.  \nonumber   
\end{align}

A natural question surrounding the covariant expansion \eqref{covexpgenD} is whether there exists an off-shell Lagrangian that can reproduce it. The authors of ref.~\cite{Campanella:2026wqt, Gambino:2024uge} find that considering a single species of spin-$1$ fields (vectors or tensors) cannot reproduce the most general quadrupole expansion. This can lead to an obstruction to the four-dimensional notion of spin-universality. However, for fully symmetric states, the obstruction comes from the stress multipole which, at order $S^{2s-2}$ can be generated by an amplitude involving massive fields of spin $2s$. One can propose then a weaker notion of spin-universality where, in a spin $2s$ amplitude, the expansion coefficients generate exactly the mass and current multipoles up to spin $S^{2s}$ and the stress currents only up to spin $S^{2s-2}$.

\par For the four-dimensional Kerr black hole, the answer has been described in \cite{Cangemi:2022bew, Cangemi:2023ysz} up to quartic interactions. The constituent fields are massive higher spin fields in the symmetric representation non-minimally coupled to gravity (or a gauge field for $\sqrt{\text{Kerr}}$). The space of possible non-minimal couplings between the higher-spin and massless fields is successfully constrained to the Kerr expansion using the closure of the Lagrangian under gauge transformations and a few extra constraints regarding the number of derivatives appearing in an interaction. However recent work by Gambino et al.\cite{Gambino:2024uge,Gambino:2025iyx} suggests that in higher dimension even minimally-coupled low spin fields, i.e. $S<2$, do not reproduce the Myers-Perry multipoles, suggesting it is unclear how to extend the Lagrangian analysis beyond $D=4$.

The Kerr amplitudes, and the generic three-point gravity amplitudes in $D=4$, can be constructed from the double-copy of two gauge theory amplitudes \cite{Kawai:1985xq,Bern:2008qj,Bern:2010ue,Chiodaroli_2022}. At three-points the double-copy procedure amounts to the product
\begin{equation}\label{eq:doublecopy}
    \mathcal{A}_{\text{gravity}}(1^{s}, 2^{s},3^{2}) \simeq \mathcal{A}_{\text{gauge}}(1^{s_{L}}, 2^{s_{L}}, 3^{1}) \mathcal{A}_{\text{gauge}}(1^{s_{R}}, 2^{s_{R}}, 3^{1})\,,
\end{equation}
where the representations of the massive states are related by $s =s_{L}+s_{R}$. Notably one can choose the simplest double copy between a scalar, $s_{L}=0$, and a spin-$s_{R}=s$ gauge theory. This decomposition of the Kerr amplitudes has motivated the study of the spin-$s$ gauge theory referred to as $\sqrt{\text{Kerr}}$-theory \cite{Arkani-Hamed:2019ymq,Chiodaroli_2022,Cangemi:2023ysz}. In higher dimensions, this scalar $\otimes$ spin-$s$ double-copy is still valid for the scattering of totally symmetric states with low spins: $s<2$. However this breaks down for $s\geq2$, when the gravity amplitude can develop non-trivial stress-multipoles \cite{Gambino:2024uge, Campanella:2026wqt}. To reproduce the stress-multipoles, the gauge theory amplitudes on the right-handside of eq.~\eqref{eq:doublecopy} must be at least spin-$1$, ie $s_{L}, s_{R}\geq 1$. For $D=5$, this can be seen clearly in the spinor helicity basis. The stress multipoles map one-to-one with the structures $([{\bm 12}\rangle\langle{\bm 1} 3\rangle[{\bm 2}3\rangle)^2$ and $(\langle{\bm 12}][{\bm 1} 3\rangle\langle{\bm 2}3\rangle)^2$, which require the massless state to be at least spin-$2$ \cite{Gambino:2024uge,Campanella:2026wqt}. Since the massless gauge boson is spin-$1$ the gauge theory amplitudes can, if $s_{L,R}\geq 1$, at most produce a single power of the structures. Therefore, for dimensions $D>4$, the Myers-Perry solution cannot be generated from a scalar $\otimes$ spin-$s$ double copy due to the presence of non-trivial stress multipoles.

\section{Scattering mixed-(A)SD states}\label{sec:5dSDscatt}
One of the main differences between four and five dimensions is the existence of more than one category of spinning states. In the previous section, we looked at the fully symmetric states, which are the direct analogue of the four-dimensional states. In this section we allow antisymmetric tensor structures to appear. Following our classification from section \ref{sec:5dGenAmp}, we can characterize these states by two integers $(s,n)$ with the number of left and right handed spinors given by $n_L=2s,\ n_R=2s-2n$ for what we call mixed-SD states and vice-versa for mixed-ASD states.

Except in section~\ref{sec:CohAmp}, the classical limits in this paper follow the ``large-charge'' reasoning; classical physics corresponds to the sector of the full quantum phase space where the quantum numbers are large to the point that the expectation values of operators behave classically \cite{Maybee:2019jus, Cristofoli:2021jas}. For fully symmetric states, there was only one spin quantum number such that there was only one choice of scaling for $s\rightarrow\infty$, and we are effectively tracing the family of amplitudes of the totally symmetric states. However, when scattering mixed-(A)SD with $n\neq0$ we have two quantum numbers, $(s,n)$, to use when defining a classical limit. These parametrise a two-dimensional space of scattering amplitudes, which we can trace through with the following three corresponding classical limits:
\begin{enumerate}
    \item send $s\rightarrow\infty$ with $\frac{n}{s}\ll1$; 
    \item send $s\rightarrow\infty$ with $s-n=\text{constant}$; 
    \item or send $s\rightarrow\infty$ with $s-n\rightarrow\infty$ and $\frac{s-n}{s}=\gamma=\text{ constant}$.
\end{enumerate}

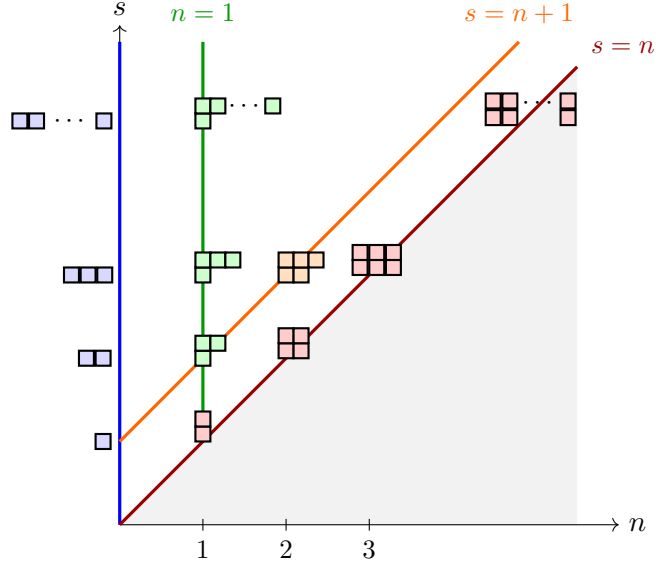
\begin{figure}
    \centering
\begin{tikzpicture}[scale=1.1]

  \fill[black!50!white, opacity=0.1]
    (0,0) -- (5.5,0) -- (5.5,5.5) -- cycle;

  \draw[->] (0,0) -- (6,0) node[right] {$n$};
  \draw[->] (0,0) -- (0,6) node[above] {$s$};

  \draw[blue, very thick] (0,0) -- (0,5.8);

  \draw[red!60!black, very thick] (0,0) -- (5.5,5.5);

  \draw[green!60!black, very thick] (1,1) -- (1,5.8);

  \draw[orange!80!red, very thick] (0,1) -- (4.8,5.8);


  \def\bsz{0.18} 

  \draw[thick, fill=blue!15] (-0.5-\bsz/2+0.3, 1-\bsz/2) rectangle +(\bsz,\bsz);

  \draw[thick, fill=blue!15] (-0.5-\bsz-0.01+0.2, 2-\bsz/2) rectangle +(\bsz,\bsz);
  \draw[thick, fill=blue!15] (-0.5+0.01+0.2, 2-\bsz/2) rectangle +(\bsz,\bsz);

  \draw[thick, fill=blue!15] (-0.5-\bsz-0.04+0.05, 3-\bsz/2) rectangle +(\bsz,\bsz); 
  \draw[thick, fill=blue!15] (-0.5+0.03,            3-\bsz/2) rectangle +(\bsz,\bsz); 
  \draw[thick, fill=blue!15] (-0.5+0.01+\bsz+0.04,  3-\bsz/2) rectangle +(\bsz,\bsz); 

  \draw[thick, fill=blue!15] (-0.5-\bsz-0.01-0.6, 4.85-\bsz/2) rectangle +(\bsz,\bsz);
  \draw[thick, fill=blue!15] (-0.5+0.01-0.6,  4.85-\bsz/2) rectangle +(\bsz,\bsz);
  \node at (0.02-0.6, 4.85) {\small$\cdots$};
  \draw[thick, fill=blue!15] (0.32-0.6, 4.85-\bsz/2) rectangle +(\bsz,\bsz);


  \draw[thick, fill=red!20] (1-\bsz/2, 1            ) rectangle +(\bsz,\bsz); 
  \draw[thick, fill=red!20] (1-\bsz/2, 1+\bsz  ) rectangle +(\bsz,\bsz); 

  \draw[thick, fill=red!20] (2-\bsz/2,          2            ) rectangle +(\bsz,\bsz); 
  \draw[thick, fill=red!20] (2-\bsz/2+\bsz, 2            ) rectangle +(\bsz,\bsz); 
  \draw[thick, fill=red!20] (2-\bsz/2,          2+\bsz  ) rectangle +(\bsz,\bsz); 
  \draw[thick, fill=red!20] (2-\bsz/2+\bsz, 2+\bsz) rectangle +(\bsz,\bsz); 

  \draw[thick, fill=red!20] (3-\bsz-0.02, 3       ) rectangle +(\bsz,\bsz);
  \draw[thick, fill=red!20] (3          , 3       ) rectangle +(\bsz,\bsz);  
  \draw[thick, fill=red!20] (3+\bsz+0.02, 3       ) rectangle +(\bsz,\bsz);
  \draw[thick, fill=red!20] (3-\bsz-0.02, 3+\bsz) rectangle +(\bsz,\bsz);
  \draw[thick, fill=red!20] (3          , 3+\bsz) rectangle +(\bsz,\bsz);
  \draw[thick, fill=red!20] (3+\bsz+0.02, 3+\bsz) rectangle +(\bsz,\bsz);

  \def\px{4.6} \def\py{4.8}
  \draw[thick, fill=red!20] (\px-\bsz-0.02, \py       ) rectangle +(\bsz,\bsz);
  \draw[thick, fill=red!20] (\px,           \py       ) rectangle +(\bsz,\bsz);
  \draw[thick, fill=red!20] (\px-\bsz-0.02, \py+\bsz+0.02) rectangle +(\bsz,\bsz);
  \draw[thick, fill=red!20] (\px,           \py+\bsz+0.02) rectangle +(\bsz,\bsz);

  \node at (\px+0.45, \py+\bsz/2+0.17) {\small$\cdots$};

  \draw[thick, fill=red!20] (\px+0.7, \py       ) rectangle +(\bsz,\bsz);
  \draw[thick, fill=red!20] (\px+0.7, \py+\bsz+0.02) rectangle +(\bsz,\bsz);


  \draw[thick, fill=green!20] (1-\bsz/2,          2-\bsz/2            ) rectangle +(\bsz,\bsz); 
  \draw[thick, fill=green!20] (1-\bsz/2,          2-\bsz/2+\bsz ) rectangle +(\bsz,\bsz); 
  \draw[thick, fill=green!20] (1-\bsz/2+\bsz, 2-\bsz/2+\bsz) rectangle +(\bsz,\bsz); 

  \draw[thick, fill=green!20] (1-\bsz/2,              3-\bsz/2            ) rectangle +(\bsz,\bsz); 
  \draw[thick, fill=green!20] (1-\bsz/2,              3-\bsz/2+\bsz ) rectangle +(\bsz,\bsz); 
  \draw[thick, fill=green!20] (1-\bsz/2+\bsz,   3-\bsz/2+\bsz ) rectangle +(\bsz,\bsz); 
  \draw[thick, fill=green!20] (1-\bsz/2+2*\bsz, 3-\bsz/2+\bsz ) rectangle +(\bsz,\bsz); 

  \draw[thick, fill=green!20] (1-\bsz/2, 4.85-\bsz/2            ) rectangle +(\bsz,\bsz);
  \draw[thick, fill=green!20] (1-\bsz/2,              4.85-\bsz/2+\bsz) rectangle +(\bsz,\bsz);
  \draw[thick, fill=green!20] (1-\bsz/2+\bsz,   4.85-\bsz/2+\bsz) rectangle +(\bsz,\bsz);
  \node at (1-\bsz/2+2*\bsz+0.25, 4.85-\bsz/2+\bsz+\bsz/2) {\small$\cdots$};
  \draw[thick, fill=green!20] (1-\bsz/2+3*\bsz+0.3, 4.85-\bsz/2+\bsz) rectangle +(\bsz,\bsz);

  \node[green!60!black, font=\small, above=4pt] at (1,5.8) {$n=1$};


  \draw[thick, fill=orange!25] (2-\bsz/2,              3-\bsz/2            ) rectangle +(\bsz,\bsz);
  \draw[thick, fill=orange!25] (2-\bsz/2+\bsz,   3-\bsz/2            ) rectangle +(\bsz,\bsz);
  \draw[thick, fill=orange!25] (2-\bsz/2,              3-\bsz/2+\bsz ) rectangle +(\bsz,\bsz);
  \draw[thick, fill=orange!25] (2-\bsz/2+\bsz,   3-\bsz/2+\bsz ) rectangle +(\bsz,\bsz);
  \draw[thick, fill=orange!25] (2-\bsz/2+2*\bsz, 3-\bsz/2+\bsz ) rectangle +(\bsz,\bsz);


  \node[orange!80!red, font=\small, above=3pt] at (4.8,5.8) {$s=n+1$};

  \foreach \x in {1,2,3}
    \draw (\x,2pt) -- (\x,-2pt) node[below, font=\small] {$\x$};


  \node[red!60!black, font=\small, above right=2pt] at (5.5,5.5) {$s=n$};

\end{tikzpicture}
    \caption{
    The representations of the massive little group in $D=5$ are labelled by $(s,n)$ for $s \geq n$.  
    The totally symmetric states correspond to the one-dimensional line $n=0$, such that $s\to\infty$ is the only classical scaling available. The other boundary of the physical region is the line $s=n$, populated by the pure (A)SD states, these also admit only one classical scaling, namely $s \to \infty$ where $s-n=0$, a constant. The interior of the region is populated by mixed tensor states, and there are multiple scalings one can use when taking a large spin limit. Taking the path traced by $n=1$, is consistent with the scaling $s \to \infty$ while $n/s \ll 1$,  and the path $s=n+1$, satisfies the scalings $s\to \infty$ while $s-n =1$. Note that while we only indicated integer $s, n$ on the graph, $s$ can be half-integer.}
    \label{fig:cllimit}
\end{figure}
Note that the last limit is a generalisation of the first case which corresponds to $\gamma=1$. It is also instructive to remark that the classical limit of three-point amplitudes involving such states has been considered in ref.~\cite{Alessio:2025nzd} for the open superstring amplitudes in arbitrary dimensions. There it was observed that if $s>0$ the classical limit of the string amplitudes is the same as the one involving fully symmetric states. However, a different multipole expansion was found for the case $s=n$, which corresponds to the states scattered above. The truncation we found is consistent with the results in ref.~\cite{Alessio:2025nzd}, as setting the spacetime dimension to five in their analysis, some of the covariant building blocks vanish. The generic Bessel functions found in the string setup are recovered by working in a higher number of dimensions. 

\par The first main result of this section is showing that the classical limit of amplitudes involving m(A)SD states spans the same space of multipole expansions as the fully symmetric states. Since these amplitudes are intrinsically chiral, we have to be careful with the action of the parity map and its behaviour under the classical limit. 
\par In the second half of the section we consider sector-flipping scattering. Although the amplitude expansion has a structure almost identical to that of the sector-preserving case, its classical limit leads to the emergence of the longitudinal part of the spin operator. We discuss how it emerges in the spinor-helicity framework and show that the classical limit leads to a single $K$-multipole contribution instead of a tower of multipoles given in powers of $k\cdot K$.
\subsection{Sector-preserving scattering}\label{sec:SDASD}

The most general amplitude between a self-dual and an anti-self-dual state is:
\begin{align} \label{eq:uneqSpinAmp}
    &\mathcal{A}(1^{(2s,2s-2n)}, 2^{(2s-2n,2s)}, 3^2) =\langle \mathbf{12}]^{2n}  \sum_{v=0}^{2s-2n} \Bigg\{(p_1 \cdot \varepsilon_{3})^2  g_{v,0,n} (\langle \mathbf{12}] [ \mathbf{12}\rangle)^{2s-2n-v} (\langle\mathbf{12}\rangle[\mathbf{12}])^{v} \nonumber\\
    +&(p_1 \cdot \varepsilon_{3}) \left(g_{v,1,n}[ \mathbf{12}\rangle \langle\mathbf{13}\rangle[\mathbf{23}\rangle + \tilde{g}_{v,1,n} \langle \mathbf{12}][\mathbf{13}\rangle\langle\mathbf{23}\rangle \right)  (\langle \mathbf{12}] [ \mathbf{12}\rangle)^{2s-2n-1-v}(\langle\mathbf{12}\rangle[\mathbf{12}])^{v}\nonumber\\
    +&\left(g_{v,2,n}[ \mathbf{12}\rangle^2 \langle\mathbf{13}\rangle^2[\mathbf{23}\rangle^2 + \tilde{g}_{v,2,n} \langle \mathbf{12}]^2[\mathbf{13}\rangle^2\langle\mathbf{23}\rangle^2 \right) (\langle \mathbf{12}] [ \mathbf{12}\rangle)^{2s-2n-2-v} (\langle\mathbf{12}\rangle[\mathbf{12}])^{v}\Bigg\}, 
\end{align}
where $\tilde{g}_{v,1,n}=0$ for $v=2s-2n$; $\tilde{g}_{v,2,n}=0$ for $v=2s-2n,\ 2s-2n-1$; $g_{v,1,n}=0$ for $v=2s-2n \text{ and } n=0$, $g_{v,2,n}=0$ for $v=2s-2n \text{ and } n=0,1$ or $v=2s-2n-1 \text{ and } n=0$. 

The main difference between this amplitude \eqref{eq:uneqSpinAmp} and the one in eq.~\eqref{eq:gen3ptAmpSpinor} where we scatter totally symmetric states is the unequal amount of massive spinors captured by the overall $\langle \mathbf{12}]^{2n} $ factor. However, as discussed in section \ref{sec:5dSOp}, this spinor structure does not have any dependence on the expectation values of the spin operator, only contributing an overall normalization factor \eqref{Spin1/2Map}. Therefore, the only difference between this scenario and the scattering of totally symmetric states is the appearance of the factor $n$ in the representation change formulae \eqref{expvalreln}. For instance, the $(p_1\cdot \varepsilon_{3})^2$ contribution from the first line in eq.~\eqref{eq:uneqSpinAmp} is:
\begin{equation}
     \langle \mathbf{12}]^{2n} \sum_{v=0}^{2s-2n}(\langle \mathbf{12}] [ \mathbf{12}\rangle)^{2s-2n-v} (\langle\mathbf{12}\rangle[\mathbf{12}])^{v}  \longrightarrow \sum_{v=0}^{2s-2n} \frac{g_{v,0,n}}{c_{0,0,v}}\lexpBig \left(\frac{k \cdot S \cdot S \cdot k}{m^2}\right)^{v} \rexpBig\,,
\end{equation}
where $c_{0,0,v} = (-2)^v (v!)^2 \binom{2s}{v}\binom{2s-2n}{v}$ is the combinatorial factor defined in eq.~\eqref{eq:combinatorialrepC} relevant for this state. The structure of the spin multipoles is unchanged, the only difference is an explicit dependence on $n$ (besides the usual $s$-dependence) in the factor, $g_{v,0,n}/c_{0,0,v}$, that we ultimately want to identify with the classical multipole coefficient, which should be independent of any quantum numbers. Note if we just sent $(s,n)$ to zero we would not recover any classical spin dependence. As illustrated in figure~\ref{fig:cllimit}, for $n\neq 0$, we have three possible classical limits.

\par Let us focus on the classical limit characterised by the scalings $s\rightarrow\infty$ with $\frac{n}{s}\ll1$. This does not specify a specific path through the $(s,n)$ space, it only requires that the number of anti-symmetrised indices is parametrically smaller than the symmetrised ones. If $n\ll s$, then the $n$ dependence in the combinatorial factors $c_{i,j,v}$ is suppressed and the factors reduce to the totally symmetric scattering scenario. Indeed, just as in the totally symmetric case, the combinatorial factors introduce a divergence in $s$, e.g. $c_{0,0,v}$ scales as $s^{(2v)}$ limit. Therefore requiring a finite classical limit demands $g_{v,0,n}\overset{s\rightarrow\infty}{\sim}s^{(2v)}$. As the $n$ dependence drops out, the required scaling for each spin multipole to survive in the classical limit is the same as the totally symmetric states and given in eq.~\eqref{eq:couplingscalings}. In this limit the range of the sums is pushed to infinity and the final classical amplitude recovers the full tower of multipoles of the $T_{\mu\nu}$ parametrising the biaxially symmetric solutions \eqref{THodgeAns}, i.e. it will also reproduce $\tilde{S}$ multipoles unless we add additional constraints that we will discuss at the end of this subsection.

\par We focus now on the second possible scaling where $s-n$ is kept finite, while $s\rightarrow\infty$. This corresponds to tracing a path through the space of scattering amplitudes where we increase the representations by increasing the number of $\zeta^{\mu\nu}$ tensor polarisations while keeping $(\varepsilon^{\mu})^{s-n}$ fixed. Intuitively, this means that we are approaching increasingly self-dual states in this limit. How is that reflected in the classical amplitude? Not much changes at the level of the combinatorial factors, they are slightly less divergent relative to the previous case such that $c_{i,j,v} \overset{s\rightarrow\infty}{\sim} s^{v+i}$. This translates to the constraint on the coupling constants being weaker,
\begin{equation}
    g_{v,0},\, \tilde{g}_{v,1}, \,\tilde{g}_{v,2}  \overset{s\rightarrow\infty}{\sim} s^{v}\,, \quad g_{v,1} \overset{s\rightarrow\infty}{\sim} s^{v+1}\,, \quad g_{v,2} \overset{s\rightarrow\infty}{\sim} s^{v+2}\,.
\end{equation}
However, the more drastic difference in this limit is that the sums in eq.~\eqref{eq:uneqSpinAmp} remain finite. Therefore we can only ever reproduce a classical solution up to a finite $2s-2n$ order in the multipole expansion from eq.~\eqref{THodgeAns}. A limiting case is where $s=n$. This corresponds to scattering states whose polarisation tensor is fully self-dual. In this case the amplitude contains only three terms
\begin{multline}
    \mathcal{A}(1^{(n,0)},2^{(0,n)},3^2)=(p_1\cdot \varepsilon_{3})^2 g_{0,0,n}\angsq^{2n}+(p_1\cdot \varepsilon_{3})g_{0,1,n}\angsq^{2n-1}\langle\mbf{1}\mbf{3}\rangle[\mbf{2}\mbf{3}\rangle\\+g_{0,2,n}\angsq^{2n-2}\langle\mbf{1}\mbf{3}\rangle^2[\mbf{2}\mbf{3}\rangle^2,
\end{multline}
but we can fix $g_{0,0,n}$ and $g_{0,1,n}$ by fixing the overall normalisation of the amplitude and imposing the universality of the spin dipole coupling. In mapping the spinor invariants to the classical expectation values, we observe that for the $s=n$ case considered above, the relations \eqref{expvalreln} give $\lexp S_{\varepsilon k}\rexp=\lexp \tilde{S}_{\varepsilon k}\rexp=\varepsilon\cdot \langle s_L\rangle\cdot  k$ and similarly for the square powers. Hence classically there is only one spin tensor satisfying $S_{\varepsilon k}=\tilde{S}_{\varepsilon k}$. We also observe that the invariants $k\cdot S\cdot S\cdot k$ do not appear. This happens for the SD classical spin tensors. The classical amplitude will only include terms up to a spin-quadrupole and spans solutions of the type 
\begin{equation}
    T_{\mu\nu}=m u_\mu u_\nu-i u_\mu (S\cdot k)_\nu-4C_{0,3}\frac{(S\cdot k)_\mu(S\cdot k)_\nu}{m},
\end{equation}
where we have related the remaining free coupling to the spin-quadrupole Wilson coefficient $g_{0,2,n}=-4n(n-1)C_{0,3}$. This family of GR solutions includes the Myers-Perry black hole whose spin vectors have the same magnitude, $a_1=a_2$. The truncation of the multipole expansion for equal-spin Myers-Perry follows from the invariant $k\cdot S\cdot S\cdot k$ being proportional to $a_1^2 - a_2^2$. These black hole solutions have an enhanced rotation group given by $U(2)$. Beyond the limiting case $s=n$, the linearised energy momentum tensor will still truncate but at some constant $(s-n)$-multipole, the physical relevance would presumably depend on the specific classical solution.

\par The third classical limit is a generalization of the first one and is characterised by the scalings $s\rightarrow\infty$, $s-n\rightarrow\infty$ and $\frac{s-n}{s}=\gamma=\text{ constant}$. This corresponds to increasing the spin-representations by fixing the number of anti-symmetrised indices $(\zeta)^n$ and increasing the rank of the symmetrised indices $(\varepsilon)^{s-n}$. As in the first case, the classical limit leads to an infinite multipole expansion as the finite sums are extended to infinite sums. The combinatorial factors still diverge, though they differ by factors of $\gamma$, i.e. $c_{0,0,v} \overset{s\to\infty}{\sim} \gamma^{v}s^{2v}$.

\par So far the classical limits generically reproduce multipoles for biaxially symmetric solutions; the multipole expansion will include multipoles involving $\tilde{S}$ unless we fine tune the parameters. In the previous section, we argued that imposing parity invariance at the spinor level was the correct constraint that restores the parity of $T_{\mu\nu}$. This constraint is particularly simple to impose on the fully symmetric amplitudes as the action of parity is closed on them. This is not the case for the sector-preserving amplitudes. To see this, let us call the parity operator $\mathbb{P}$. More details on its action and definition are in Appendix \ref{appconv}. The parity reversed amplitude is:

\begin{equation} 
    \mathbb{P}(\mathcal{A}(1^{(2s,2s-2n)}, 2^{(2s-2n,2s)}, 3^2)) = \mathcal{A}((\mathbb{P}1)^{(2s-2n,2s)}, (\mathbb{P}2)^{(2s,2s-2n)}, (\mathbb{P}3)^2)
\end{equation}
where on the right hand side $\mathbb{P}1, \mathbb{P}2, \mathbb{P}3$ are momenta defined with $p_{i=1,2; 4} \to - p_{i=1,2; 4}$ for the massive fields and $k_4 \to - k_4$ for the massless one. As a consequence of parity, the two $SU(2)$ copies in the little group also swap. As such, the overall number of square and angle spinors inside the amplitude swaps. 

At this point there seems to be an apparent conflict. The parity operation does not impose a relation between the coefficients of a single amplitude as the spinor blocks differ between the initial amplitude and the final one due to the angle to square swap. Its behaviour is closer to the four-dimensional case where parity relates different helicity amplitudes. However, we saw that in the classical limit, we can recover parity-invariant solutions starting from this class of amplitudes. We can then ask: what constraint should be imposed on the amplitude coefficients such that they lead to a parity-invariant multipole expansions? 

\par
\begin{figure}[ht]
\centering

\begin{minipage}[c]{0.28\textwidth}
\centering
\begin{tikzpicture}[>=stealth, thick, scale=1.8]

\node (A) at (0,0) {$\mathcal{A}$};
\node[anchor=west] at ($(A.north east)+(-0.12,-0.1)$)
  {\scriptsize $\mathrm{symmetric}$};

\draw[->]
  (A.west)
  .. controls +(-1.2,1.0) and +(-1.2,-1.0) ..
  node[pos=0.5,left] {$\mathbb{P}$}
  (A.west);

\end{tikzpicture}
\end{minipage}
\hfill
\begin{minipage}[c]{0.68\textwidth}
\centering
\begin{tikzpicture}[>=stealth, thick]
  \node (L) at (0,0) {$\mathcal{A}(|p_i\rangle,|p_i]; p_i)$};
  \node (R) at (5.6,0) {$\mathcal{A}(|p_i],|p_i\rangle; \mathbb{P}p_i)$};
  \node (C) at (2.8,-3.1) {classical amplitude};

  \draw[<->] (L) -- node[above] {$\mathbb{P}$} (R);
  \draw[->] (L) -- node[midway, sloped, above] {$s \to \infty$} (C);
  \draw[->] (R) -- node[midway, sloped, above] {$s \to \infty$} (C);
\end{tikzpicture}
\end{minipage}
\caption{Parity implements constraints on amplitudes in different ways depending on the scattered states. \textit{Left}: An amplitude described by fully symmetric states maps to another amplitude in the same class under parity. Thus parity invariance imposes a constraint. \textit{Right}: The amplitudes describing the scattering of a mSD to a mASD state are not closed under parity as the square and angle spinors change. Parity is only a constraint in the classical limit.}
\label{paritydiagram}

\end{figure}
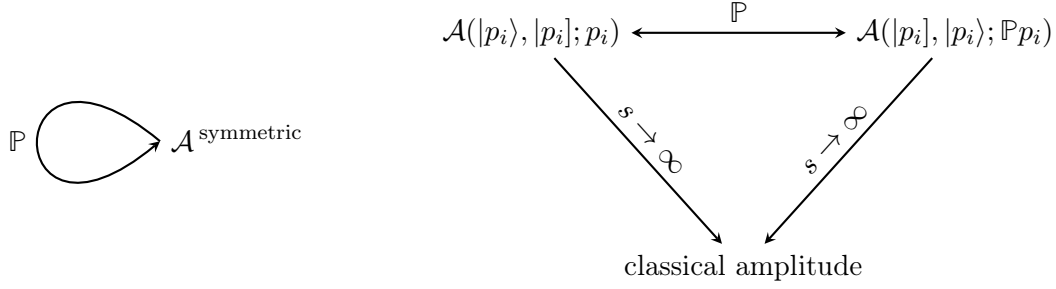

 The resolution comes from the behaviour of these amplitudes in the various classical limits. For now we consider the first, where $n\ll s$ and both $\mathcal{A}$ and $\mathbb{P}(\mathcal{A})$ map to similar expansions. Then, while we cannot impose a relationship between $\mathcal{A}$ and $\mathbb{P}(\mathcal{A})$ while keeping $s$ finite, we can impose:
\begin{equation}\label{paritycondition}
    (\mathcal{A}-\mathbb{P}(\mathcal{A}))|_{s\rightarrow\infty}=0.
\end{equation}
We summarized the action and constraints imposed by parity in Figure \ref{paritydiagram}.
\newline
In order to see the constraint on the coefficients, we focus on the $p_1\cdot \varepsilon_{3}$ terms in eq.~\eqref{eq:uneqSpinAmp}. In the classical limit, there will be a term multiplying $S_{\varepsilon k}$ and one for $\tilde{S}_{\varepsilon k}$. We summarize the different coefficients that can appear in Table \ref{paritycompare}. We then see that the condition \eqref{paritycondition} requires $g_{v,1,n}=-\tilde{g}_{v,1,n}+\mathcal{O}(\frac{n}{s})$, hence the terms involving the Hodge dual $\tilde{S}$ vanish in the classical limit. A similar analysis can be applied to the $g_{v,2,n},\ \tilde{g}_{v,2,n}$ coefficients.

\begin{table}[ht]
\centering
\caption{Parity action on $p_1\cdot \varepsilon_{3}$ coefficients}

\label{paritycompare}
\renewcommand{\arraystretch}{2} 
\setlength{\tabcolsep}{14pt}      

\begin{tabular}{|c|c|c|}
    \hline
     & 
     $\displaystyle
     \lexpBig 
     \frac{S_{\varepsilon k}}{m}
     \left(\frac{k \cdot S \cdot S \cdot k}{m^2}\right)^v
     \rexpBig$ 
     & 
     $\displaystyle
     \lexpBig 
     \frac{\tilde{S}_{\varepsilon k}}{m}
     \left(\frac{k \cdot S \cdot S \cdot k}{m^2}\right)^v
     \rexpBig$ 
     \\
     \hline

    $\mathcal{A}$ 
    & 
    $\displaystyle
    \left(
    \frac{g_{v,1,n}}{2s-v}
    -
    \frac{\tilde{g}_{v,1,n}}{2s-2n-v}
    \right)$ 
    & 
    $\displaystyle
    \left(
    \frac{g_{v,1,n}}{2s-v}
    +
    \frac{\tilde{g}_{v,1,n}}{2s-2n-v}
    \right)$
    \\
    \hline

    $\mathbb{P}(\mathcal{A})$ 
    & 
    $\displaystyle
    \left(
    \frac{g_{v,1,n}}{2s-v}
    -
    \frac{\tilde{g}_{v,1,n}}{2s-2n-v}
    \right)$ 
    & 
    $\displaystyle
    -\left(
    \frac{g_{v,1,n}}{2s-v}
    +
    \frac{\tilde{g}_{v,1,n}}{2s-2n-v}
    \right)$
    \\
    \hline
\end{tabular}

\end{table}

Indeed, looking at Table \ref{paritycompare}, the discussion regarding parity invariance also applies when we scale $s\rightarrow\infty$, $s-n\rightarrow\infty$ and $\frac{s-n}{s}=\gamma=\text{ constant}$. The right parity condition is still eq.~\eqref{paritycondition}. However, the $\gamma\neq 1$ does impose modified requirements on the couplings, e.g. $\frac{g_{v,1,n}}{\tilde{g}_{v,1,n}}\sim -\gamma^{-1}+\mathcal{O}(s^{-1})$.
\par In the second classical limit we can still impose the condition \eqref{paritycondition}. For $s-n\neq 0$, the classical limit will have both $S$ and $\tilde{S}$ contributions, but with a finite multipole expansion. The parity conditions are those from the Table \ref{paritycompare} for $v<2s-2n$ but require $g_{2s-2n,1,n}=0$ as it will bring a contribution linear in $\tilde{S}\,(k\cdot S\cdot S\cdot k)^{2s-2n}$ that cannot be cancelled otherwise. We note that for $s=n$ the solution is SD and $S\propto \tilde{S}$ so the solution is already parity covariant, but as soon as $n<s$, the classical solution moves away from the SD point and will involve an odd number of Levi-Civita tensors.

\subsection{Sector-flipping scattering}\label{sec:SDSD}

In the previous subsection we used states that obey $n_{R,2}=n_{L,1}$ and $n_{L,2}=n_{R,1}$, meaning that they are of the same type in our all-incoming momenta conventions. However, we argued in section~\ref{sec:5dStates} that we could scatter states such that $n_{R,2}=n_{R,1}$ and $n_{L,2}=n_{L,1}$ instead. As with the previous conventions, the most general such amplitude is given by:
\begin{align}\label{SDSD}
    &\mathcal{A}(1^{(2s,2s-2n)},2^{(2s,2s-2n)},3^2)= \ang^{2n}\sum_{v=0}^{2s-2n}\Bigg\{(p_1\cdot \varepsilon_{3})^2\ \mathsf{g}_{v,0,n}  (\angsq \sqang)^{2s-2n-v}(\ang\sq)^{v} \nonumber\\
    +&(p_1\cdot \varepsilon_{3})\ ( \mathsf{g}_{v,1,n}\angsq\ASD  + \tilde{\mathsf{g}}_{v,1,n}\sqang\SD  )(\angsq \sqang)^{2s-2n-v-1}(\ang\sq)^{v}\nonumber\\
    +&( \mathsf{g}_{v,2,n} \angsq^2(\ASD)^2 + \tilde{\mathsf{g}}_{v,2,n}\sqang^2 (\SD)^2 )(\angsq\sqang)^{2s-2n-v-2}(\ang\sq)^{v}\Bigg\}.\end{align}
where the constraints on the coefficients $\mathsf{g}_{v,i,n}$ are the same as those on $g_{v,i,n}$ below eq.~\eqref{eq:uneqSpinAmp}.

\par The structure of the scattering amplitude, given in spinor variables, is not so different from the previous case \eqref{sec:SDASD}, only now the mismatch in the spinors of the two massive legs is encoded in the $\langle{\bm 1 \bm 2}\rangle^{2n}$ factor out front. However, this small change in building blocks markedly changes the physics in the classical limit. As discussed in section \ref{sec:5dSOp}, this spinor variable captures non-trivial degrees of freedom of the Lorentz generator in the direction of $p^{\mu}$,
\begin{equation}
    \langle{\bm 1 \bm 2}\rangle = \frac{1}{2} k \cdot [\bar{\kappa}\rangle\,.
\end{equation}

The rest of the amplitude structure can be fully expressed in terms of spin-tensor expectation values such that the amplitude, when expressed in terms of lowest-representation building blocks, is 
\begin{align}\label{SDSD}
    \mathcal{A}(1^{(2s,2s-2n)},2^{(2s,2s-2n)},3^2)= &\left(\frac{k \cdot [\bar{\kappa}\rangle}{2}\right)^{2n}\, \sum_{v=0}^{2s-2n}\Bigg\{ \left(\frac{k {\cdot} \la s_L \ra {\cdot} [s_{R}] {\cdot} k}{8m^2}\right)^{v} \times \nonumber\\
    \Bigg( &(p_1\cdot \varepsilon_{3})^2\ \mathsf{g}_{v,0,n} -p_1 {\cdot} \varepsilon_{3}\, \frac{\mathsf{g}_{v,1,n}\,\varepsilon_{3} {\cdot} \la s_L \ra {\cdot} k -\tilde{\mathsf{g}}_{v,1,n}\,\varepsilon_{3} {\cdot}  [s_{R}] {\cdot} k 
    }{2}\nonumber\\
    +&\frac{\mathsf{g}_{v,2,n}\, (\varepsilon_{3} {\cdot} \la s_L \ra {\cdot} k)^2 -  \tilde{\mathsf{g}}_{v,2,n}\,(\varepsilon_{3} {\cdot}  [s_{R}] {\cdot} k)^2 }{4}\Bigg)\Bigg\}.
\end{align}

Since we still need to change the representation, we cannot read off the classical amplitude yet, however we can already note some key features that will survive in the classical limit. Firstly, the behaviour under parity follows a similar logic to the sector-preserving amplitudes above; for finite $s$ parity relates two different amplitudes, but the classical limit will require a relation between $\mathsf{g}_{v,i}$ and $\tilde{\mathsf{g}}_{v,i}$ in the classical limit. Secondly note that the $[\bar{\kappa}\rangle$ factor does not appear in the summand, so there is only a single power $n$ that appears. The representation change operation does not change the number of insertions of each operator, see eq.~\eqref{eq:Krepchanges} in appendix \ref{appconv}, so we can expect the final classical amplitude to inherit this and only reproduce a single $K^{\mu}$ multipole.

\par We now look at the classical limit scalings from the previous subsection. Consider the first scaling, $s \to \infty$ and $n/s \ll 1$. The combinatorial factor due to representation change is modified by the appearance of the $[\bar{\kappa}\rangle$ factor and are defined in appendix~\ref{appconv} in eq.~\eqref{eq:combinatoricd}. For example the combinatorial factor related to the $\mathsf{g}_{v,0,n}$ terms scales as  $d_{0,0,n} \overset{s \to \infty}{\sim} s^{2n+2v}$, such that all the couplings in this case will need to scale by an additional $s^{2n}$ in order to be classically relevant. If this constraint is satisfied, the classical amplitude will reproduce the freedom in the following classical stress-energy tensor
\begin{align}
   T_{\mu\nu}(k) = &m u_\mu u_\nu \Bigg( 1 + \sum_{n'=1}^{\infty} \frac{C_{n',n,1}}{(2n')!\,n!}\left(\frac{k \cdot S \cdot S \cdot k}{-m^2}\right)^{n'}\left(\frac{k\cdot K}{im}\right)^{n} \Bigg) \nonumber\\
   -i&u_{(\mu}(S \cdot k)_{\nu)}\Bigg( 1 + \sum_{n'=1}^{\infty}\frac{2 C_{n',n,2}}{(2n'+1)!\,n!}\left(\frac{k \cdot S \cdot S \cdot k}{-m^2}\right)^{n'}\left(\frac{k\cdot K}{im}\right)^{n} \Bigg) \nonumber\\
   - &\frac{( S\cdot k)_\mu( S\cdot k)_\nu}{m} \sum_{n'=0}^{\infty}\frac{4 C_{n'+1,n,3}}{(2n'+2)!\,n!}\left(\frac{k \cdot S \cdot S \cdot k}{-m^2}\right)^{n'}\left(\frac{k\cdot K}{im}\right)^{n}\,,
\end{align}
where we have imposed a parity constraint on the amplitude such that there are no $\tilde{S}$-dependent multipoles. The analysis in the case of the scaling $s, (s-n) \to \infty$ with $\gamma = (s-n)/s=$ constant generates the same classical amplitude as long as the scalings of the couplings are modified to account for $\gamma$ suppressions.

\par However scaling $s, n\to \infty$ such that $s-n=$ constant in this scenario cannot generate anything consistent beyond the trivial mass monopole and dipole given that any higher spin-tensor multipoles would include a divergent $(k \cdot K)^{\infty}$ multipole.

Therefore no matter what classical scaling is used we can only parameterise trivial stress-energy tensors or those with boost multipoles at a fixed $n$-th multipole. This is a feature of scattering fixed spin states, and in the next section we will see that we are able to recover a full classical expansion in $K^\mu$ by constructing coherent states that allow us to scatter a range of spin-states simultaneously.

\section{Scattering coherent states}\label{sec:KVecScat}

So far we have seen that scattering amplitudes between massive states that satisfy the constraints $n_{L,1} = n_{R,2} = 2s$ and $n_{R,1}=n_{L,2} = 2s-2n$ can be expressed in expectation values of only the spin-tensor operator and its Hodge dual. Thus they span the set of linearised energy-momentum tensors for a massive worldline with only spin degrees of freedom. However in the previous section we also considered amplitudes between a state and its dual, i.e. the spin-quantum numbers of the massive legs satisfied $n_{L,1} = n_{L,2} = 2s$ and $n_{R,1}=n_{R,2} = 2s-2n$ instead. We observed that their classical limit introduces an additional degree of freedom, the longitudinal part of the angular momentum $K^\mu$. The states chosen in these amplitudes are part of a more general class of states that satisfy
\begin{equation}\label{genstates}
    n_{L,1}=n_{R,2}+\Delta S = 2s\,, \qquad n_{R,1}=n_{L,2}-\Delta S=2s-2n,
\end{equation}
where $\Delta S$ characterises how the mismatch between the left and right little group components of the two massive fields. The case treated previously corresponds to the case where $\Delta S= 2n$. In this more general class of amplitudes we can now study other $\Delta S$ values. For example, the interaction $\zeta^{\mu\nu}h_{\mu\tau}\partial_\nu\varepsilon_\tau$, which is between a massive spin-1 vector $(n_{L,1}=1,n_{R,1}=1)$ and a massive spin-1 tensor $(n_{L,2}=2, n_{R,2}=0)$ interacting with a graviton, would correspond to $\Delta S = 1$.   

\par The general scattering amplitudes characterising the scattering for states in eq.~\eqref{genstates}, where $\Delta S\geq 2n$ can be written compactly as\footnote{The amplitudes for the cases where $\Delta S < 2n$ correspond to sending $\Delta S \leftrightarrow 2n$ in all the terms above, including the end range of the sum, but leaving the $\ang^{\Delta S}$ prefactor unchanged.}
\begin{multline}\label{AmpShift}
    \mathcal{A}(1^{(n_L,n_R)},2^{(n_R+\Delta S,n_L-\Delta S)},3^2)=\ang^{\Delta S}\sqang^{\Delta S-2n} \sum\limits_{v=0}^{\Delta S}(\ang\sq)^{v} (\angsq \sqang)^{2s-\Delta S-v-2}\times\\
    \times\Bigg\{\mathfrak{g}_{\Delta S, 0,v}(p_1\cdot \varepsilon_{3})^2 (\angsq \sqang)^{2}
    +(p_1\cdot \varepsilon_{3}) \angsq \sqang\left(\mathfrak{g}_{\Delta S, 1,v}\angsq\ASD+\mathfrak{g}_{\Delta S, 2,v}\SD \sqang\right)  \\
    + \left(\mathfrak{g}_{\Delta S, 3,v}(\angsq\ASD)^2+\mathfrak{g}_{\Delta S, 4,v}(\sqang\SD)^2
    \right) \Bigg\}.
\end{multline}
Similarly to the previous section, the prefactor $\ang^{\Delta S}$ will be rewritten as a fixed power $\Delta S$ of the expectation of $K^\mu$ and is independent of the summation. In order to recover an infinite multipolar expansion in the classical boost vector $K^{\mu}$ we need to introduce a sum over possible spin shifts $\Delta S$. This naturally leads us to considering coherent states of the massive spinning states, as these will include a summation over all possible field configurations. Coherent spin states were set up and studied for $D=4$ massive spinning states in ref.~\cite{Aoude:2021oqj} following the construction of Schwinger in ref.~\cite{Schwinger:1952dse}. In the next section we will set up a construction in a similar flavour to ref.~\cite{Aoude:2021oqj}. In principle these scattering amplitudes will include scenarios where $n_{L,1}+n_{R,1}\neq n_{L,2}+n_{R,2}$, but we will argue that these are suppressed in the classical limit following a similar argument as the authors used in ref.~\cite{Aoude:2021oqj} in order to claim spin-changing amplitudes were suppressed in $D=4$.

\subsection{Coherent state amplitudes}\label{sec:CohAmp}

Before introducing the coherent states, we will revisit the fixed spin-states in $D=5$ that we have been studying so far. As discussed extensively in sections \ref{sec:littlegroupsec} and \ref{sec:5dSH}, these states $|p; n_L,n_R\ra$ are labelled by their momentum $p$ and two spin-quantum numbers which correspond to the two $SU(2)$ copies in the massive little group. We can build these general fixed spin-states from the zero-spin state $|p; 0,0\ra$  using Schwinger's construction \cite{Schwinger:1952dse, Atkins:1971zy, Radcliffe:1971ayi, Perelomov:1977}. In order to do so we will define the oscillator modes of the two $SU(2)$ copies, $a_a$ and $b_{\dot{a}}$. These satisfy the usual oscillator algebra
\begin{equation}
    [a^a,a^\dagger_b]=\delta^a_b,\  [b^{\dot{a}},b^\dagger_{\dot{b}}]=\delta^{\dot{a}}_{\dot{b}},
\end{equation}
where all other commutators vanish. Then by acting with the creation operators we can construct a general massive spinning state
\begin{equation}
    |p;n_L, n_R\ra =\frac{(z \cdot a^\dagger)^{n_L}(w \cdot b^\dagger)^{n_R}}{\sqrt{(n_L)!(n_R)!}}|p;0,0\ra.
\end{equation}
We have absorbed the little group indices with the usual spinors $z^a, w^{\dot{a}}$. As discussed in section \ref{sec:5dSH}, the on-shell degrees of freedom for $|p;n_L, n_R\ra$ correspond to the massive spinor variables $|{\bm p}\ra^{n_L} |{\bm p}]^{n_R}$.

We can define the spin and boost operators that act on $|p;n_L, n_R\ra$ by absorbing the lowest weight spin and boost operators defined in \eqref{eq:LRspinops} and \eqref{eq:kappaops} by absorbing the little group indices with the relevant oscillators
\begin{align}
    \mathsf{S}_{L}^{\mu\nu} =a^\dagger_{a}s_{L}^{\mu\nu,a}{}_{b} a^{b}\,, \qquad \mathsf{S}_{R}^{\mu\nu} &=b^\dagger_{\dot{a}} s_{R}^{\mu\nu,\dot{a}}{}_{\dot{b}} b^{\dot{b}}\,, \qquad \mathsf{K}^{\mu} =a^\dagger_{a} \kappa^{\mu,a}{}_{\dot{b}} b^{\dot{b}}\,, \qquad \bar{\mathsf{K}}^{\mu} =b^\dagger_{\dot{a}} \bar{\kappa}^{\mu,\dot{a}}{}_{\dot{b}} a^{b}\,.
\end{align}
These operators satisfy the spin operator identities, which are explained in Appendix \ref{spinApp}. The expectation value of $\mathsf{S}_{L}^{\mu\nu}$ with respect to two finite spin states follows from the commutation algebra of the oscillators such that
\begin{equation}\label{eq:fixedSpinExps}
    \la p;\bar{n}_L, \bar{n}_R| \mathsf{S}_L^{\mu\nu} |p;n_L, n_R\ra = n_L \la s_{L}^{\mu\nu}\ra (\bar{z}\cdot z)^{n_L-1}(\bar{w}\cdot w)^{n_R} \delta_{n_{L},\bar{n}_L} \delta_{n_{R},\bar{n}_R} \,.
\end{equation}
The result is written in terms of the expectation of the $(1,0)$-rep. spin tensor, and vanishes if the states have mismatched quantum numbers. The expectation of $\mathsf{S}_{R}^{\mu\nu}$ behaves similarly, meanwhile the expectation values of $\mathsf{K}$ and $\bar{\mathsf{K}}$ require a mismatch of one between the quantum numbers, e.g.
\begin{equation}\label{eq:fixedKExps}
    \la p;\bar{n}_L, \bar{n}_R| \mathsf{K}^{\mu} |p;n_L, n_R\ra = \sqrt{(n_L+1)n_R}\, \la \kappa^{\mu}] (\bar{z}\cdot z)^{n_L}(\bar{w}\cdot w)^{n_R-1} \delta_{n_{L}+1,\bar{n}_L} \delta_{n_{R}-1,\bar{n}_R} \,.
\end{equation}
We can now define our coherent states 
\begin{align}\label{CohSpinExp}
    |p;z,w \ra&=e^{-|z|^2/2-|w|^2/2}e^{z^\al a_\al^\dagger+w^{\dot{\al}}b_{\dot{\al}}^\dagger}|p;0,0\ra\nonumber\\
    &=e^{-|z|^2/2-|w|^2/2}\sum_{n_{L,R}=0}^{\infty}\frac{1}{\sqrt{n_L!\,n_R!}}|p; n_L, n_R\ra\,,
\end{align}
which are inspired by constructions that are well-suited to describe classical angular momentum in four-dimensions \cite{Atkins:1971zy, Radcliffe:1971ayi,Perelomov:1974yw, Aoude:2021oqj}. 

\par From now on, in expectation values of operators $\mathcal{O}$, we will take our initial state to be the coherent state $|p;z,w \ra$ while our final state is $| p;\bar{z}, \bar{w}\ra^{\dagger}$,
\begin{equation}
    \langle \mathcal{O}\rangle_{z,w} := \la p;\bar{z}, \bar{w}| \mathcal{O} | p; z,w\ra
\end{equation}
i.e. the initial states have the same momentum but the little groups are distinguished by the LG spinors. Using the fixed-spin expectation values in eq.~\eqref{eq:fixedSpinExps},~\eqref{eq:fixedKExps} and given our choice of normalisation for the coherent states, we find the following simple expectation values
\begin{equation}\label{cohexpvalS}
    \la \mathsf{S}_L^{\mu\nu}\ra_{z,w}= \la s_{L}^{\mu\nu}\ra\,,\quad \la \mathsf{S}_R^{\mu\nu}\ra_{z,w}=[s_R^{\mu\nu}]\,,\quad \la \mathsf{K}^\mu\ra_{z,w} = [\kappa^{\mu}\ra\,,\quad  \la \bar{\mathsf{K}}^\mu\ra_{z,w} = \la\kappa^{\mu}]\,.
\end{equation}
One can also check that the algebra obeyed by these operators is the usual Poincaré one.

\par In the coherent state picture, the classical limit is not implemented by scaling the spin quantum numbers, indeed here we resum them. Instead, we need to reinsert $\hbar$ dependence, where we had previously set $\hbar =1$, and keep leading terms in the $\hbar \to 0$ limit \cite{Aoude:2021oqj}. One place where this dependence must be restored is in the expectation values of the spin and boost operators. This can be done by dimensional analysis such that we need to rescale $\la s_{L}^{\mu\nu}\ra, [s_{R}^{\mu\nu}], [\kappa^{\mu}\ra,\la\bar{\kappa}^{\mu}]$ by a factor of $\hbar$. It then follows that more insertions of the spin and boost operators have leading order behaviour
\begin{equation}\label{cohexpsplit}
    \la \mathsf{S}^{\mu\nu} \mathsf{S}^{\rho\tau}\ra_{z,w}=\la \mathsf{S}^{\mu\nu}\ra_{z,w}\la \mathsf{S}^{\rho\tau}\ra_{z,w}+\mathcal{O}(\hbar),
\end{equation}
and likewise for $\mathsf{K}, \bar{\mathsf{K}}$. Therefore the expectation values in eq.~\eqref{cohexpvalS} have the right classical behaviour at leading order in $\hbar \to 0$. However demanding the expectation values in eq.~\eqref{cohexpvalS} are finite as $\hbar \to 0$ demands that each external polarisation data scales as $|z|, |w|\sim \mathcal{O}(\hbar^{-1/2})$ such that $\la \mathsf{S}_L^{\mu\nu}\pm \mathsf{S}_R^{\mu\nu} \ra \rightarrow S_{\pm,\text{cl}}^{\mu\nu}$ and similarly for $K^\mu$.

\par Having defined the spin operator expectation values, the goal is to start from the spinor-helicity amplitude between coherent states and consider its classical limit. The result will be a sum over different spin configurations, 
\begin{equation}\label{eq:cohAmp}
    \mathcal{A}(1^{z,w},2^{\bar{z},\bar{w}},3^2) = \sum_{n_{L,R},\bar{n}_{L,R}=0}^{\infty} \frac{e^{-|z|^2-|w|^2}}{\sqrt{n_L! \, n_R! \,\bar{n}_L! \,\bar{n}_R! }}\mathcal{A}(1^{(n_L, n_R)},2^{(\bar{n}_R, \bar{n}_L)},3^2)\,.
\end{equation}
The fixed spin amplitudes in the summand,
\begin{equation}
    \mathcal{A}(1^{(n_L, n_R)},2^{(\bar{n}_R,\bar{n}_L)},3^2)\,,
\end{equation}
covering a broader class of amplitudes than we have considered so far. The most general scattering amplitude we have considered so far is given in eq.~\eqref{AmpShift} and is only valid for the cases $n_{L,1}+n_{R,1}= \bar{n}_L + \bar{n}_R$.\footnote{Note here we have slightly abused notation as we identify the quantum numbers $(n_{L,2}, n_{R,2})$ for incoming particle $p_2$ with the quantum numbers $(\bar{n}_L, \bar{n}_R)$ of the outgoing final state $\bar{p}$ where $n_{L,2} = \bar{n}_R$ and $n_{R,2} = \bar{n}_L$ due to the flip in the $SU(2)$'s induced by the reflection. See section \ref{sec:5dSOp} for more details on this.} We refer to scattering that satisfies this constraint as ``total spin-conserving''. In the upcoming subsection \ref{sec:CohAmpMatching} we will explore the space of classical amplitudes that arises if we only allow ``total-spin'' conserving amplitudes. 

\par However, in principle, the sum in eq.~\eqref{eq:cohAmp} also includes scenarios when $n_{L,1}+n_{R,1}\neq n_{L,2}+n_{R,2}$, where there is a ``total spin'' mismatch between the incoming and the outgoing states. We expect that the worldline EFT associated to these amplitudes would need more than just the spin and boost degrees of freedom which are spin-conserving since in $D=4$ these amplitudes encode absorption effects \cite{Aoude:2023fdm}. We can show that, if we fix some scaling of the coupling coefficients, the non-conserving amplitudes are suppressed relative to conserved case in the classical limit.

\par We can refer to eq.~(3.44) in ref.~\cite{Aoude:2024jxd} for the scattering amplitudes for non-total-spin-conserving cases. For now, let us focus on a particular contribution to this amplitude where the kinematic data of the graviton only appears as $(p_1\cdot \varepsilon_{3})^2$. We also pick the configuration where $n_{L}=n_{R}\equiv n_1$ and $\bar{n}_{L}=\bar{n}_{R}\equiv n_2$ with $n_1>n_2$ and $\Delta S=n_1-n_2$. Note that this configuration does not obey eq.~\eqref{genstates}. Then we will have the following contribution
\begin{equation}
    (p_1\cdot \varepsilon_{3})^2\sum_{n_1=0}^\infty \sum_{n_2=0}^\infty \sum_{2v=0}^{n_2}\frac{\mathfrak{g}_{v,n_1,n_2}}{(n_2)!(n_1)!}\angsq^v\sqang^v \sq^{-v+n_2}\ang^{-v+n_1}\left(\frac{\bar{u}_{1,\dot{a}} w^{\dot{a}}}{\bar{u}_{2,a} \bar{z}^a}\right)^{n_1-n_2}.
\end{equation}
where the $u,\bar{u}$ variables are a basis for the little group space introduced in \cite{Cheung:2009dc} and used in \cite{Chiodaroli_2022,Pokraka:2024fao}. We do not need their explicit definition here other than that the ratio $\frac{\bar{u}_{1,\dot{a}} w^{\dot{a}}}{\bar{u}_{2,a} \bar{z}^a}$ is finite as $\hbar 
\to 0$. Indeed, when boosting to the rest frame of the first field, there are three spinor invariants that are finite in this limit: $a_1=\sq,\ a_2=\ang$ and $a_3=\frac{\bar{u}_{1,\dot{a}} w^{\dot{a}}}{\bar{u}_{2,a} \bar{z}^a}$, while $X = \angsq \sqang \sim \mathcal{O}(\hbar^{-1})$. We can then reshuffle the sum in terms of these variables such that
\begin{equation}
     (p_1\cdot \varepsilon_{3})^2\sum_{n_2=0}^\infty (a_1 a_2)^{n_2}\sum_{\Delta S=1}^{\infty}\frac{(a_1 a_3)^{\Delta S}}{(n_2)!(\Delta S+n_2)!}\sum_{2v=0}^{n_2}\mathfrak{g}_{v,\Delta S{ +} n_2,n_2}X^{2v}(a_1a_2)^{-v},
\end{equation}
The advantage of separating the sum as above is that, we can bound the coefficients $\mathfrak{g}_{v,\Delta S{ +} n_2,n_2}\leq \max_{v}(\mathfrak{g}_{v,\Delta S{ +} n_2,n_2})$ \cite{Aoude:2021oqj} such that one can bound the sum in $v$ by the following resummation
\begin{equation}
    \frac{1}{1-\frac{X^2}{a_1 a_2}} \sim \mathcal{O}(\hbar)\,.
\end{equation}
This term can be factored out of the other nested sums and scales with $\hbar$ due to the non-trivial scaling of $X$. This extra scaling is generic for the non-total-spin-conserving contributions. Suppose then that the scaling of the amplitude coefficients $\mathfrak{g}$ with $\hbar$ is the same across all the amplitudes in eq.~\eqref{eq:cohAmp}. We remark then that among the generic amplitude contributions, the ones coming from non-total-spin-conserving amplitudes will have extra factors of $\hbar$ as above and hence will be subleading in the classical limit. We will then ignore the effect of these amplitudes in the classical limit. However, we note that such amplitudes appear in four dimensions as well \cite{Aoude:2021oqj} and their effects have been linked to the inelastic observables in ref.~\cite{Aoude:2024jxd}.

\subsection{Matching to spacetime solutions}\label{sec:CohAmpMatching}

Here we will restrict our analysis of the coherent state amplitudes such that we only allow ``total-spin'' conserving amplitudes to appear in the summand,
\begin{align}
    \mathcal{A}(1^{z,w},2^{\bar{z},\bar{w}},3^2) e^{|z|^2+|w|^2} &= \!\!\!\sum_{\substack{n_{L,R}=0\\\bar{n}_{L,R}=0}}^{\infty} \frac{\delta_{n_L+n_R,\, \bar{n}_L + \bar{n}_R}}{\sqrt{n_L! \, n_R! \,\bar{n}_L! \,\bar{n}_R! }}\mathcal{A}(1^{(n_L, n_R)},2^{(\bar{n}_R, \bar{n}_L)},3^2) 
\end{align}
We expect these to reconstruct the stress-energy tensor in eq.~\eqref{TNonAxi} including the $K^\mu$ contributions. 
For simplicity let us consider only the terms proportional to $(p_1 \cdot \varepsilon_{3})^2$ in eq.~\eqref{AmpShift}. Note that we have identified $n_L = \bar{n}_L - \Delta S = 2s$ and $n_R = \bar{n}_R + \Delta S = 2s -2n$,
\begin{multline}
    \mathcal{A}(1^{z,w},2^{\bar{z},\bar{w}},3^2)\Big|_{(p_1 \cdot \varepsilon_{3})^2} = e^{-|z|^2-|w|^2}\sum_{n_L, n_R=0}^\infty \sum_{\Delta S=0}^{n_L} \mathcal{N}_L^{-1}\mathcal{N}_R^{-1} \ang^{\Delta S} \times\\
    \sum_{v=0}^{\min(n_R,n_L-\Delta S)}\!\!\mathfrak{g}_{\Delta S,0,v}  \angsq^{n_L-v-\Delta S}\sqang^{n_R-v}(\ang\sq)^{v}
\end{multline}
where we package the factorials into $\mathcal{N}_L = \sqrt{(n_L)!(n_L-\Delta S)!}$ and $\mathcal{N}_R = \sqrt{(n_R)!(n_R+\Delta S)!}$\,. Note that the only $n_L$ dependence is $\mathcal{N}^{-1}_L \angsq^{n_L}$ and the only $n_R$ dependence is $\mathcal{N}^{-1}_R \sqang^{n_R}$. However before we can resum, we need to change the order of the summations,
\begin{equation}
    \sum_{n_L, n_R=0}^\infty \sum_{\Delta S=0}^{n_L}\sum_{v=0}^{\min(n_R,n_L-\Delta S)} \to  \sum_{\Delta S=0}^\infty \sum_{v=0}^\infty \sum_{n_L=\Delta S+v}^\infty \sum_{n_R=v}^\infty\,.
\end{equation}
While we do not have an exact closed form for the $n_{L,R}$ sums we can resum the leading order in the classical limit,
\begin{align}\label{eq:resumnLnR}
    \sum_{n_L=\Delta S+v}^\infty \frac{\angsq^{n_L}}{\sqrt{(n_L)!(n_L-\Delta S)!}}&=e^{\angsq}\angsq^{\Delta S}(1+\mathcal{O}(\hbar)),\nonumber \\
     \sum_{n_R=v}^\infty \frac{\sqang^{n_R}}{\sqrt{(n_R)!(n_R+\Delta S)!}}&=e^{\sqang}\sqang^{-\Delta S}(1+\mathcal{O}(\hbar)).
\end{align}
We can then translate the leading order of the amplitude in spinor-helicity to an amplitude in the expectation value of the spin operators as before using eqs.~\eqref{Spin1/2Map} and \eqref{cohexpvalS}. Note that the exponential factors in eq.~\eqref{eq:resumnLnR} cancel against the normalisation factor $e^{-|z|^2-|w|^2}$. The classical contribution from this $(p_1 \cdot \varepsilon_{3})^2$ term in the amplitude is then
\begin{multline}
    \mathcal{A}(1^{z,w},2^{\bar{z},\bar{w}},3^2)\Big|_{(p_1 \cdot \varepsilon_{3})^2}\sim\\\sum_{\Delta S=0}^\infty \sum_{v=0}^{\infty}\frac{\mathfrak{g}_{\Delta S,0,v}}{|z|^{2v+\Delta S }|w|^{2v+\Delta S}}\left(\frac{\bar{k}\cdot \la \mathsf{S}_L\ra_{z,w}\cdot \la \mathsf{S}_R\ra_{z,w}\cdot \bar{k}}{2}\right)^v \left(\frac{\la \bar{k}\cdot \bar{\mathsf{K}} \ra_{z,w}}{4}\right)^{\Delta S}.
\end{multline}
We can observe that this is exactly the type of double sum that appears in the stress-energy tensor for non-biaxially symmetric solutions given in eq.~\eqref{TNonAxi}. However, in order to have a finite classical limit, we need to scale $\mathfrak{g}_{\Delta S,0,v}\sim \mathcal{O}(\hbar^{ 2v+\Delta S})$. This is not surprising as in the infinite spin limit taken in definite spin states we also needed to give the coefficients an explicit spin dependence in order to balance the powers of spin coming from the factorials. The procedure works similarly for the other contributions to the amplitude. Below we will just give the explicit map between the spinor-helicity coefficients and the stress-energy tensor ones:
\begin{align}
    C_{n,n',1}&=\lim_{\hbar\rightarrow 0}\mathcal{N}\frac{\mathfrak{g}_{n',0,n}}{|z|^{2n+n'}|w|^{2n+n'}},\ C_{n,n',2}=\lim_{\hbar\rightarrow 0}\frac{-i\mathcal{N}}{16(2n+1)}\frac{\mathfrak{g}_{n',1,n}+\mathfrak{g}_{n',2,n}}{|z|^{2n+n'}|w|^{2n+n'} },\nonumber\\
    \tilde{C}_{n,n',2}&=\lim_{\hbar\rightarrow 0}\frac{-i\mathcal{N}}{16(2n+1)}\frac{\mathfrak{g}_{n',1,n}-\mathfrak{g}_{n',2,n}}{|z|^{2n+n'}|w|^{2n+n'}},\ 
    C_{n,n',3}=\lim_{\hbar\rightarrow 0}\frac{-4\mathcal{N}}{128(2n+1)(2n+2)}\frac{\mathfrak{g}_{n',3,n}+\mathfrak{g}_{n',4,n}}{|z|^{2n+n'}|w|^{2n+n'}},\nonumber\\
    \tilde{C}_{n,n',3}&=\lim_{\hbar\rightarrow 0}\frac{-4\mathcal{N}}{128(2n+1)(2n+2)}\frac{\mathfrak{g}_{n',3,n}-\mathfrak{g}_{n',4,n}}{|z|^{2n+n'}|w|^{2n+n'}},
\end{align}
with
\begin{equation}
    \mathcal{N}=\frac{(-1)^n i^{n'}(2n)!(n')!}{2^n 4^{n'}}.
\end{equation}

\section{Conclusion}\label{sec:Concl}

In this work, we focused on a class of three-point scattering amplitudes between two massive spinning states and a graviton that satisfies some notion of spin conservation. We studied the amplitudes in various bases: using spinor variables, expectation values of spin-operators, and finally coherent states. The latter two bases allow us to identify a classical amplitude which defines a space of possible linearised gravitational solutions. 

We were interested in investigating how this space of solutions depends on the states involved in the scattering.  In $D=5$, we have more than just totally symmetric massive spin states; the massive states are labelled by the quantum numbers $(n_L,n_R)$ for the two $SU(2)$ copies. We found that any scattering that preserved the quantum numbers in the left and right $SU(2)$ copies spans the space of linearised biaxially-symmetric spinning solutions. These are described by a massive spinning worldline coupled to gravity, where the spin degrees of freedom are encoded in the classical spin tensor and, if the theory is parity violating, its Hodge dual. These cases include scattering totally symmetric states $(n_L=n_R)$ but also scattering of non-trivial tensor states so long as the spin quantum numbers are unchanged in the final state.

\par However, scattering that allows some internal spin change in the final state introduces a dependence on the longitudinal part of the spin tensor, $K^\mu$, in the classical expansion. For fixed-spin scattering, the classical amplitude generates a single multipole order $(K)^{n}$. However we recovered the full multipolar expansion for non-biaxially-symmetric solutions when considering the scattering of coherent spin states.

\par As explicit examples, we identified the amplitudes whose leading order in spin behaviour reproduces the Myers-Perry black hole. We also extended the Myers-Perry analysis to arbitrary dimensions using an expansion of the amplitude in building blocks involving Lorentz covariant structures available in any number of dimensions. We also did a brief analysis of the black ring solution, which has a rich multipolar structure. However, our analysis does not indicate that there is a link between this structure and the richer little group structure in $D=5$ amplitudes. Indeed, this solution can be generated by scattering totally symmetric states that have polynomial couplings in $\nu$, the ratio of the periods of the ring.
\par While the main two black-hole examples studied were Myers-Perry and the black ring,  the formalism of this paper can describe asymptotically-flat linearised gravitational solutions in five or higher dimensions. As such, there are more intricate solutions such as the Black Saturn \cite{Elvang:2007rd} or multi-horizon solutions. To linear order, these can look like a superposition of Myers-Perry black holes or black rings and are captured by our results. While, the analysis in this paper does not determine the non-linear completion, once the Wilson coefficients in the worldline action are fixed, one can perform an analysis similar to ref.~\cite{Mougiakakos:2020laz} to reconstruct the metric to all orders in $G$.

\par There are several possible further directions to explore. In this work we worked with general amplitudes and saw, as was the case in $D=4$, that there is a large degeneracy in the class of amplitudes that map to the same black hole solution. In $D=4$, there has been substantial analysis of which amplitudes and QFTs are best suited to describe the Kerr black hole, either through the high-energy behaviour of the amplitude itself \cite{Arkani-Hamed:2017jhn} or consistency of the effective higher-spin Lagrangians \cite{Cangemi:2022bew}. There are instances where similar methods have been applied in $D=5$; see refs.~\cite{Pokraka:2024fao, Gambino:2025iyx,Campanella:2026wqt} but further work is needed.

\par One could also study supergravity solutions in five-dimensions where the degrees of freedom introduced by allowing the $\tilde{S}^{\mu\nu}$ might appear more naturally. However, in order to incorporate their effects, we need to extend our massless field content to include the supergravity multiplets. Once this is achieved, one can also study black holes with different asymptotic expansions, such as asymptotically Kaluza-Klein black holes \cite{Harmark:2005pp,Katona:2023uaj}.

\par Alternatively one could attempt to study features of $D=5$ GR solutions from the amplitudes perspective. Notably, Myers-Perry, along with several other solutions, can be generated from a Newman-Janis-type shift in five dimensions \cite{Erbin:2016lzq}.

\section*{Acknowledgements}
We are grateful to Tim Adamo, Rafael Aoude, Fabian Bautista,  Francesco Campanella, Henrik Johansson, Cynthia Keeler, James Lucietti, Panagiotis Marinellis, Donal O'Connell, Julio Parra-Martinez and Paolo Pichini for useful discussions. We would also like to acknowledge the hospitality of the Henri Poincaré Institute in the course of this work. IS is supported by an STFC studentship.

\appendix
\section{Conventions and Spinor Identities}
\label{appconv}
We work in mostly minus signature $\eta^{\mu\nu} = \textrm{diag}(1,-1,\dots,-1)$ regardless of dimension. All amplitudes are written in all-incoming momentum conventions. Therefore the three-point kinematics for scattering between two massive particles $p_{i=1,2}$ and a massless particle $k$ satisfy
\begin{equation}
    p_1 + p_2 + k =0 \,, \quad p_i^2 = m^2 \,, \quad k^2 =0\,, \quad p_i \cdot k = 0\,.
\end{equation}

We use the following representation for the 5D Clifford algebra,

\begin{alignat}{3}
    \Gamma^0 &=\begin{pmatrix}
        \phantom{-}0& \phantom{-}0& \phantom{-}1 & \phantom{-}0\\
        \phantom{-}0& \phantom{-}0& \phantom{-}0& \phantom{-}1\\
        \phantom{-}1 &\phantom{-}0&\phantom{-}0&\phantom{-}0 \\
        \phantom{-}0 & \phantom{-}1 & \phantom{-}0 & \phantom{-}0 
    \end{pmatrix} \,, \quad &&\Gamma^1 &&= \begin{pmatrix}
        \phantom{-}0& \phantom{-}0& \phantom{-}0 & \phantom{-}1\\
        \phantom{-}0& \phantom{-}0& \phantom{-}1& \phantom{-}0\\
        \phantom{-}0&-1&\phantom{-}0&\phantom{-}0 \\
        -1 & \phantom{-}0 & \phantom{-}0 & \phantom{-}0 
    \end{pmatrix}\,,\quad
    \Gamma^2 = \begin{pmatrix}
        \phantom{-}0& \phantom{-}0& \phantom{-}0 & -i\\  \phantom{-}0& \phantom{-}0& \phantom{-}i& \phantom{-}0\\ 
        \phantom{-}0 &\phantom{-}i&\phantom{-}0&\phantom{-}0 \\ 
        -i & \phantom{-}0 & \phantom{-}0 & \phantom{-}0 
    \end{pmatrix} \,,\nonumber\\
    \Gamma^3 &=  \begin{pmatrix}
        \phantom{-}0& \phantom{-}0& \phantom{-}1 & \phantom{-}0\\
        \phantom{-}0& \phantom{-}0& \phantom{-}0& -1\\
        -1 &\phantom{-}0&\phantom{-}0&\phantom{-}0 \\
        \phantom{-}0 & \phantom{-}1 & \phantom{-}0 & \phantom{-}0 
    \end{pmatrix} \,, \qquad &&\Gamma^4 &&= \begin{pmatrix}
        -i& \phantom{-}0& \phantom{-}0 & \phantom{-}0\\
        \phantom{-}0& -i& \phantom{-}0& \phantom{-}0\\
        \phantom{-}0 &\phantom{-}0&\phantom{-}i&\phantom{-}0 \\
        \phantom{-}0 & \phantom{-}0 & \phantom{-}0 & \phantom{-}i 
    \end{pmatrix} = -\Gamma^0\Gamma^1\Gamma^2\Gamma^3
\end{alignat}
The Gamma matrices are defined with indices $(\Gamma^{\mu})_{A}{}^{B}$ with $A,B = 1,\dots, 4$ which can be raised and lowered via 
\begin{equation}
    C_{AB} = \begin{pmatrix}
        0 & 1 & 0 & 0 \\ -1 & 0 & 0 & 0 \\ 0 & 0 & 0 & 1 \\ 0 & 0& -1 & 0
    \end{pmatrix} = - C^{AB}\,,
\end{equation}
such that $C^{AC}(\Gamma^{\mu})_{C}{}^{D}C_{DB} = (\Gamma^{\mu, T})^{A}{}_{B}$. The five-dimensional Levi-Civita is defined with $\epsilon^{01234}=1$ and is proportional to the trace
\begin{equation}\label{eq:GammaIds}
    \text{Tr}[\Gamma^{\mu}\Gamma^{\nu}\Gamma^{\rho}\Gamma^{\sigma}\Gamma^{\tau}] = 4 \epsilon^{\mu\nu\rho\sigma\tau}
\end{equation} We also define the anti-symmetrised tensor $\Gamma^{\mu \nu} = \frac{1}{2}[\Gamma^{\mu}, \Gamma^{\nu}]$ and note the following contractions will be needed for later
\begin{align}
    (\Gamma_{\mu})_{A}{}^{B}(\Gamma^{\mu})_{C}{}^{D} =& 2 \, \delta_{A}{}^{D} \delta_{C}{}^{B}- \delta_{A}{}^{B} \delta_{C}{}^{D} - 2 \, C_{AC}C^{BD}\nonumber\\ 
    (\Gamma^{\mu\nu})_A{}^C(\Gamma_{\mu\nu})_B{}^D=&-4 C_{AB}C^{CD}-4\delta_B^C\delta_A^D
\end{align}
The massive momentum $p_{\mu} = (p_0, p_1,p_2,p_3,p_4)$ is parameterised
\begin{equation}
    {}_{A}|p^{a}\rangle  = \frac{\sqrt{m}}{\sqrt{2}}{\begin{pmatrix}
        - \frac{p_1 - i p_2}{m} &  \frac{m - i p_4}{p_0 - p_3}\\
        -  \frac{p_0 - p_3}{m} & 0 \\
        0 & 1 \\
       - \frac{m + i p_4}{m} & \frac{p_1 + i p_2}{p_0 - p_3}
    \end{pmatrix}_{A}}^{a}\,, \quad {}_{A}|p^{\dot{a}}]  = i \frac{\sqrt{m}}{\sqrt{2}}{\begin{pmatrix}
         \frac{p_1 - i p_2}{m} & - \frac{m - i p_4}{p_0 - p_3}\\
         \frac{p_0 - p_3}{m} & 0 \\
        0 & 1\\
       - \frac{m - i p_4}{m} & \frac{p_1 + i p_2}{p_0 - p_3}
    \end{pmatrix}_{A}}^{\dot{a}}\,.
\end{equation}
Note that ${}_{A}|(-p)^{a}\rangle = (-i) {}_{A}|p^{\dot{a}}]$ so dotted and undotted indices swap under reflection, as expected since the $SU(2)$ copies flip. We can define a valid parity operators $\mathbb{P} = \Gamma_4$ which sends $x_4 \to - x_4$, this action on the spinors is
\begin{equation}
    \mathbb{P} |p^a\ra = |(\mathbb{P}p)^{\dot{a}}]\,,\qquad \mathbb{P} |p^{\dot{a}}] =- |(\mathbb{P}p)^{a}\ra
\end{equation}
where the spinors on the right-hand side are defined with respect to the momentum $(\mathbb{P}p)_{\mu} = (p_0,p_1,p_2,p_3,-p_4)$. 

We absorb the little-group indices with the spinors $z_{a}, w_{\dot{a}}$ to parameterise the little-group of the initial state and $\bar{z}_{a}, \bar{w}_{\dot{a}}$ to parameterise the final state. The resulting massive spinors satisfy the following identities
\begin{alignat}{3}
    \slashed{p} |{\bm p} \rangle &= m | {\bm p} \rangle\,, \quad && \langle \bar{{\bm p}} {\bm p} \rangle &&= -m \, \bar{z}\cdot z\nonumber\\
    \slashed{p} |{\bm p} ] &=- m | {\bm p} ]\,, \quad &&[\bar{{\bm p}} {\bm p} ] &&= m \, \bar{w}\cdot w\,,
\end{alignat}
where we introduce shorthand for the contraction of the little-group spinors with the left/right $SU(2)$ two-tensor, $\bar{z}\cdot z = \epsilon^{ab} \bar{z}_a z_b  = \bar{z}_1 z_2 - \bar{z}_2 z_1$ and $\bar{w}\cdot w = \epsilon^{\dot{a}\dot{b}} \bar{w}_{\dot{a}} w_{\dot{b}}  = \bar{w}_1 w_2 - \bar{w}_2 w_1$.

The Lorentz generator decomposes into
\begin{equation}
    \mathbb{M}^{\mu\nu} = \mathbb{S}^{\mu\nu} + 2\frac{p^{[\mu} \mathbb{K}^{\nu]}}{m}\,,
\end{equation}
where we use the projectors $\mathbb{P}^{\mu}{}_{\nu} = \delta^{\mu}{}_{\nu} - \frac{p^{\mu}p_{\nu}}{m^2}$ to define the SSC satisfying spin tensor
\begin{align}
    \mathbb{S}^{\mu\nu} &= \mathbb{P}^{\mu}{}_{\rho}\mathbb{P}^{\nu}{}_{\sigma}\mathbb{M}^{\rho\sigma}\,,
\end{align}

If we use the spin-$1/2$ Lorentz generators, $\mathbb{M}^{\mu\nu}= \Gamma^{\mu\nu}$ then the boost generator has the form $\mathbb{K}^{\mu} = \frac{\slashed{p}\Gamma^{\mu}}{m}$. All the possible contractions of the spin-$1/2$ Lorentz generators can then be written as 
\begin{align}
    \langle \mathbb{M}^{\mu\nu} \rangle &=  \langle \mathbb{S}^{\mu\nu} \rangle = \frac{1}{m} \langle \bar{{\bm p}}| \Gamma^{\mu\nu}|{\bm p}\rangle\,, \quad  [\mathbb{M}^{\mu\nu}] =  [\mathbb{S}^{\mu\nu}] = -\frac{1}{m} [\bar{{\bm p}}| \Gamma^{\mu\nu}|{\bm p}]\\
    [\mathbb{M}^{\mu\nu} \rangle &=  \frac{p^{[\mu}}{m}[ \mathbb{K}^{\nu]} \rangle =- \frac{p^{[\mu}}{m^2} [\bar{{\bm p}}| \Gamma^{\nu]}|{\bm p}\rangle \,, \quad    \langle \mathbb{M}^{\mu\nu}] =  \frac{p^{[\mu}}{m}\langle \mathbb{K}^{\nu]}] = \frac{p^{[\mu}}{m^2} \langle \bar{{\bm p}}| \Gamma^{\nu]}|{\bm p}]
\end{align}

We define the operators that act on the little group by stripping off the little-group spinors. The building blocks for the higher rep. operators are the spin-$1/2$ ones
\begin{alignat}{3}
    (s^{\mu\nu}_L)^{a}{}_{b} &= \frac{1}{m}  \langle p^{a}| \Gamma^{\mu\nu}|p_{b}\rangle \,, \qquad &&(s^{\mu\nu}_R)^{\dot{a}}{}_{\dot{b}} &&= -\frac{1}{m}  [ p^{\dot{a}}| \Gamma^{\mu\nu}|p_{\dot{b}}] \\
    (\kappa^\mu)^{a}{}_{\dot{b}}&=\frac{1}{m} \langle p^{a}| \Gamma^{\mu}|p_{\dot{b}}] \,&&(\bar{\kappa}^\mu)^{\dot{a}}{}_{b}&&=\frac{1}{m} [ p^{\dot{a}}| \Gamma^{\mu}|p_{b}\rangle\,.
\end{alignat}
Note $\langle \mathbb{K}^{\mu}] = \langle {\kappa}^{\mu}] $, $[\mathbb{K}^{\mu}\rangle = -[ \bar{\kappa}^{\mu}\rangle $.
The higher-weight representations are built by tensoring in identity matrices while fully symmetrising the indices, for example,
\begin{equation}
    (S^{\mu\nu}_L)^{\Vec{a}}{}_{\Vec{b}} = (s^{\mu\nu}_L)^{(a_1}{}_{(b_1} \delta^{a_2}{}_{b_2} \dots \delta^{a_{n})}{}_{b_n)}\,.
\end{equation}

If our initial and final states are the same little group irrep, labelled by $(n_L,n_R)$ we can construct the Lorentz generator acting on the little group as 
\begin{align}\label{LorentzgenLittlgrp}
    \lexp \mathbb{M}^{\mu\nu} \rexp^{\Vec{a} \Vec{\dot{a}}}{}_{\Vec{b} \Vec{\dot{b}}} &= \begin{pmatrix}
       (S^{\mu\nu}_L)^{\Vec{a}}{}_{\Vec{b}} & p^{[\mu}(\kappa^{\nu]})^{\Vec{a}}{}_{\Vec{\dot{b}}} \\
        p^{[\nu}(\bar{\kappa}^{\mu]})^{\Vec{\dot{a}}}{}_{\Vec{b}} &(S^{\mu\nu}_R)^{\Vec{\dot{a}}}{}_{\Vec{\dot{b}}}
    \end{pmatrix}\,,
\end{align}
which factorises over the two copies of the $SU(2)$ since $\Vec{a},\Vec{b}$ are multi-indices of length $n_L$ over the left $SU(2)$, and $\Vec{\dot{a}},\Vec{\dot{b}}$ are multi-indices of length $n_R$ over the right $SU(2)$. See Appendix \ref{spinApp} for more details on the operator algebra. 

Using the Gamma identities in eq.~\eqref{eq:GammaIds}, we can decompose massive spinors into the following 
\begin{align}\label{eq:MSdecomp}
    {}_{A}|  {\bm p}\rangle \langle\bar{{\bm p}}|^{B} &= \frac{1}{4}\left[ -(\slashed{p}_{A}{}^{B} +m \delta_{A}{}^{B}) \,\bar{z}\cdot z  - m/2 \langle S_L^{\mu\nu}  \rangle (\Gamma_{\mu\nu})_{A}{}^{B} \right]\\
    {}_{A}| {\bm p}][ \bar{{\bm p}}|^{B} &= \frac{1}{4}\left[ -(\slashed{p}_{A}{}^{B} - m \delta_{A}{}^{B})\,\bar{w}\cdot w  + m/2 [ S_R^{\mu\nu}] (\Gamma_{\mu\nu})_{A}{}^{B}\right]\\
    {}_{A}| {\bm p}\rangle [ \bar{{\bm p}}|^{B} &= \frac{1}{4} \left[ m [\bar{\kappa}^{\mu} \rangle (\Gamma_{\mu})_{A}{}^{B}  + p^{\mu} [ \bar{\kappa}^{\nu} \rangle (\Gamma_{\mu\nu})_{A}{}^{B}\right]\\
    {}_{A}| {\bm p}]\langle\bar{{\bm p}}|^{B} &= \frac{1}{4}  \left[ m \langle{\kappa}^{\mu}](\Gamma_{\mu})_{A}{}^{B}  - p^{\mu}\langle {\kappa}^{\nu}] (\Gamma_{\mu\nu})_{A}{}^{B}\right]\,.
\end{align}

Massless spinors are parameterised by the angle ket spinor,
\begin{equation}
    {}_{A}|k^{I}\rangle  ={\begin{pmatrix}
        i k_1 + k_2 & \frac{- k_4}{k_0 - k_3}\\
        i k_0 - i k_3 & 0 \\
        0 & -i \\
       - k_4 & \frac{i k_1 - k_2}{k_0 - k_3}
    \end{pmatrix}_{A}}^{I}\,.
\end{equation}
The bra spinors are defined $\langle k^{I}|^{A} = C^{AB} {}_B |k^{I}\rangle$ and $SU(2)$ index is lowered by $\epsilon_{IJ}$, where $\epsilon_{12}=-1$. The spinors satisfy $\slashed{k}|k^{I}\rangle = 0 = \langle k^{I}|\slashed{k}$ and $\langle k^{I} k^{J} \rangle =0$. 

We absorb the massless little group $SU(2)$ index with the little-group spinors $u_{I =1,2}$. The massless polarisations are defined with respect to an auxiliary null momentum $r^{\mu}$,
\begin{equation}
    \varepsilon(k) = \frac{\langle k^{I}| \slashed{r} \Gamma^{\mu} | k^{J}\rangle }{2 k \cdot r} u_{I}u_{J}\,.
\end{equation}
When we expand the polarisation as a polynomial in $u_{I}$ we can identify the coefficients of the three monomials $\{u_{1}^2, u_{1}u_{2}, u_{2}^2\}$ with the three physical massless polarisations in $D=5$. The massless spinors satisfy 
\begin{equation}
    |k^{I}\ra\la k^{J}| u_{I}u_{J} = \frac{1}{2} \slashed{k} \slashed{\varepsilon}_k\,.
\end{equation}

The spinors of leg $2$ are parametrised in terms of the spinors of $p_1$ and we distinguish the little group by using barred little-group spinors 
\begin{align}
    |{\bm p}_2 \rangle &= i \left(|\bar{{\bm p}}_1] - \frac{1}{2m} \slashed{k}|\bar{{\bm p}}_1]\right)\,, \quad
    |{\bm p}_2] = i \left(|\bar{{\bm p}}_1\rangle + \frac{1}{2m} \slashed{k}|\bar{{\bm p}}_1\rangle\right)\,.
\end{align}
These spinors satisfy all the same identities as those defined for $p_1$. 

The amplitudes in the spin-operator basis are functions of the spin-$1/2$ building blocks 
\begin{equation}
    \{p \cdot \varepsilon_k, \varepsilon_k \cdot \langle s_{L}\rangle \cdot k,  \varepsilon_k \cdot [ s_{R} ] \cdot k, k \cdot \langle \kappa ], k \cdot [\bar{\kappa}\rangle , k \cdot \langle s_{L}\rangle \cdot [s_{R}] \cdot k\}\,.
\end{equation}

However there are quadratic relations between these building blocks, inherited from commutators that the Lorentz generators satisfy. In particular
\begin{equation}
    k \cdot \langle \kappa ] \, k \cdot [\bar{\kappa}\rangle  = \frac{1}{2} k \cdot \langle s_{L}\rangle \cdot [s_{R}] \cdot k\,.
\end{equation}

We choose to use this identity such that we maximise the spin-tensors in the final form of the amplitude. For example when we scatter totally symmetric states the final amplitude is only expressed in terms of spin-tensors as expected. However, scattering that involves some internal spin change between the $SU(2)$ copies will require at least one factor of $k \cdot \langle \kappa ]$ or $k \cdot [\bar{\kappa}\rangle $.

The scattering amplitudes of massive states satisfying $n_L = \bar{n}_L$, $n_{R} = \bar{n}_R$ and $n_L+ n_R =2s$ are only functions of the following monomials 
\begin{equation}
    (\varepsilon \cdot \langle s_{L}\rangle\cdot k)^{i} (\varepsilon \cdot [s_{R}]\cdot k)^{j}(k \cdot \la s_L \ra \cdot [s_{R}] \cdot k)^{n} \,, \quad 0 \leq i+j \leq 2\,, \, 0 \leq  n \leq \lfloor s-(i+j)/2\rfloor
\end{equation}
when expanded in the operator basis. Note not all these monomials are independent, indeed there is the identity
\begin{equation}
    (\varepsilon \cdot \langle s_{L}\rangle \cdot k)(\varepsilon \cdot [s_{R}]\cdot k)=\frac{(p\cdot \varepsilon)^2}{2m^2} (k \cdot \la s_L \ra \cdot [s_{R}] \cdot k)\,.
\end{equation}

In order to change the representation to the $(n_L, n_R)$-operators we need the following
\begin{align}
    \lexp (k \cdot S \cdot S \cdot k)^{v} \rexp = &\phantom{\big[}c_{0,0,v} (k \cdot \la s_L \ra \cdot [s_{R}] \cdot k)^{v}\nonumber\\
    \lexp S_{\varepsilon k} (k \cdot S \cdot S \cdot k)^{v} \rexp = &\big[ c_{1,0,v} (\varepsilon \cdot \langle s_{L}\rangle \cdot k) + c_{0,1,v}  (\varepsilon \cdot [s_{R}]\cdot k)\big] (k \cdot \la s_L \ra \cdot [s_{R}] \cdot k)^{v} \nonumber \\
    \lexp (S_{\varepsilon k})^2(k \cdot S \cdot S \cdot k)^{v} \rexp = &\big[ c_{2,0,v} (\varepsilon \cdot \langle s_{L}\rangle \cdot k)^2+ 2 c_{1,1,v}  (\varepsilon \cdot \langle s_{L}\rangle \cdot k)(\varepsilon \cdot [s_{R}]\cdot k) \nonumber\\
    + &\phantom{\big[}c_{0,2,v}  (\varepsilon \cdot [s_{R}]\cdot k)^2\big] (k \cdot \la s_L \ra \cdot [s_{R}] \cdot k)^{v}\,,
\end{align}
where the combinatorics is captured by 
\begin{equation}
    c_{i,j,v} = (-1)^{j} 2^{v}(v+i)!(v+j)!\binom{n_L}{v+i}\binom{n_R}{v+j} \,.
\end{equation}
Note that we have set $-\bar{z}\cdot z = 1 = \bar{w}\cdot w$, this is equivalent to normalising our initial and final states 
\begin{equation}
   \{ |{\bm p}\rangle^{n_L}|{\bm p}]^{n_R},\langle{\bm p}|^{n_L}[{\bm p}|^{n_R}\} \to \left\{\frac{|{\bm p}\rangle^{n_L}|{\bm p}]^{n_R}}{\langle \bar{{\bm p}} {\bm p} \rangle^{n_L/2} [\bar{{\bm p}} {\bm p}]^{n_R/2}},\frac{\langle{\bm p}|^{n_L}[{\bm p}|^{n_R}}{\langle \bar{{\bm p}} {\bm p} \rangle^{n_L/2} [\bar{{\bm p}} {\bm p}]^{n_R/2}}\right\}
\end{equation}

We also study amplitudes between states that are different irreps, in particular those satisfying $n_L = \bar{n}_L + 2n$ and $n_{R} = \bar{n}_R - 2n$ and $n_L +n_R = 2s$. For $n>0$ the amplitude is a polynomial in the following monomials
\begin{equation}
    k \cdot [\bar{\kappa} \rangle^{2n} (k \cdot \la s_L \ra \cdot [s_{R}] \cdot k)^{v} (\varepsilon \cdot \langle s_{L}\rangle\cdot k)^{i} (\varepsilon \cdot [s_{R}]\cdot k)^{j}\,,
\end{equation}
for $\quad 0 \leq i+j \leq 2$ and $0 \leq  v \leq \lfloor s-n -(i+j)/2\rfloor$. So in this case we need the following representation change formulae
\begin{align} \label{eq:Krepchanges}
    \lexp (k \cdot K)^{2n}(k \cdot S \cdot S \cdot k)^{v} \rexp = &\phantom{\big[}d_{0,0,v} k \cdot [\bar{\kappa} \rangle^{2n}(k \cdot \la s_L \ra \cdot [s_{R}] \cdot k)^{v}\nonumber\\
    \lexp S_{\varepsilon k}  (k \cdot K)^{2n} (k \cdot S \cdot S \cdot k)^{v} \rexp = &\big[ d_{1,0,v} (\varepsilon \cdot \langle s_{L}\rangle \cdot k) + d_{0,1,v}  (\varepsilon \cdot [s_{R}]\cdot k)\big]\nonumber\\
    &\times k \cdot [\bar{\kappa} \rangle^{2n}(k \cdot \la s_L \ra \cdot [s_{R}] \cdot k)^{v} \nonumber \\
    \lexp (S_{\varepsilon k})^2  (k \cdot K)^{2n} (k \cdot S \cdot S \cdot k)^{v} \rexp = &\big[ d_{2,0,v} (\varepsilon \cdot \langle s_{L}\rangle \cdot k)^2+ 2 d_{1,1,v}  (\varepsilon \cdot \langle s_{L}\rangle \cdot k)(\varepsilon \cdot [s_{R}]\cdot k) \nonumber\\
    + &\phantom{\big[}d_{0,2,v}  (\varepsilon \cdot [s_{R}]\cdot k)^2\big] k \cdot [\bar{\kappa} \rangle^{2n} (k \cdot \la s_L \ra \cdot [s_{R}] \cdot k)^{v}\,,
\end{align}
with
\begin{equation}\label{eq:combinatoricd}
    d_{i,j,v}=(-1)^{j}2^v(v+i+2n)!(v+j)!\binom{n_L}{v+i+2n}\binom{n_R}{v+j}.
\end{equation}

\section{Spin in the rest frame}\label{app:RFspin}

In the rest-frame $p_{\mu}= (m ,0,0,0,0)$ the spinors reduce to
\begin{equation}
    {}_{A}|{\bm p}\rangle  = \frac{\sqrt{m}}{\sqrt{2}}\begin{pmatrix}
        z_2\\
        - z_1\\
        z_2 \\
       - z_1
    \end{pmatrix}_{A}\,, \quad {}_{A}|{\bm p}]  = \frac{i\sqrt{m}}{\sqrt{2}}\begin{pmatrix}
        -w_2\\
         w_1\\
        w_2 \\
       -w_1
    \end{pmatrix}_{A}\,.
\end{equation}

We can evaluate the spin operators in the rest-frame, where we can construct a classical spin-tensor $S^{ij}$ such that $S^{12} =a_1$ is the magnitude of spin in the $1-2$-plane and $S^{34}=a_2$ is the magnitude of spin in the $3-4$-plane. 

We can then choose to align our little group on the spin-operators by picking the little-group spinors
\begin{equation} \label{eq:RFauxspinors}
    \bar{z}_a = |z|\begin{pmatrix}
    -i \\ 0
\end{pmatrix}_a\,,\quad z_a = |z| \begin{pmatrix}
    0 \\ -i
\end{pmatrix}_a\,,\quad \bar{w}_{\dot{a}} =  |w|\begin{pmatrix}
    -i \\ 0
\end{pmatrix}_{\dot{a}}\,,\quad w_{\dot{a}} =  |w|\begin{pmatrix}
    0 \\ -i
\end{pmatrix}_{\dot{a}}\,,
\end{equation}
where  $|z|^2 = -\bar{z}_a \epsilon^{ab} z_b$, $|w|^2 =-\bar{w}_{\dot{a}} \epsilon^{\dot{a}\dot{b}} w_{\dot{b}}$.

In this frame we can identify the spin-operators evaluated in the left $SU(2)$ with the anti-self-dual classical spin tensor
\begin{equation}
    \langle \mathbb{S}^{ij} \rangle = S^{ij} - \frac{1}{2} \epsilon^{ij}{}_{kl}S^{kl} = (a_1-a_2)\begin{pmatrix}
        0 & 1 & 0 & 0 \\
        -1 & 0 & 0 & 0 \\
        0 & 0 & 0 & 1 \\
        0 & 0 & -1 & 0
    \end{pmatrix}^{ij}\,,
\end{equation}
where the little-group spinors are normalised $|z|^2 = i( a_1 - a_2)$.
Likewise the spin-operator evaluated in the right $SU(2)$ can be identified with self-dual classical spin tensor
\begin{equation}
    [ \mathbb{S}^{ij} ] = S^{ij} + \frac{1}{2}\epsilon^{ij}{}_{kl}S^{kl} = (a_1+a_2)\begin{pmatrix}
        0 & 1 & 0 & 0 \\
        -1 & 0 & 0 & 0 \\
        0 & 0 & 0 & -1 \\
        0 & 0 & 1 & 0
    \end{pmatrix}^{ij}\,,
\end{equation}
where $|w|^2 = i (a_1+a_2) $.

We can also consider the boost vector in this frame $\langle \mathbb{K}^{i}]$ 
\begin{align}
    \langle \mathbb{K}^{i}] = |z||w| \begin{pmatrix}
        0 \\0\\ i\\-1
    \end{pmatrix}\,, \qquad [ \mathbb{K}^{i}\rangle = |z||w| \begin{pmatrix}
        0 \\0\\ i\\1
    \end{pmatrix}\,,
\end{align} 

\section{Spin operator algebra}\label{spinApp}

In this appendix we check that the operators (\ref{LorentzgenLittlgrp}) satisfy the usual commutation relations
\begin{align}\label{LorAlg}
    \lexp [\mathbb{M}^{\mu\nu} , \mathbb{M}^{\rho\sigma} ]\rexp= \lexp \mathbb{M}^{\mu\sigma} \rexp \eta^{\nu \rho}-\lexp \mathbb{M}^{\nu\sigma} \rexp \eta^{\mu \rho}+\lexp \mathbb{M}^{\nu\rho} \rexp \eta^{\mu \sigma}-\lexp \mathbb{M}^{\mu\rho} \rexp \eta^{\nu \sigma}\,.
\end{align}
Since $S_L$ and $S_R$ are transverse, their algebra reduces to the four-dimensional $SO(4)$ algebra which is obeyed by construction. One can also check that $[S_L^{\mu\nu},S_R^{\rho\tau}]=0$. We have to check then the commutators involving $\kappa^\mu,\ \bar{\kappa}^\mu$. Focusing on the upper-left block of $\langle \!\langle \mathbb{M}^{\mu\nu}\rexp$ we have:
\begin{align}
    (\kappa^\mu \bar{\kappa}^\nu-\kappa^\nu \bar{\kappa}^\mu)^a{}_b=&\frac{1}{2m^2}\la p^a(\Gamma^\mu (\slashed{p}-m)\Gamma^\nu-\Gamma^\nu(\slashed{p}-m)\Gamma^\mu|p_b\ra\nonumber\\
    =&\frac{1}{m^2}(\la p^a|p^\nu \Gamma^\mu-p^\mu \Gamma^\nu|p_b\ra-2m\la p^a|\Gamma^{\mu\nu}|p_b\ra)\nonumber\\
    =&-2 (S_L^{\mu\nu})^a{}_b.
\end{align}
Similarly, for the dotted block we get $(\bar{\kappa}^\mu\kappa^\nu-\bar{\kappa}^\nu\kappa^\mu)^{\dot{a}}{}_{\dot{b}}=-2(S_R^{\mu\nu})^{\dot{a}}{}_{\dot{b}}$. Organizing $\kappa^\mu, \bar{\kappa}^\mu$ in an off-diagonal matrix, we have:
\begin{equation}
    \left[\begin{pmatrix}
        0 & \kappa^\mu \\
        \bar{\kappa}^\mu & 0
    \end{pmatrix},\begin{pmatrix}
        0 & \kappa^\nu \\
        \bar{\kappa}^\nu & 0
    \end{pmatrix} \right]=-2\begin{pmatrix}
        S_L^{\mu\nu}  & 0\\
        0 & S_R^{\mu\nu} 
    \end{pmatrix}.
\end{equation}
We now check the commutator between $S^{\mu\nu}_{L,R}$ and $K^\mu$:
\begin{align}
    (S_{L}^{\mu\nu}\kappa^\rho-\kappa^\rho S_R^{\mu\nu})^a{}_{\dot{b}}=&\frac{2p^\rho}{m^2}\la p^a|\Gamma^{\mu\nu}|p_{\dot{b}}]+2(\eta^{\nu\rho}{\kappa^\mu}^a{}_{\dot{b}}-\eta^{\mu\rho}{\kappa^\nu}^a{}_{\dot{b}})\nonumber\\
    =&2(P^{\nu\rho}\kappa^\mu-P^{\mu\rho}\kappa^\nu)^a{}_{\dot{b}},
\end{align}
where $P^{\mu\nu}$ is the projector on the space orthogonal to $p^\mu$. The lower-left block works in a similar manner. This checks that the operators obey:
\begin{equation}
    \left[\begin{pmatrix}
        S_L^{\mu\nu} & 0 \\
        0 & S_R^{\mu\nu}
    \end{pmatrix},\begin{pmatrix}
        0 & \kappa^\rho\\
        \bar{\kappa}^\rho & 0
    \end{pmatrix}\right]=2\left(P^{\nu\rho}\begin{pmatrix}
        0 & \kappa^\mu\\
        \bar{\kappa}^\mu & 0
    \end{pmatrix}-P^{\mu\rho}\begin{pmatrix}
        0 & \kappa^\nu\\
        \bar{\kappa}^\nu & 0
    \end{pmatrix}\right).
\end{equation}
When further normalized to $\mathbb{M}^{\mu\nu}\rightarrow\frac{\hbar}{2}\mathbb{M}^{\mu\nu}$, the operators obey then (\ref{LorAlg}). All the relations above were true for the fundamental spinor representations, but a similar proof can be used to extend to arbitrary spin representations. For operators acting on $n_L$ and $n_R$ copies of $V_{L,R}$, in the new normalization we have:
\begin{equation}
    S_L^{\mu\nu}S_{L,\mu\nu}=-n_L\left(n_L+2\right)\mathbb{1},\ S_R^{\mu\nu}S_{R,\mu\nu}=-n_R\left(n_R+2\right)\mathbb{1},\ \kappa^\mu\bar{\kappa}_\mu=-(n_L+n_R+n_Ln_R)\mathbb{1}.
\end{equation}
For the operators in eq.~\eqref{cohexpvalS}, we can adapt the proof above. The idea is that the operators are products of the form $\mathcal{O}_X=a^\dagger X a$, where $a,a^\dagger$ are some creation and annihilation operators satisfying the usual commutation relations and $X$, a matrix operator such as $s_L^{\mu\nu,a}{}_b$. Then schematically the commutators satisfy the following relation
\begin{equation}
    [\mathcal{O}_X,\mathcal{O}_Y]=\mathcal{O}_{[X,Y]}.
\end{equation}
Since we already know that the algebra in eq.~$\eqref{LorAlg}$ is obeyed, this means that the commutator of the operators in eq.~\eqref{cohexpvalS} also inherit the same algebra.

\section{Spinor Reduction to 4D}\label{app4Dred}
In this appendix, we will outline how the four-dimensional spinor variables can be generated from a dimensional reduction of the five-dimensional variables. Suppose we reduce over the $x^4$ dimension, the four-dimensional Gamma matrices are simply a subset $\mu=0,1,2,3$ of the five-dimensional ones defined in section~\ref{sec:5dSH} and can be written in terms of the Pauli matrices $(\sigma^{\mu})^{\alpha\dot{\alpha}} = (\mathbb{1},\sigma^i) = (\sigma_{\mu})_{\dot{\alpha}\alpha}$. 

The Dirac equation for the 5D Dirac spinor ${}_A|{p^{a}}\ra$ then reduces to two four-dimensional equations,
\begin{equation}
    p^{\alpha\dt}\tilde{\lambda}_{p \dt}^{a}=(m-ip_4)\,\lambda_p^{\alpha a}\,,\qquad p_{\dt \alpha}\lambda_p^{\alpha a}=(m+ip_4)\,\tilde{\lambda}_{p\dt}^{a}\,, \qquad \ _A|p^a\rangle=\begin{pmatrix}
        \lambda_p^{\alpha a}\\
        \tilde{\lambda}_{p\dt}^{a}\end{pmatrix}\,,
\end{equation}
where we have decomposed ${}_A|{p^{a}}\ra$ into its Weyl spinors. The Weyl spinors can be rescaled by $\lambda\rightarrow\lambda e^{-i\theta/2},\ \tilde{\lambda}\rightarrow\tilde{\lambda}e^{i\theta/2}$, with $\theta$ being the phase of the complex mass $m-ip_4=m_4e^{i\theta}$. We treat the other independent 5D Dirac spinor ${}_A|{p^{\dot{a}}}]$ in a similar manner. We define the Weyl spinors ${}_A|p^{\dot{a}}]=\begin{pmatrix} \chi_p^{\alpha \dot{a}} &\tilde{\chi}_{p\dt}^{\dot{a}} \end{pmatrix}^{T}$ such that we have the following identities 
\begin{align}
    p^{\al\dt}\tilde{\lambda}_{p\dt}^{a}&=m_4\lambda_p^{\al a} \,, \quad p_{\dt \al}\lambda_p^{\al a}=m_4\tilde{\lambda}_{p\dt}^{a}\,, \quad
    p^{\al\dt}\tilde{\chi}_{p\dt}^{\dot{a}}=-m_4\chi_p^{\al \dot{a}} \,, \quad m_4\lambda_p^{\al a}\chi_p^{\al \dot{a}}=-m_4\tilde{\chi}_{p\dt}^{\dot{a}} \,.
\end{align}

Besides reducing the spacetime groups we need to also reduce the massive five-dimensional little group $SU(2) \times SU(2)$ to the four-dimensional little group, which is just $SU(2)$. The two five-dimensional little groups reduce to the diagonal between the two $SU(2)_{L,R}$. This is achieved through a rotation matrix $U^I_a$, $U_{\dot{a}}^I$, where $I$ is an index for the four-dimensional massive little group. This transformation relates the $\lambda$ and $\chi$ spinors as $\lambda^{\al a} U_a^I=\chi^{\al \dot{a}} U_{\dot{a}}^I$. We will choose the matrices, 
\begin{equation}
    U^{I}_a =-\frac{i}{\sqrt{2}} \delta^{I}{}_a\,, \qquad U^{I}_{\dot{a}} =\frac{1}{\sqrt{2}} \delta^{I}{}_{\dot{a}}\,
\end{equation}
such that we can parameterise our four-dimensional spinors as a linear combination of these variables
\begin{align}
    {}^{\alpha}|p^{I}\ra_{4D} \tilde{z}_I &= \lambda_{p}^{\al a} z_a + \chi_{p}^{\al\dot{a}} w_{\dot{a}}\,, \qquad
    {}_{\dot{\alpha}}|p^{I}]_{4D}  \tilde{z}_I = \tilde{\lambda}_{p\dt}^{a} z_a - \tilde{\chi}_{p\dt}^{\dot{a}} w_{\dot{a}}\,,
\end{align}
where our choice of rotation matrices above requires the identification $\tilde{z}_{I} = \frac{1}{\sqrt{2}}(w_{\dot{a}} - i z_a)$. Note that the angle and square spinors are not independent in four-dimensions and satisfy $p_{\dt \alpha} {}^{\alpha}|p^{I}\ra_{4D} \tilde{z}_I  = m_4 {}_{\dot{\alpha}}|p^{I}]_{4D} $.

The reduction of the massless momentum, $k^{\mu}$, follows a similar logic and we parameterise the four-dimensional spinors as
\begin{equation}
    {}_{\alpha}|k\ra_{4D} = \lambda_{k, \alpha}^{I} \delta_{I 1}\,,\qquad |k]_{4D} = \tilde{\lambda}_k^{\dot{\alpha}I} \delta_{I 2}\,, \qquad \text{where } |k^I\ra = \begin{pmatrix}
         \lambda_{k, \alpha}^{I} \\ \tilde{\lambda}_k^{\dot{\alpha}I}
    \end{pmatrix}\,,
\end{equation}
where $k_4=0$ is a necessary constraint to ensure the momentum is null in four-dimensions.

\section{Multipole expansion in five dimensions}\label{sec:5dMultipoles}
In this appendix we complete the classical picture of section \ref{sec:5dCWL} with the analysis of the linearized five-dimensional spacetime metrics sourced by the stress-energy tensors from section \ref{sec:5dCWL}. Having a concrete parametrization of the linearised metric allows us to extract multipole data  from the known gravitational solutions. 

\subsection{General form}
The stress-energy tensor expressions $T^{\mu\nu}(\Vec{k})$ are related to linearised metric perturbations to Einstein's equations by a Fourier transform, which in five dimensions is:
\begin{equation}
    h_{\mu\nu}(\Vec{x})=\frac{\kappa}{2}\int\frac{\mathrm{d}^4\vec{k}}{(2\pi)^4}\frac{e^{-i\Vec{k}\cdot \Vec{x}}}{\vec{k}^2}P_{\mu\nu,\rho\tau}T^{\rho\tau}(\Vec{k}).
\end{equation}
Here $P_{\mu\nu,\rho\tau}=\Pi_{\mu(\rho}\Pi_{\sigma)\nu}-\frac{1}{3}\Pi_{\mu\nu}\Pi_{\rho\sigma}$  is the transverse projector with $\Pi_{\mu\nu}=\eta_{\mu\nu}-\frac{k_\mu k_\nu}{k^2}$. The multipole towers in $T^{\mu\nu}(\Vec{k})$ will map to a generalization of Thorne's ACMC formalism \cite{Thorne:1980ru}. If we restrict to a metric defined by the stress-energy tensor without any Levi-Civita factors \eqref{TAxiAns}, the general metric expression is \cite{Bianchi:2024shc}:
\begin{align}\label{acmcmetric}
    h_{00}&=\frac{8}{3\pi r^2}\sum_{l=0}^\infty \frac{Gm\rho(r)}{r^l} \left(M_{A_l}^{(l)}N_{A_l}+\mathcal{O}(N_{A_{l-1}})\right)+\mathcal{O}(G^2), \nonumber\\
    h_{0i}&=\frac{4}{\pi r^2}\sum_{l=1}^\infty \frac{Gm \rho(r)}{r^l}\left(J_{i,A_l}^{(l)}N_{A_l}+\mathcal{O}(N_{A_{l-1}})\right)+\mathcal{O}(G^2),\\
    h_{ij}&=\frac{8}{3\pi r^2}\sum_{l=2}^\infty\frac{Gm\rho(r)}{r^l}\left(\tilde{G}_{ij,A_l}^{(l)}N_{A_l}+\mathcal{O}(N_{A_{l-1}})\right)+\mathcal{O}(G^2), \nonumber
\end{align}
where $A_{l}=a_1a_2...a_l$, $N_{A_l}=\frac{x_{a_1}...x_{a_l}}{r^l}$ and $\tilde{G}_{ij,A_l}^{(l)}$ is related to the stress multipole as:
\begin{equation}
    G_{ij,A_l}^{(l)}=\tilde{G}_{ij,A_l}^{(l)}+\frac{1}{2}\delta_{ij}\left(M_{A_l}^{(l)}-\tilde{G}_{kk,A_l}^{(l)}\right).
\end{equation}
The tensors $M,\ J,\ G$ are the mass, current and stress multipole moments and are related to the $C_{n,i}$ coefficients as:
\begin{align}\label{multTowers}
    M_{A_{2l}}^{(2l)}&=\frac{(4l)!!}{2}\left(-4\frac{C_{l,3}}{(2l)!}+2\frac{C_{l,1}}{(2l)!}\right)(-S\cdot S)_{A_{2l}}|_{\text{STF}},\nonumber\\
    J_{i,A_{2l+1}}^{(2l+1)}&=\frac{(4l+2)!!}{(2l+1)!}C_{l,2}S_{i,a_1}(-S\cdot S)_{A_{2l}}|_{\text{ASTF}},\\
    G_{ij,A_{2l}}^{(2l)}&=-12\frac{(4l)!!}{(2l)!}C_{l,3}S_{i,a_1}S_{j,a_2}(-S\cdot S)_{A_{2l-2}}|_{\text{RSTF}}\nonumber.
\end{align}
Here STF means the symmetric trace-free part of the product of $(-S\cdot S)^{a_1a_2}$. Similarly, ASTF means symmetric in $a$s, antisymmetric in $(i,a_1)$, traceless with respect to all indices and with vanishing antisymmetrisation over any three indices. Finally, RSTF implies STF in $a$s, displaying the symmetries of the Riemann tensor in $(i,a_1,j,a_2)$, traceless in all indices and vanishing antisymmetrisation in any three indices. 
\par We see that the mass and stress multipoles are non-zero only for even indices and vice-versa for the current moments. This means that the metric \eqref{acmcmetric} has biaxial symmetry. This should not be a surprise as, using only the constrained $S^{\mu\nu}, k^{\mu}$ data, we cannot construct tensors that break the biaxial symmetry. We also observe that in the expansion \eqref{acmcmetric}, the multipole moments multiply the angular pieces of the highest degree. Lower angular pieces can also appear in the asymptotic expansion, but they can be shifted away by gauge transformations and we captured their contribution in $\mathcal{O}(N_{A_{l-1}})$. The expansion above did not rely on $T^{\mu\nu}(\Vec{k})$ being five-dimensional, and it can easily be extended to higher dimensions by modifying the prefactors in eq.~\eqref{acmcmetric} and the combinatorial factors in eq.~\eqref{multTowers}. We will see an example of this when matching a specific covariant expansion of an amplitude in arbitrary number of dimensions to Myers-Perry is $D>5$.
\par We can apply the same procedure to the extra towers coming from allowing $\tilde{S}^{\mu\nu}$ in eq.~\eqref{THodgeAns}). They lead to two more towers of current and stress multipoles:
\begin{align}
    \tilde{J}_{i,A_{2l+1}}^{(2l+1)}&=\frac{(4l+2)!!}{(2l+1)!}\tilde{C}_{l,2}\tilde{S}_{i,a_1}(-S\cdot S)_{A_{2l}}|_{ASTF},\nonumber\\
    \tilde{G}_{ij,A_{2l}}^{(2l)}&=-12\frac{(4l)!!}{(2l)!}\tilde{C}_{l,3}\tilde{S}_{i,a_1}S_{j,a_2}(-S\cdot S)_{A_{2l-2}}|_{RSTF}.
\end{align}

\subsection{Black ring multipole moments}\label{BRapp}
In this appendix we introduce the black ring multipole structure. As discussed in the main text, the black ring is an asymptotically flat black hole in five dimensions that has horizon topology $S^2\times S^1$. Similarly to the MP black hole, it is characterized by its mass and two angular momenta. For simplicity we will focus on the case with a single angular momentum. The metric is given by:
\begin{equation}
    \begin{aligned}
d s^{2}= & -\frac{A(y)}{A(x)}\left(d t-C \mathcal{R} \frac{1+y}{A(y)} d \varphi_{1}\right)^{2} \\
& +\frac{\mathcal{R}^{2}}{(x-y)^{2}} A(x)\left(-\frac{B(y)}{A(y)} d \varphi_{1}^{2}-\frac{d y^{2}}{B(y)}+\frac{d x^{2}}{B(x)}+\frac{B(x)}{A(x)} d \varphi_{2}^{2}\right),
\end{aligned}
\end{equation}
with
\begin{equation}
    A(x)=1+\lambda x,\ B(x)=(1-x^2)(1+\nu x),\ C=\sqrt{\lambda(\lambda-\nu)\frac{1+\lambda}{1-\lambda}}.
\end{equation}
The free parameters are $\mathcal{R}$ and $\lambda, \nu$, being constrained by $0<\nu\leq \lambda<1$. The angular momentum is in the $\varphi_1$ direction. One can generalise the construction to include two angular momenta \cite{Pomeransky:2006bd}. In order to avoid conical singularities, we require that the gravitational self-attraction of the black hole is compensated by its centrifugal force and impose the equilibrium condition:
\begin{equation}
    \lambda=\frac{2\nu}{1+\nu^2}.
\end{equation}
The parameter $\nu$ characterizes the ratio between the $S^2$ and $S^1$ describing the horizon. In order to perform the expansion and multipole extraction procedure, we need to rewrite the metric in a more suitable set of coordinates. This is achieved by the transformation \cite{Heynen:2023sin}:
\begin{equation}
    x=-\frac{r^2-2\mathcal{R}^2\frac{1-\lambda}{1-\nu}\cos^2\theta}{r^2},\ y=-\frac{r^2+2\mathcal{R}^2\frac{1-\lambda}{1-\nu}\sin^2\theta}{r^2},\ (\phi_1,\phi_2)=\frac{1-\nu}{\sqrt{1-\lambda}}(\varphi_1,\varphi_2).
\end{equation}
At equilibrium, the physical mass and angular momentum are given by:
\begin{equation}
    M=\frac{3\pi}{2}\frac{\mathcal{R}^2\nu}{(1+\nu^2)(1-\nu)},\ J=\frac{\pi\mathcal{R}^3\nu}{2}\sqrt{\frac{2(1+\nu)^3}{(1-\nu)^3(1+\nu^2)^3}}.
\end{equation}
The computation of the multipole moments follows standard methods outlined in \cite{Heynen:2023sin}. Unlike Myers-Perry the form factors will not be merely numbers, but turn out to be rational functions depending on $\nu$. Below we computed the form factor coefficients up to order $S^{12}$. In the present form, a resummation pattern is not obvious. The $C_{i,2}$ coefficients are:
{\small
\begin{align}
      & C_{1,2}=-\frac{3!}{2}\frac{9(1+3\nu^2)}{16(1+\nu)^3}\quad C_{2,2}=\frac{5!}{2}\frac{27(5 - \nu + 33 \nu^2 - 3 \nu^3 + 46 \nu^4)}{1280(1+\nu)^6}\nonumber \\& C_{3,2}=-\frac{7!}{2}\frac{81(175 - 90 \nu + 1869 \nu^2 - 612 \nu^3 + 5697 \nu^4 - 
 834 \nu^5 + 4995 \nu^6)}{1433600 (1 + \nu)^9}\nonumber\\&
 C_{4,2}=\frac{9!}{2}\frac{243(735 - 665 \nu + 11183 \nu^2 - 7289 \nu^3 + 55133 \nu^4 - 
 22275 \nu^5 + 104957 \nu^6 - 19179 \nu^7 + 65560 \nu^8)}{321126400 (1 + \nu)^{12}}\nonumber \\
 &C_{5,2}=-\frac{11!}{2}\frac{\splitfrac{243(24255 - 32760 \nu + 490125 \nu^2 - 505872 \nu^3 + 
 3446942 \nu^4 - 2504320 \nu^5 + 10584290 \nu^6 - 
 4767920 \nu^7}{ + 14263987 \nu^8 - 2956232 \nu^9 + 
 6794625 \nu^{10})}}{282591232000 (1 + \nu)^{15}}\nonumber\\
 &C_{6,2}=\frac{13!}{2}\frac{\splitfrac{2187 (825825 - 1518825 \nu + 21197325 \nu^2 - 30707845 \nu^3 + 
   198320306 \nu^4 - 215313826 \nu^5 + 867782450 \nu^6}{ - 
   662136770 \nu^7 + 1880664685 \nu^8 - 898185461 \nu^9 + 
   1922995521 \nu^{10} - 430537785 \nu^{11} + 729193600 \nu^{12})}}{3232843694080000 (1 + \nu)^{18}}
\end{align}}
The $C_{i,3}$ coefficients are:
\begin{align}
    &C_{1,3}=\frac{2!}{4}\frac{3\nu(3+\nu^2)}{8(1+\nu)^3},\quad C_{2,3}=-\frac{4!}{4}\frac{9\nu(18 - 9 \nu + 59 \nu^2 - 3 \nu^3 + 15 \nu^4)}{320(1+\nu)^6}\nonumber\\
    &C_{3,3}=\frac{6!}{4}\frac{81\nu(36 - 39 \nu + 254 \nu^2 - 128 \nu^3 + 390 \nu^4 - 
 33 \nu^5 + 80 \nu^6)}{35840(1+\nu)^9}\nonumber\\
 &C_{4,3}=-\frac{8!}{4}\frac{81\nu(480 - 825 \nu + 5573 \nu^2 - 5629 \nu^3 + 17825 \nu^4 - 
 8891 \nu^5 + 17263 \nu^6 - 1871 \nu^7 + 2955 \nu^8)}{5734400 (1 + \nu)^{12}}\nonumber\\
 &C_{5,3}=\frac{10!}{4}\frac{\splitfrac{243 \nu (100800 - 240975 \nu + 1705882 \nu^2 - 
   2615982 \nu^3 + 8870190 \nu^4 }{- 8353460 \nu^5 + 
   17444850 \nu^6 - 8528082 \nu^7 + 12066922 \nu^8 - 
   1515645 \nu^9 + 1763100 \nu^{10})}}{70647808000 (1 + \nu)^{15}}
\end{align}
The $C_{i,1}$ coefficients are:
\begin{align}
    &C_{1,1}=2!\frac{3(-6 + 3 \nu - 18 \nu^2 + \nu^3)}{16(1+\nu)^3},\quad C_{2,1}=-4!\frac{9(-45 + 51 \nu - 333 \nu^2 + 163 \nu^3 - 426 \nu^4 + 30 \nu^5)}{1280(1+\nu)^6}\nonumber\\
    &C_{3,1}=6!\frac{81(-175 + 320 \nu - 2246 \nu^2 + 2278 \nu^3 - 6933 \nu^4 + 
 3266 \nu^5 - 5310 \nu^6 + 400 \nu^7)}{358400(1+\nu)^9}\nonumber\\
 & C_{4,1}=-8!\frac{\splitfrac{81(-11025 + 28245 \nu - 211713 \nu^2 + 332093 \nu^3 - 
 1126975 \nu^4 + 1040315 \nu^5}{ - 2045915 \nu^6 + 
 922303 \nu^7 - 1079828 \nu^8 + 82740 \nu^9)}}{321126400 (1 + \nu)^{12}}\nonumber\\
 & C_{5,1}=10!\frac{\splitfrac{243 (-72765 + 241710 \nu - 1925790 \nu^2 + 4102342 \nu^3 - 
   15297772 \nu^4 + 21125510 \nu^5}{ - 47542830 \nu^6 + 
   40332550 \nu^7 - 58748087 \nu^8 + 25299812 \nu^9 - 
   23091540 \nu^{10} + 1763100 \nu^{11})}}{141295616000 (1 + \nu)^{15}}
\end{align}
The black ring reduces to the Myers-Perry black hole with one angular momentum in the limit $\mathcal{R}\rightarrow 0$, $\nu\rightarrow 1$ with the following identifications for the mass and angular momentum:
\begin{equation}
    M=\frac{2\mathcal{R}^2}{1-\nu},\ a^2=2\mathcal{R}^2\frac{\nu(1+\nu)}{(1-\nu)(1+\nu^2)}.
\end{equation}
It is straightforward to check that the numerical expressions for the form factors reduce to the Myers-Perry ones in the limit. 

\section{Covariant expansion maps}
\label{Appcov}

The map between the covariant expansion blocks and the expectation value ones in arbitrary number of dimensions is:
\begin{align}
    \langle (k {\cdot} S {\cdot} S {\cdot} k)^n \rangle &= (-1)^{n} \frac{S!}{(S-n)!}\frac{(2S+ D -5)!!}{(2S+ D - 5 - 2n)!!} (\bm{\varepsilon}_1 {\cdot} k \bar{\bm{\varepsilon}}_1 {\cdot} k)^{n} (\bar{\bm{\varepsilon}}_1 {\cdot}\bm{\varepsilon}_1)^{S-n}\nonumber \\
    \langle (\varepsilon_{3} {\cdot} S {\cdot} k) \, (k {\cdot} S {\cdot} S {\cdot} k)^n \rangle &= (-1)^{n+1} \frac{S!}{(S-n-1)!}\frac{(2S+ D -5)!!}{(2S+ D - 5 - 2n)!!} (\bar{\bm{\varepsilon}}_1 {\cdot} f_3 {\cdot} \bm{\varepsilon}_1 )(\bm{\varepsilon}_1 {\cdot} k \bar{\bm{\varepsilon}}_1 {\cdot} k)^{n} (\bar{\bm{\varepsilon}}_1 {\cdot}\bm{\varepsilon}_1)^{S-n-1}\nonumber \\
    \langle (\varepsilon_{3} {\cdot} S {\cdot} k)^2 \, (k {\cdot} S {\cdot} S {\cdot} k)^n \rangle &= (-1)^{n} \frac{S!}{(S-n-2)!}\frac{(2S+ D-5)!!}{(2S+ D - 5 - 2n)!!} (\bar{\bm{\varepsilon}}_1 {\cdot} f_3 {\cdot} \bm{\varepsilon}_1 )^2(\bm{\varepsilon}_1 {\cdot} k \bar{\bm{\varepsilon}}_1 {\cdot} k)^{n} (\bar{\bm{\varepsilon}}_1 {\cdot}\bm{\varepsilon}_1)^{S-n-2}\nonumber \\
    &-(-1)^{n} \frac{S!}{(S-n-1)!}\frac{(2S+ D -5)!!}{(2S+ D - 5 - 2n)!!} (p_1 {\cdot} \varepsilon_{3})^2 (\bm{\varepsilon}_1 {\cdot} k \bar{\bm{\varepsilon}}_1 {\cdot} k)^{n} (\bar{\bm{\varepsilon}}_1 {\cdot}\bm{\varepsilon}_1)^{S-n} .
\end{align}
In five dimensions, we can write an explicit map between the spinor-helicity blocks and the covariant ones. The identities needed for that are:
\begin{align}
    {\bm \varepsilon}_1 \cdot {\bm \varepsilon}_2 &=\frac{1}{4m^2}( -\langle \mathbf{1} \mathbf{2} ] [\mathbf{1} \mathbf{2} \rangle + \langle \mathbf{1} \mathbf{2} \rangle [\mathbf{1} \mathbf{2} ]),\ 
    {\bm \varepsilon}_1\cdot k\ {\bm \varepsilon}_2\cdot k=-2[\mbf{12}]\langle \mbf{12}\rangle,\nonumber\\
     \bm{\varepsilon}_1\cdot f_3\cdot \bm{\varepsilon}_2 &=\frac{1}{2m^2} (\langle \mbf{12}][\mbf{1}3\rangle\langle \mbf{2}3\rangle+[ \mbf{12}\rangle[\mbf{2}3\rangle\langle \mbf{1}3\rangle)-\frac{1}{4m^2}(p_1\cdot \varepsilon_{3} )\langle \mbf{12}\rangle [\mbf{12}],\nonumber\\
    \epsilon^{\mu\nu\rho\tau\sigma}\bm{\varepsilon}_{1\mu}\bm{\varepsilon}_{2,\nu} p_{1,\rho} f_{3,\tau\sigma}&=\frac{1}{2m}\left([\mbf{1}3\rangle\langle \mbf{2}3\rangle\langle\mbf{12}]-\langle \mbf{1}3\rangle[\mbf{2}3\rangle[\mbf{12}\rangle\right).
\end{align}
Even though not explicitly considered in the bulk of the paper, we can extend the maps to tensor polarisations too:
\begin{align}\label{CovMapSD}
    \text{Tr}[{\bm\zeta}^{+}_{1}{\bm\zeta}^{-}_{2}]=-\frac{1}{4m^2}\langle \mbf{12}]^2\,,\quad k\cdot {\bm\zeta}^{+}_{1}\!\!\cdot {\bm\zeta}^{-}_{2}\!\!\cdot k=0\,,\quad \text{Tr}[f_{3}{\bm\zeta}^{+}_{1}{\bm\zeta}^{-}_{2}]=\frac{1}{4m^2}\langle \mbf{12}]\langle \mbf{1}3\rangle[ \mbf{2}3\rangle.
\end{align}

\bibliographystyle{JHEP}
\bibliography{references}

\end{document}